\documentclass[a4paper]{article}

\usepackage[table]{xcolor} 
\usepackage{setspace} 
\usepackage{amsfonts}
\usepackage{verbatim}
\usepackage{graphicx}
\usepackage{lscape} 
\usepackage{subfigure}
\usepackage{multirow}

\usepackage{amsfonts, amsmath, hanging, hyperref, parskip, times}
\usepackage[centering]{geometry} 

\newtheorem{proposicion}{Proposition}[section]

\newtheorem{definition}{Definition}[section]
\newtheorem{example}{Example}[section]
\newtheorem{property}{Property}[section]

\hypersetup{
  colorlinks,
  linkcolor=blue,
  urlcolor=blue
}

\renewcommand{\title}[1]{\begin{center}{\bf \LARGE #1}\end{center}}

\newcommand{\keywords}{\paragraph{Keywords:}}

\begin{document}
\pagestyle{plain}  

\begin{doublespace}
\title{\textit{Distribution and Symmetric Distribution} \\[8pt] Regression Model for Histogram-Valued Variables}
\end{doublespace}

\begin{center}
  {\bf S\'{o}nia Dias$^{1,^\star}$, Paula Brito$^{2}$}
\end{center}

\begin{affiliations}
\noindent
1. Escola Superior de Tecnologia e Gest\~{a}o do Instituto Polit\'{e}cnico Viana do Castelo \&  LIAAD-INESC Porto LA, Universidade do Porto, Portugal \\[-2pt]
2. Faculdade de Economia \& LIAAD-INESC Porto LA, Universidade do Porto, Portugal \\[-2pt]
$^\star$Contact author: \href{mailto:foo@bar.com}{sdias@estg.ipvc.pt}
\end{affiliations}

\begin{abstract}

Histogram-valued variables are a particular kind of variables studied in \textit{Symbolic Data Analysis} where to each
entity under analysis corresponds a distribution that may be represented by a histogram or by a quantile function.
Linear regression models for this type of data are necessarily more complex than a simple generalization of the classical model: the parameters cannot be negative still the linear relationship between the variables must be allowed to be either direct or inverse.
In this work we propose a new linear regression model for histogram-valued variables that solves this  problem, named \textit{Distribution and Symmetric Distribution Regression Model}. To determine the parameters of this model it is necessary to solve a quadratic optimization problem, subject to non-negativity constraints on the unknowns; the error measure between the predicted and observed distributions uses the Mallows distance. As in classical analysis, the model is associated with a goodness-of-fit measure whose values range between 0 and 1. Using the proposed model, applications with real and simulated data are presented.

\end{abstract}

\keywords data with variability; linear regression; Symbolic Data Analysis; quantile functions;\\ Mallows distance.

\section{Introduction}\label{s1.1}

Classical multivariate statistics studies data tables that summarize observations made on ``statistical units" (individuals); each row of the table represents one individual and each of these individuals is characterized by different variables (in columns). The ``values" attained by the variables may be real values if the variable represents the measurement of a quantity (quantitative variables) or a category if the variable is qualitative. As an example, let us have classical quantitative variables such as the age, weight and height of a particular football player. The observations of these data are typically represented in classical data tables, but how can we represent the result of the weight of the football player if we don't know his exact weight? And what if we are interested in studying the age, weight and height not of one single player but of a football team?  In the first situation, the individuals are described by attributes whose associated values are quantitative values that cannot be ``measured" with precision. In cases like this, we are in the presence of imprecise data. In the second situation we are interested in describing one class of individuals. The ``best values" attained by the variables that characterize each class are not real values or categories but sets of ``values", intervals or distributions.
Even though data with variability or uncertainty may be represented by the same type of elements, the meaning of these elements is different. For example, the interval $[80,82]$ may mean that the weight of one football player is between $80$ and $82$ Kg. On the other hand, the interval $[75,80]$ may represent the weights of all players from a given football team. In the first situation the interval represents the imprecision of the weight value, whereas in the second situation the interval considers the variability of weight values in the football team.

In this research we will focus on situations where variability in data description occurs. The classical solution to analyze these data is to reduce the collection of records associated to each individual or class of individuals to one value, this may be the mean, mode or maximum/minimum; however, with this option the variability across the records is lost.
In alternative to the classical analysis applied to these kind of data, Diday \cite{diday88} introduced \textit{Symbolic Data Analysis}, where the term \textit{symbolic data} refers precisely to data with variability.
To understand the concept of symbolic data it is important to assess where variability comes from.
The variability of the data might emerge due to the aggregation of observations \cite{tesear08} that can be contemporary, if the records are collected in the same temporal instant or the temporal instant is not relevant, and temporal if the time is the aggregation criterion and the records are grouped along one unit of time, for example one day. In both situations, the initial data or micro-data, are organized in classical data tables where each individual, termed first-level unit, is described by classical variables. Depending on the type of aggregation, the construction of the symbolic data table is different.
When the aggregation is temporal, the entities under analysis are the original first-level units, now characterized by sets of values originating from the records collected over a unit of time. In situations where the aggregation is contemporary, the entities -  higher-level units - are classes of individuals (sets of first-level units) grouped according to specific characteristics. In this situation, the variables describing both the higher-level and the respective first-level units are the same; however the ``values" that the variables take for each higher-level unit are now sets of values or functions obtained from the respective first-level units.

Following the definition of Bock and Diday \cite{bodi00}, a symbolic variable $Y$ is a mapping $Y: E \rightarrow \mathbb{B}$ defined on a set $E$ of statistical entities ($E=\Omega=\left\{1,2,\ldots,m\right\}$ when the individuals are first-level units or $E=\left\{C_{1},C_{2},\ldots \right\}$ with $C_{j}\subseteq \Omega$ when the individuals are higher-level units) and which takes its values in a set $\mathbb{B}$.
Henceforth in this work, when we use the term unit, we will be referring to a first-level unit or to a higher-level unit, according to the kind of prior aggregation of the micro-data used to build the symbolic data table.

Similarly to the classical case, symbolic variables can also be classified as quantitative or qualitative, according to the nature of the elements of $\mathbb{B}$. For quantitative symbolic variables, each unit is allowed to take a single value (single-valued variables); a finite set of values (multi-valued variables); an interval (interval-valued variables); or a mapping that can be a probability/frequency/weight distribution (modal-valued variables).

In this paper, we will be dealing with a particular type of modal-valued variables, the \textit{histogram-valued variables}. In this case, the values attained by the variable for each unit are empirical frequency distributions or, more specifically, histograms, where the values in each subinterval are assumed to be uniformly distributed. If we consider a symbolic variable where all units are associated to one only interval of real numbers (uniformly distributed) with probability/frequency/weight equal to one, then we are in the presence of \textit{interval-valued variables}.

As an example, consider a symbolic data table containing information about patients (adults) attending healthcare centers, during a fixed period of time. In healthcare centre A, the age of patients ranged from 25 to 53 years old, in healthcare centre B, it ranged from 33 to 68 years old and in healthcare centre C, the age of patients ranged from 20 to 75 years old, so that the age is an interval-valued variable. Now consider another variable which records the waiting time for consultations. In this case, information is recorded for 5 time lengths (0 to 15 minutes, 15 to 30 minutes,...), and the corresponding symbolic variable is therefore a histogram-valued variable (see Table \ref{HCcentres}). Notice that in this example the entities under analysis are the healthcare centers (higher-level units), for each of which we have aggregated information (contemporary aggregation), and NOT the individual patients attending each centre (first-level units).

\renewcommand{\arraystretch}{1.5}
\begin{table}[h!]
  \centering
\begin{tabular}{|c|c|c|} \hline
Healthcare centers & Age & Waiting Time (minutes) \\ \hline
A & $\left[25,53\right]$  & $\left\{\left[0,15\right[,0; \left[15,30\right[, 0.25; \left[30,45\right[, 0.5; \left[45,60\right[,0; \ge 60, 0.25 \right\}$ \\ \hline
B & $\left[33,68\right]$ & $\left\{\left[0,15\right[,0.25; \left[15,30\right[, 0.25; \left[30,45\right[, 0.25; \left[45,60\right[,0.25; \ge 60, 0 \right\}$ \\ \hline
C & $\left[20,75\right]$  &  $\left\{\left[0,15\right[,0.33; \left[15,30\right[, 0; \left[30,45\right[, 0.33; \left[45,60\right[,0; \ge 60, 0.33 \right\}$ \\ \hline
  \end{tabular}
 \caption{Data for three healthcare centers.}\label{HCcentres}
 \end{table}

Symbolic Data Analysis has achieved considerable development since the eighties of last century (see, for instance, \cite{bidi07}, \cite{bidi03}, \cite{bodi00}, \cite{dinoir08}, \cite{noirbri11}). Recently, there has been a growing interest in the analysis of histogram-valued variables, though still more research is developed for interval-valued variables. The methods proposed so far for the former are indeed, frequently, a generalization of their counterparts for the latter.
The main definitions of descriptive statistics for one, two or more histogram-valued variables have already been studied.
Billard and Diday \cite{bidi03} defined mean, observed and relative frequency, empirical density function, empirical joint density function; for variance and covariance two definitions were proposed \cite{bidi07}; \cite{bidi03}; \cite{bidi02}; Irpino and Verde \cite{irve06} defined distribution functions and joint distribution functions.

The first definitions and methods for histogram-valued variables are generally obtained from the application of the classic concepts to the midpoints of the histograms' subintervals, using the respective weights. Furthermore, although the symbolic variables' values are distributions and not real numbers, the results of the application of these concepts are real numbers. For example, the mean of $m$ observations of the histogram-valued variable, proposed by Billard and Diday \cite{bidi03}, is a real number. It should be noticed, however, that in recent years other works have been put forward where the ``results" are already distributions.
For example, Irpino and Verde \cite{irve06} present an alternative definition of mean for histogram-valued variables, which produces a mean distribution, that they termed by \textit{barycentric histogram}.

Work with histogram-valued variables has been recently reported in different domains, such as Principal Component Analysis \cite{ro00}, \cite{ropa04}; Cluster Analysis \cite{irve06}; Time series \cite{arma09} and Linear Regression \cite{bidi02}, \cite{irve10}.

The first linear regression model for histogram-valued variables was a generalization of the first model proposed for interval-valued variables by Billard and Diday \cite{bidi07}, \cite{bidi00}. Other models have also been proposed for interval-valued variables \cite{netocar10}, \cite{netocar08}; however, these models present some limitations: firstly, they are based on differences between real values and do not appropriately quantify the closeness between intervals; then, the elements predicted by the models may fail to build an interval; the most recent model imposes non-negativity constraints on the coefficients, therefore forcing a direct linear relationship. These limitations prevent a generalization of the models to histogram-valued variables, so that alternative models are being developed (see, e.g., \cite{irve12}, \cite{irve10}). Our goal is to propose a linear regression model for histogram-valued variables allowing predicting distributions from other distributions, without forcing a direct linear relationship.

The development of non-descriptive methods for Symbolic Data Analysis is still an open research topic for almost all kinds of symbolic variables. Notice, however, papers recently published proposing probabilistic models for interval-valued variables \cite{brsi11},\cite{netocar11}.

The remaining of the paper is organized as follows. Section 2 introduces histogram-valued variables in more detail, and presents a short study about the space of the quantile functions. In Section 3, the problem of defining a linear regression model for histogram-valued variables is addressed. A model and a respective goodness-of-fit measure are proposed. Section 4 reports results of a simulation study and two examples that illustrate the application of the model. Finally, Section 5 concludes the paper, pointing out directions for future research.

\section{Symbolic Data Analysis: histogram data}\label{s2}
\subsection{Histogram-valued variables} \label{ss2.1}

Consider a symbolic variable $Y: E \rightarrow \mathbb{B}.$ The set of units $E$ may be $E=\Omega=\left\{1,2,\ldots,m\right\}$ when the individuals are first-level units or $E=\left\{C_{1},C_{2},\ldots \right\}$ with $C_{j}\subseteq \Omega$ when the individuals are higher-level units. Consider also the quantitative (single-value) variable $\dot{Y}$ defined on a set $\Omega.$ If the aggregation of the observations is temporal, to each unit $j \in \Omega$ corresponds the empirical distribution of the values that $\dot{Y}$ takes within a certain unit of time. If the aggregation is contemporary, to each unit $j$ corresponds the empirical distribution of $\dot{Y}$ in $C_{j}.$ As histograms are a usual representation of empirical distributions, this kind of symbolic variables are termed \textit{histogram-valued variables}. More generally we can define histogram-valued variables as follows:

\begin{definition}\label{def2.1}
 $Y$ is a histogram-valued variable when to each unit $j$ corresponds a empirical distribution $Y(j),$ that can be represented by a histogram \cite{bodi00}, \cite{bidi03}:
\begin{equation}\label{eq2.1}
H_{Y(j)}=\left\{\left[\underline{I}_{Y(j)_1},\overline{I}_{Y(j)_1}\right[,p_{j1};\left[\underline{I}_{Y(j)_2},\overline{I}_{Y(j)_2}\right[,p_{j2};\ldots;
\left[\underline{I}_{Y(j){n_{j}}},\overline{I}_{Y(j){n_{j}}}\right],p_{jn_{j}}\right\}
\end{equation}

where $\underline{I}_{Y(j)_i}$ and $\overline{I}_{Y(j)_i}$ represent the lower and upper bound of the interval $i;$ $p_{ji}$ is the frequency associated to the subinterval $\left[\underline{I}_{Y(j)_{i}},\overline{I}_{Y(j)_{i}}\right[$ with $i \in \left\{1,2,\ldots,n_{j}\right\},$ $n_{j}$  is the number of subintervals for the $j^{th}$ unit, $j=1,\ldots,m$ $\displaystyle\sum\limits_{i=1}^{n_{j}} p_{ij}=1,$ $\underline{I}_{Y(j)_i} \leq \overline{I}_{Y(j)_{i}}$ and $\overline{I}_{Y(j)_i} \leq \underline{I}_{Y(j)_{i+1}}.$

Alternatively, $Y(j)$ can be represented by the inverse of the cumulative empirical distribution function, also called quantile function $\Psi_{Y(j)}^{-1}$ \cite{irve06}:

\begin{center}
\begin{equation}\label{eq2.3}
\Psi_{Y(j)}^{-1}(t)=\left\{\begin{array}{lll}
                     \underline{I}_{Y(j)_{1}}+\frac{t}{w_{j1}} a_{Y(j)_{1}} & if & 0 \leq t < w_{j1} \\
                    \underline{I}_{Y(j)_{2}} +\frac{t-w_{j1}}{w_{j2}-w_{j1}} a_{Y(j)_{2}} & if & w_{j1} \leq t < w_{j2} \\
                     \vdots & & \\
                   \underline{I}_{Y(j)_{n_{j}}} +\frac{t-w_{jn_{j}-1}}{1-w_{jn_{j}-1}} a_{Y(j)_{n_j}} & if & w_{jn_{j}-1} \leq t \leq
                   1
 \end{array}
  \right.
\end{equation}
\end{center}

where $\quad w_{jl}=\left\{\begin{array}{ccc}
                  0 & if & l=0 \\
                 \displaystyle \sum_{h=1}^{l} p_{jh} & if & l=1,\ldots,n_{j}\\
                \end{array}
              \right.$ and $\quad a_{Y(j)_i}=\overline{I}_{Y(j)_i}-\underline{I}_{Y(j)_i}$ with $i\in\{1,\ldots,n_{j}\}$; \\[5pt] $n_{j}$ is the number of subintervals in $Y(j).$

Or, considering the subintervals of the histograms defined by the centers $c_{Y(j)_{i}}$ and half-ranges $r_{Y(j)_{i}},$ the representation of the $Y(j)$ can be given by
\begin{equation}\label{eq2.1B}
H_{Y(j)}=\left\{\left[c_{Y(j)_1}-r_{Y(j)_1},c_{Y(j)_1}+r_{Y(j)_1}\right[,p_{j1};\ldots;
\left[c_{Y(j){n_{j}}}-r_{Y(j){n_{j}}},c_{Y(j){n_{j}}}+r_{Y(j){n_{j}}}\right],p_{jn_{j}}\right\}
\end{equation}
or
\begin{center}
\begin{equation}\label{eq2.3B}
\Psi_{Y(j)}^{-1}(t)=\left\{\begin{array}{lll}
                     c_{Y(j)_{1}}+\left(2\frac{t}{w_{j1}}-1\right) r_{Y(j)_{1}} & if & 0 \leq t < w_{j1} \\
                    c_{Y(j)_{2}} +\left(2\frac{t-w_{j1}}{w_{j2}-w_{j1}}-1\right) r_{Y(j)_{2}} & if & w_{j1} \leq t < w_{j2} \\
                     \vdots & & \\
                   c_{Y(j)_{n_{j}}} +\left(2\frac{t-w_{jn_{j}-1}}{1-w_{jn_{j}-1}}-1\right) r_{Y(j)_{n_j}} & if & w_{jn_{j}-1} \leq t \leq
                   1
 \end{array}
  \right.
\end{equation}
\end{center}

Any of these representations of the empirical distribution that each unit takes can be termed \textit{histogram value}.
Henceforth, when we use the term distribution, we are referring to an empirical distribution of a continuous variable. Furthermore, it is also assumed that within each subinterval $\left[\underline{I}_{Y(j)_{i}},\overline{I}_{Y(j)_{i}}\right[$ the values for the variable $Y$ for each unit $j=1,\ldots,m$, are uniformly distributed.

If any of the weights $p_{ji}$ with $i>1$ is nullo, the function $\Psi_{Y(j)}$  doesn't have inverse with domain between 0 and 1. Consequently the function $\Psi_{Y(j)}^{-1}$ is not continuous and has $n_{j}-1$ pieces. In this case it is not possible to calculate the value of $\Psi_{Y(j)}^{-1}(w_{ji-1})$ but only $\lim_{t\rightarrow w_{ji-1}^{-}}\Psi_{Y(j)}^{-1}(t)$ and $\lim_{t\rightarrow w_{ji-1}^{+}}\Psi_{Y(j)}^{-1}(t).$
\end{definition}

When $n_{j}=1$ and for each unit $j,$ $Y(j)$ takes values only on the interval $\left[\underline{I}_{Y(j)},\overline{I}_{Y(j)}\right[$  with frequency $p_{j}=1,$ the histogram-valued variable is then reduced to the particular case of an interval-valued variable. In this case, the quantile function is given by

\begin{center}
\begin{equation}\label{eq2.4}
\Psi_{Y(j)}^{-1}(t)=\begin{array}{lll}
                    \underline{I}_{Y(j)}+ \left(\overline{I}_{Y(j)}- \underline{I}_{Y(j)}\right)t , & with & 0 \leq t \leq 1 .
\end{array}
\end{equation}
\end{center}

When we work with histogram-valued variables, it is important to note that for different observations, the number of subintervals in the histograms or the pieces in functions may be different; the subintervals of histograms $H_{Y(j)}$ are considered ordered and disjoint, and if this is not the case, it must be possible to rewrite them in the required form \cite{tesewi89}, \cite{tesear08}.

\begin{example} \label{ex1}
Consider the histograms $$H_{X}=\left\{\left[1,3\right[, 0.1;\left[3,5\right [, 0.6; \left[5,8\right], 0.3\right\} $$ and $$H_{Y}=\left\{\left[0,1\right[, 0.8;\left[1,4\right], 0.2\right\} $$ that caracterize an unit for the histogram-valued variables $X$ and $Y,$ respectively. These histograms are represented in \textit{Figure \ref{fig1}}:

\begin{figure}[h!] \label{fig1}
  \centering
    \includegraphics[width=0.9\textwidth]{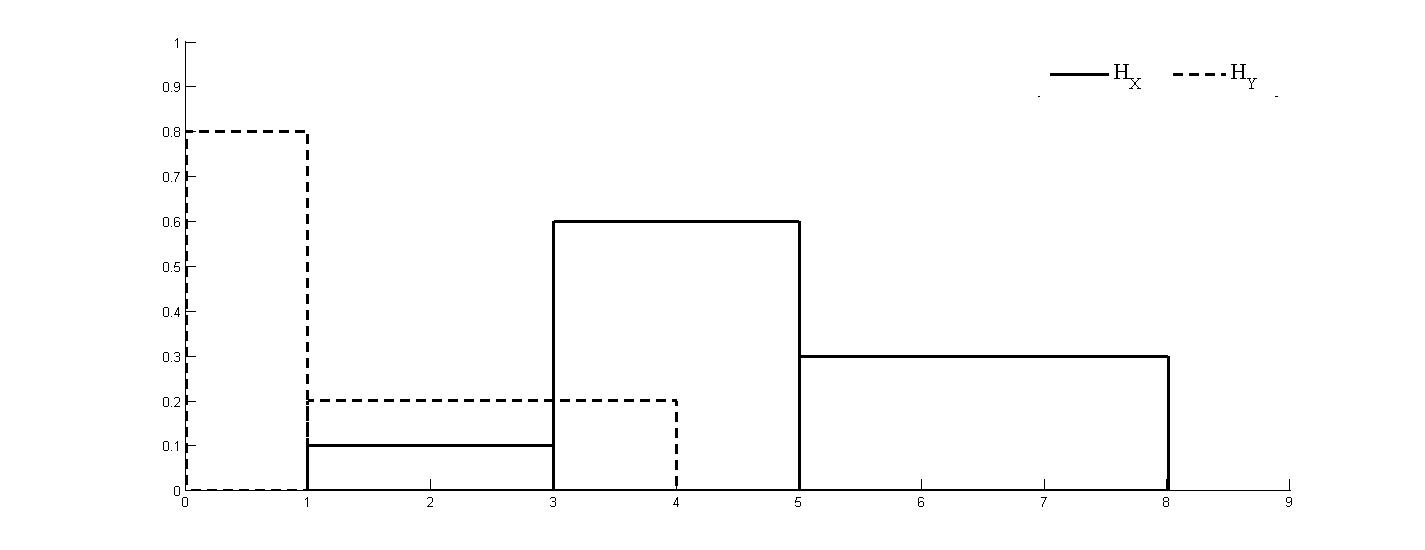}
      \caption{Representation of the histograms $H_{X}$ and $H_{Y}$ in Example \ref{ex1}.}\label{fig1}
\end{figure}

Alternatively, these histograms can be represented by their quantile functions (see \textit{Figure \ref{fig2}}):
\small{
$$
\Psi_{X}^{-1}(t)=\left\{\begin{array}{lll}
                     1+\frac{t}{0.1}\times 2  & \quad if \quad  & 0 \leq t <0.1 \\
                    3 +\frac{t-0.1}{0.6} \times 2 &  \quad if \quad & 0.1 \leq t < 0.7 \\
                   5 +\frac{t-0.7}{0.3} \times 3 & \quad if \quad & 0.7 \leq t \leq 1
 \end{array}
  \right.
\qquad
\Psi_{Y}^{-1}(t)=\left\{\begin{array}{lll}
                     \frac{t}{0.8}  & \quad if \quad & 0 \leq t <0.8 \\
                    1 +\frac{t-0.8}{0.2}\times 3  & \quad if \quad & 0.8 \leq t \leq 1
 \end{array}
  \right.
$$}

\begin{figure}[h!] \label{fig2}
  \centering
    \includegraphics[width=0.9\textwidth]{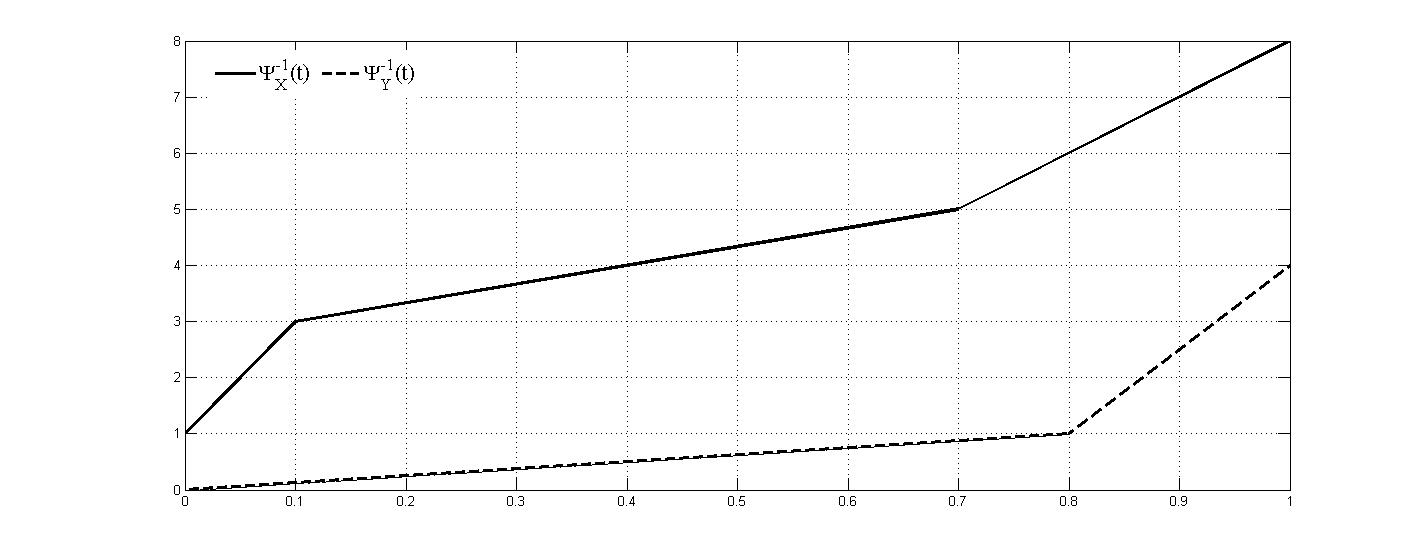}
      \caption{Representation of the quantile functions $\Psi^{-1}_{X}$ and $\Psi^{-1}_
      {Y}$ in Example \ref{ex1}.}\label{fig2}
\end{figure}
\end{example}

It is important to bear in mind that in a histogram the lower bound of each subinterval is always less than or equal to the upper bound, $\underline{I}_{Y(j)i} \leq \overline{I}_{Y(j)i}$ and the upper bound of the following subinterval is always greater or equal to the previous, $\overline{I}_{Y(j)i}\leq\underline{I}_{Y(j)i+1}$. Consequently, the quantile function that represents the empirical distribution is always a non-decreasing function in the domain $[0,1].$

Many concepts and methods for histogram-valued variables have been defined using the representation of their realizations in the form of histograms \cite{bidi07}, \cite{bidi03}. Only in more recent studies have these variables' values been represented as quantile functions \cite{arma09}, \cite{irve06}, \cite{irve08}, \cite{irve07}.
When the distributions are represented as histograms, the choice of the arithmetic becomes crucial. The complexity of the arithmetics \cite{colombo80}, \cite{tesewi89} that have been proposed so far for histograms was arguably the reason why the distributions started being represented as quantile functions. If we represent the distribution that each unit takes on a histogram-valued variable by a quantile function, then operations are simplified because, as quantile functions are piecewise functions, the adequate arithmetic for them is a function arithmetic. In this work the option is to represent the distributions by quantile functions. However, this representation raises other questions.

To operate with quantile functions, it is necessary to define all functions involved with an equal number of pieces or, equivalently, to rewrite all correspondent histograms with the same number of subintervals. For this, it may be necessary to apply the procedure defined by Irpino and Verde \cite{irve06}.
In addition, it is important to avoid that the number of subintervals for each histogram becomes ``too" large (which could happen by applying the process proposed by Irpino and Verde \cite{irve06}), in which case the distributions that represent the data would be meaningless.
To prevent this situation, we may consider the suggestion of Colombo \cite{colombo80} who encountered similar problems, and has considered advantageous to work with equiprobable histograms (histograms of equal probability subintervals).

\vspace{-0.2cm}
\subsection{The space of quantile functions} \label{ss2.2}

Quantile functions are a particular kind of functions. If we consider the set of the functions defined from $\mathbb{R}$ in $\mathbb{R},$ $\mathcal{F}(\mathbb{R},\mathbb{R})$ and the usual operations defined in $\mathcal{F}:$  addition $(f+g)(x)=f(x)+g(x), \forall x\in \mathbb{R}$ and  product of a function by a real number $(\lambda f)(x)= \lambda f(x), \forall x \in \mathbb{R},$ and $\lambda \in \mathbb{R},$ it follows that $(\mathcal{F},+,.)$ is a vector space. However, if we consider the particular case of the set of the quantile functions, $\mathcal{E}([0,1],\mathbb{R}),$  defined on $[0,1]$, we don't have a subspace of the vector space  $(\mathcal{F},+, .).$ Analyzing the behavior of these operations it is possible to understand why $\mathcal{E}([0,1],\mathbb{R}),$ with the usual operations, does not verify the vector space definition.

Consider the quantile functions $\Psi_{X}^{-1}(t)$ and $\Psi_{Y}^{-1}(t)$ defined according to (\ref{eq2.3}) in\textit{ Definition \ref{def2.1}}  both with $n$ subintervals, after having been rewritten in accordance with the process described in \cite{irve06}. These functions represent the distributions that the histogram-valued variables $X$ and $Y$ take for one unit. The addition of these quantile functions leads to the function

$$
\Psi_{X}^{-1}(t)+\Psi_{Y}^{-1}(t)=\left\{\begin{array}{lll}
                     \underline{I}_{X_{1}} + \underline{I}_{Y_{1}}+\frac{t}{w_{1}} (a_{X_{1}}+a_{Y_{1}}) & if & 0 \leq t < w_{1} \\
                    \underline{I}_{X_{2}}+ \underline{I}_{Y_{2}} +\frac{t-w_{1}}{w_{2}-w_{1}} (a_{X_{2}}+a_{Y_{2}}) & if & w_{1} \leq t < w_{2} \\
                     \vdots & & \\
                   \underline{I}_{X_{n}}+\underline{I}_{Y_{n}} +\frac{t-w_{n-1}}{1-w_{n-1}} (a_{X_{n}}+a_{Y_{n}}) & if & w_{n-1} \leq t \leq
                   1
 \end{array}
  \right.
$$

When we add two quantile functions we obtain a non-decreasing function. In this case both the slope and the $y$-intercept of the resulting function are influenced by the two functions.

The particular case of the addition of a quantile function $\Psi_{X}^{-1}(t)$ with a real number $\alpha$ is the function

$$
\left(\Psi_{X}^{-1}+\alpha\right)(t)=\left\{\begin{array}{lll}
                     \underline{I}_{X_{1}}+\alpha +\frac{t}{w_{1}} a_{X_{1}} & if & 0 \leq t < w_{1} \\
                    \underline{I}_{X_{2}}+\alpha +\frac{t-w_{1}}{w_{2}-w_{1}} a_{X_{2}} & if & w_{1} \leq t < w_{2} \\
                     \vdots & & \\
                   \underline{I}_{X_{n}}+\alpha +\frac{t-w_{n-1}}{1-w_{n-1}} a_{X_{n}} & if & w_{n-1} \leq t \leq
                   1
 \end{array}
  \right.
$$

In this case, only the $y$-intercept is affected by the operation, we have  a translation up when adding a real positive number $\alpha$ and a translation down when the real number $\alpha$ is negative.

The multiplication of the quantile function $\Psi_{X}^{-1}(t)$ by a real number $\lambda$ leads to the function

$$
\lambda \Psi_{X}^{-1}(t)=\left\{\begin{array}{lll}
                     \lambda \underline{I}_{X_{1}} +\frac{t}{w_{1}} (\lambda a_{X_{1}}) & if & 0 \leq t < w_{1} \\
                    \lambda\underline{I}_{X_{2}} +\frac{t-w_{1}}{w_{2}-w_{1}} (\lambda a_{X_{2}}) & if & w_{1} \leq t < w_{2} \\
                     \vdots & & \\
                   \lambda \underline{I}_{X_{n}} +\frac{t-w_{n-1}}{1-w_{n-1}} (\lambda a_{X_{n}}) & if & w_{n-1} \leq t \leq
                   1
 \end{array}
  \right.
$$

In this case, both the slope and the $y$-intercept are affected by $\lambda.$ If $\lambda$ is positive we will have a non-decreasing function but if
 $\lambda$ is negative we will obtain a decreasing function that cannot be a quantile function, because quantile functions must always be non-decreasing functions. It is for this reason that the $\mathcal{E}([0,1],\mathbb{R}),$ is a semi-vectorial space.

The following example illustrates this situation.

\begin{example}\label{ex3}
Consider the distribution represented by the quantile function $\Psi_{X}^{-1}(t)$ presented in \textit{Example \ref{ex1}}. If we multiply the quantile function $\Psi_{X}^{-1}(t)$ by the positive real number $2,$ we obtain a non-decreasing function but if we multiply the quantile function $\Psi_{X}^{-1}(t)$ by the negative real number $-1$ the resulting function is not a non-decreasing function. The following functions and representations in \textit{Figure \ref{fig9}} illustrate this situation.

$$
2\Psi_{X}^{-1}(t)=\left\{\begin{array}{lll}
                     2+\frac{t}{0.1}\times 4  & \quad if \quad  & 0 \leq t <0.1 \\
                    6 +\frac{t-0.1}{0.6} \times 4 &  \quad if \quad & 0.1 \leq t < 0.7 \\
                   10 +\frac{t-0.7}{0.3} \times 6 & \quad if \quad & 0.7 \leq t \leq 1
 \end{array}
  \right.
$$

$$-\Psi_{X}^{-1}(t)=\left\{\begin{array}{lll}
                     -1+\frac{t}{0.1}\times (-2)  & \quad if \quad  & 0 \leq t <0.1 \\
                    -3 +\frac{t-0.1}{0.6} \times (-2) &  \quad if \quad & 0.1 \leq t < 0.7 \\
                   -5 +\frac{t-0.7}{0.3} \times (-3) & \quad if \quad & 0.7 \leq t \leq 1
 \end{array}
  \right.
$$

\begin{figure}[h!]\label{fig9}
  \centering
    \includegraphics[width=0.9\textwidth]{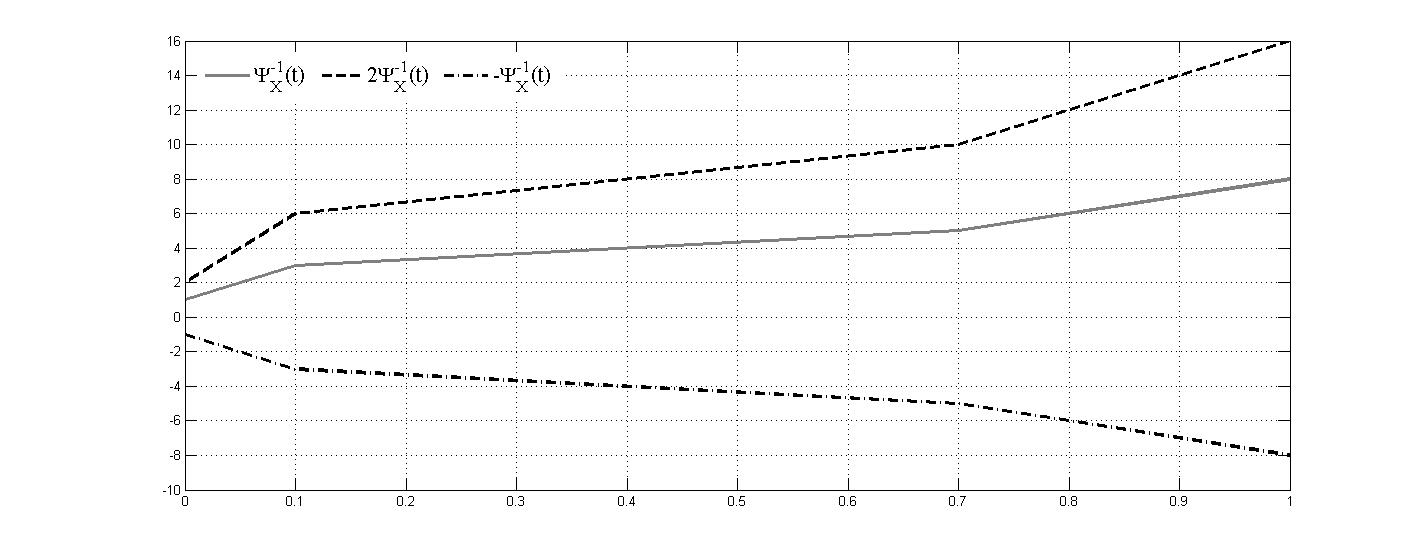}
      \caption{Representation of the functions $\Psi_{X}^{-1}(t),$ $2\Psi_{X}^{-1}(t),$ $-\Psi_{X}^{-1}(t)$ in Example \ref{ex3}.}\label{fig9}
\end{figure}
\end{example}

In conclusion, $\mathcal{E}([0,1],\mathbb{R}),$ is not a vector space because the elements of this space do not have symmetric elements. If we have a quantile function $\Psi_{X}(t),$ the function $-\Psi_{X}^{-1}(t)$ is not a non-decreasing function and consequently cannot be a quantile function. However if we consider the distributions represented by histograms and use the histograms arithmetic proposed by Colombo \cite{colombo80} it is possible to obtain a new histogram, that is the symmetric of the histogram $H_{X}.$ The histogram $-H_{X}$ is the symmetric of the histogram $H_{X}$ if $-H_{X}$ and $H_{X}$ are symmetric in relation to the $yy-$axis.

\begin{figure}[h!]
  \centering
    \includegraphics[width=0.9\textwidth]{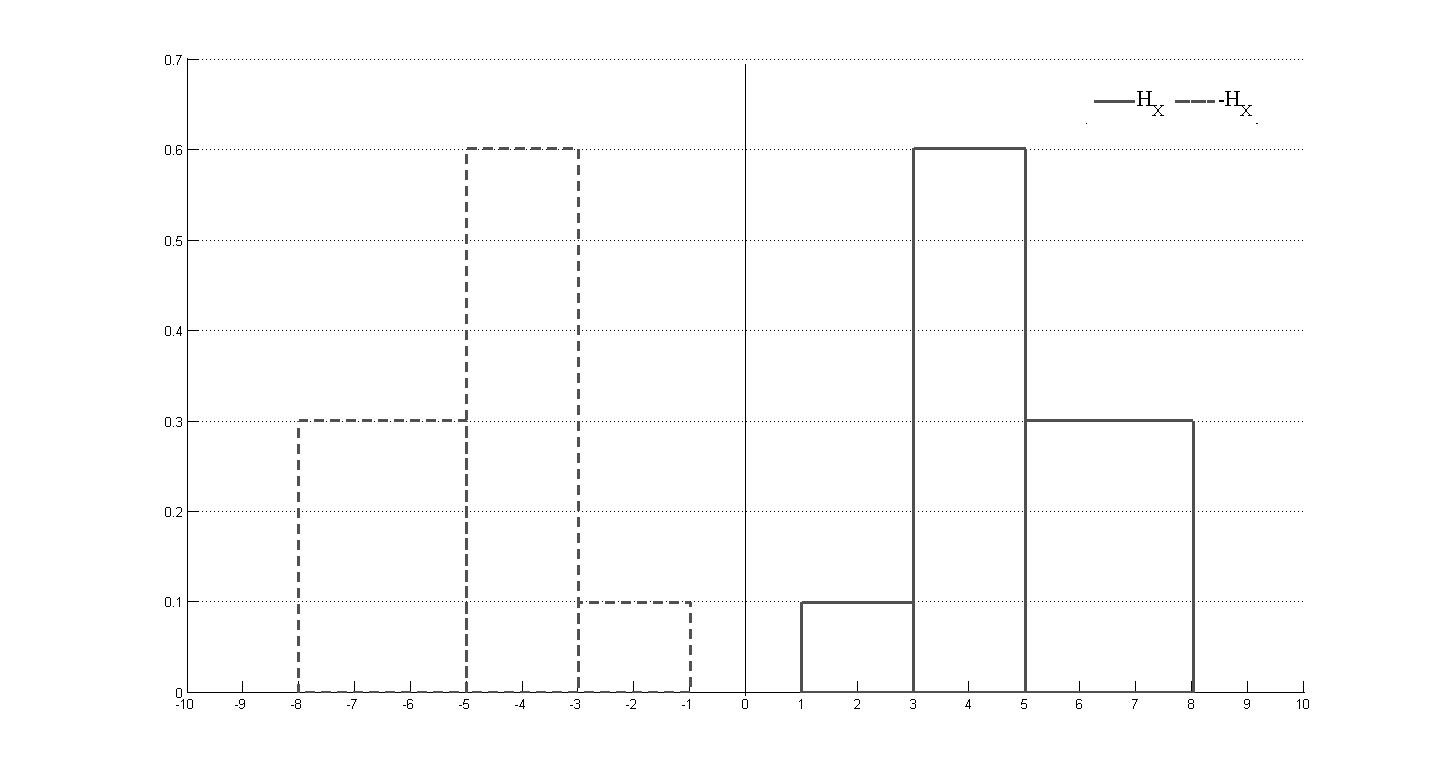}
      \caption{Representation of the histogram $H_{X}$ in Example \ref{ex1} and the respective symmetric histogram $-H_{X}$. }
      \label{fig9B}
\end{figure}

As an example of the situation above, \textit{Figure \ref{fig9B}} represents the histogram $H_{X}$ in Example \ref{ex1} and the respective symmetric histogram.

\begin{figure}[h!]
  \centering
    \includegraphics[width=0.9\textwidth]{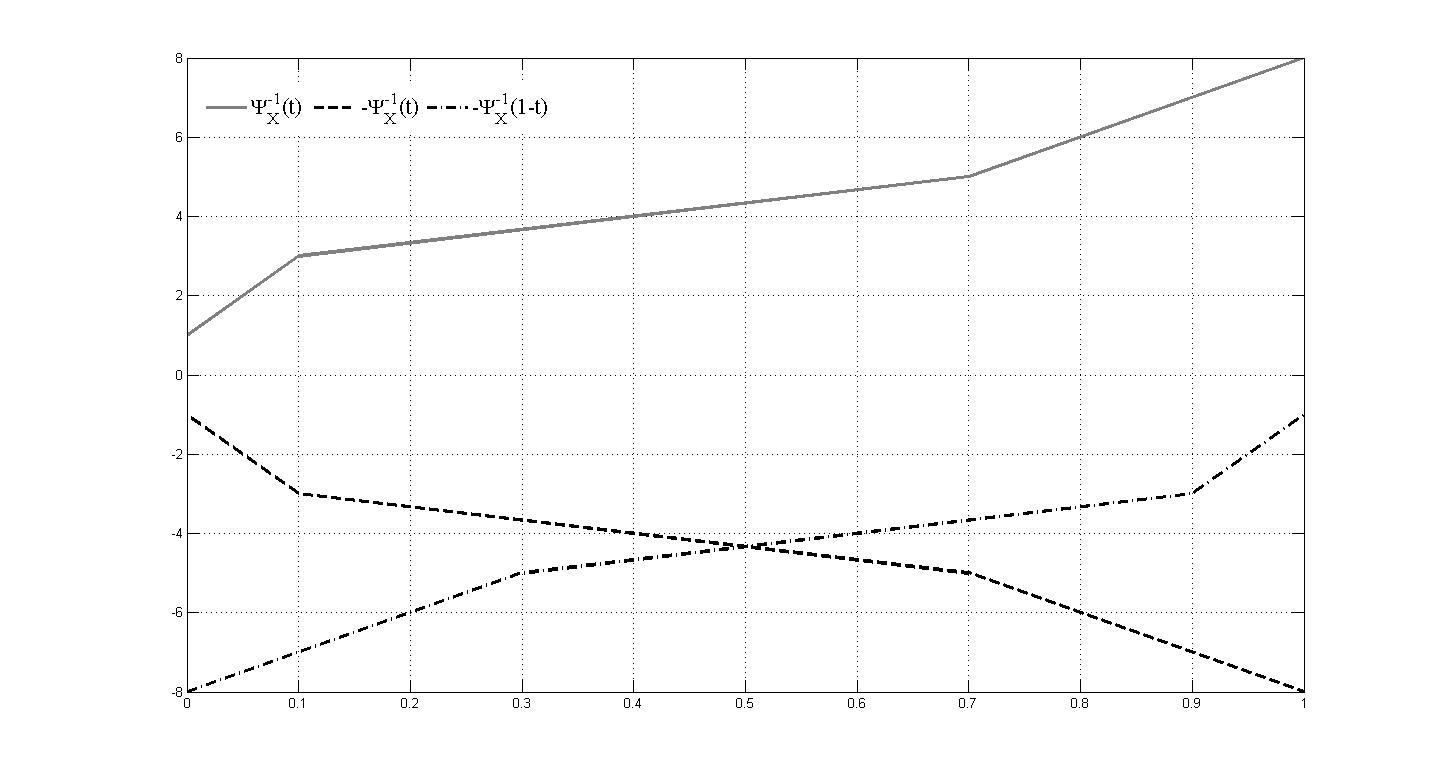}
      \caption{Representation of the functions $\Psi_{X}^{-1}(t),$ $-\Psi_{X}^{-1}(t),$ and $-\Psi_{X}^{-1}(1-t),$ in Example \ref{ex3}. }
      \label{fig10}
\end{figure}

It is obviously possible to define the quantile function that represents the distribution of the histogram $-H_{X}.$ This quantile function is $-\Psi_{X}^{-1}(1-t)$ with $t \in [0,1]$ and is not the function obtained by multiplying the quantile function $\Psi_{X}^{-1}(t)$ by $-1.$ \textit{Figure \ref{fig10}} shows that the function $-\Psi_{X}^{-1}(t)$ in Example \ref{ex3} is different from the quantile function $-\Psi_{X}^{-1}(1-t)$ that corresponds to the histogram $-H_{X}.$

To conclude this section, it is important to underline some conclusions about the function $-\Psi_{X}^{-1}(1-t),$ $t \in [0,1]$:
\begin{itemize}
	\item As it is required for quantile functions, $-\Psi_{X}^{-1}(1-t)$ is a non-decreasing function;
	\item  $\Psi_{X}^{-1}(t)-\Psi_{X}^{-1}(1-t)$ is not a null function, as expected, but is a quantile function with null (symbolic) mean \cite{bidi03};
	\item the functions $-\Psi_{X}^{-1}(1-t)$ and $\Psi_{X}^{-1}(t)$ are linearly independent, providing that $-\Psi_{X}^{-1}(1-t) \neq \Psi_{X}^{-1}(t)$;
	\item $-\Psi_{X}^{-1}(1-t) = \Psi_{X}^{-1}(t)$ only when the histogram $H_{X}$ is symmetric with respect to the $yy-$axis
\end{itemize}

\section{Linear Regression Model for histogram-valued variables}\label{s3.1}

The first linear regression model for histogram-valued variables was proposed by Billard and Diday \cite{bidi02}. This model is a generalization of the \textit{Center Model} \cite{bidi00} defined by the same authors for interval-valued variables but, with this model it is possible that the predicted results are not histogram values. Because of this, recently same studies have emerged in an attempt to find new proposals for a linear regression model for this kind of variables. A recent model has been proposed by Verde and Irpino \cite{irve10}.

Our main goal in this work is to propose a linear regression model for histogram-valued variables. More precisely, to provide a linear regression model that considers data with variability and allows predicting histogram values.

To define this model, three problems need to be solved:
\begin{itemize}
  \item Find an error measure to quantify the difference between the observed and predicted distributions represented by histograms or quantile functions;
  \item Define a linear regression model for histogram-valued variables that allows predicting histograms or their quantile functions from other histograms or quantile functions, without forcing a direct linear relationship;
  \item Measure the goodness-of-fit of the model.
\end{itemize}

\subsection{Error measure} \label{ss3.1}

In classical linear regression, to quantify the error between the observed values $y_{j}$ and the predicted values $\widehat{y}_{j}$ the difference between two real numbers, $e_{j}=y_{j}-\widehat{y}_{j}$ is used. In this case, the model to estimate the values $\widehat{y}_{j}$ minimizes the quantity $\displaystyle\sum\limits_{j=1}^m (y_{j}-\widehat{y}_{j})^2$.
However, due to the complexity of histogram-valued variables, the error between the observed and predicted distributions requires a different approach.

In their work about forecasting time series, applied to histogram-valued variables, Arroyo and Mat\'{e} \cite{arma09}, \cite{tesear08} also needed to measure the error between the observed and forecasted distributions. Therefore, they sought for a good measure to analyze the similarity between two distributions. Firstly, they considered the possibility of computing the difference between two distributions represented by their respective histograms using the histograms' arithmetic. However, this option turned out to be of little use. As we have seen, it is not easy to operate with the  histograms arithmetic and some results are not as expected. This shows that it is not adequate to analyze the similarity between distributions with this concept. The options of those authors were to use dissimilarity measures for distributions and they opted for the Wasserstein and Mallows distance \cite{mall72}, \cite{arma09} to measure the diference between the observed and forecasted distributions. The justification for the choice of the Wasserstein and Mallows distance was the fact that they are distances and thus present interesting properties for error measurement: positive definiteness, symmetry, and triangle inequality condition. On the other hand, for Arroyo and Mat\'{e} \cite{arma09}, \cite{tesear08}, the Mallows distance is the one that better adjusts to the concept of distance as assessed by the human eye. This distance was also used in other works such as Irpino and Verde \cite{irve06}, where the Mallows distance is used to determine the \textit{barycentric histogram} and is then successfully applied to cluster histogram data. The same authors used this distance in their linear regression model for histogram-valued variables \cite{irve10}.

In using the Wasserstein and Mallows distances, the distributions taken by the histogram-valued variables are represented by their quantile functions. These distances are defined as follows:

\begin{definition}\label{def3.1}
Given two quantile functions $\Psi_{X(j)}^{-1}(t)$ and $\Psi_{Y(j)}^{-1}(t)$ that represent the distributions that the histogram-valued variables $X$ and $Y$ take at unit $j$, the Wasserstein distance is defined as:
\begin{center}
\begin{equation}\label{eq3.1}
D_W(\Psi_{X(j)}^{-1}(t),\Psi_{Y(j)}^{-1}(t))=\int_{0}^{1}\left|\Psi_{X(j)}^{-1}(t)-\Psi_{Y(j)}^{-1}(t)\right|dt
\end{equation}
\end{center}
and the Mallows distance:
\begin{center}
\begin{equation}\label{eq3.2}
D_M(\Psi_{X(j)}^{-1}(t),\Psi_{Y(j)}^{-1}(t))=\sqrt{\int_{0}^{1}(\Psi_{X(j)}^{-1}(t)-\Psi_{Y(j)}^{-1}(t))^2dt}
\end{equation}
\end{center}
\end{definition}

Instead of using the quantile functions that represent the distributions, Irpino and Verde \cite{irve06} rewrote the Mallows distance using the histograms, more specifically the centre and half-range of their subintervals. The square of the Mallows distance can be also defined as follows:

\begin{property}\label{pr3.1}
Consider two histogram-valued variables $X$ and $Y.$ The distributions that these variables take for a given unit $j,$ can be represented by the quantile functions $\Psi_{X(j)}^{-1}(t)$ and $\Psi_{Y(j)}^{-1}(t)$ or the histograms $H_{X(j)}$ and $H_{Y(j)}.$
The square of the Mallows distance between these distributions is given by
$$
D^{2}_M(\Psi_{X(j)}^{-1}(t),\Psi_{Y(j)}^{-1}(t))=\sum_{i=1}^{n}p_{i}\left[(c_{X(j)_{i}}-c_{Y(j)_{i}})^2+\frac{1}{3}(r_{X(j)_{i}}-r_{Y(j)_{i}})^2\right]
$$
where, relatively to the histogram-valued variables $X$ or $Y$ for unit $j:$
\begin{itemize}
  \item $c_{X(j)_{i}}=\frac{\overline{I}_{X(j)_{i}}+\underline{I}_{X(j)_{i}}}{2}$ and $c_{Y(j)_{i}}=\frac{\overline{I}_{Y(j)_{i}}+\underline{I}_{Y(j)_{i}}}{2}$ are the centers of the intervals $i,$ with  $i \in \left\{1, \ldots, n \right\};$
  \item $r_{X(j)_{i}}=\frac{\overline{I}_{X(j)_{i}}-\underline{I}_{X(j)_{i}}}{2}$ and $r_{Y(j)_{i}}=\frac{\overline{I}_{Y(j)_{i}}-\underline{I}_{Y(j)_{i}}}{2}$ are the half-ranges of the intervals $i,$ with $i \in \left\{1,\ldots, n \right\}.$
\end{itemize}

\end{property}

It seems therefore appropriate to choose the Wasserstein or Mallows distance to measure the similarity between the observed and predicted distributions by the linear regression model. Because of the properties of the absolute value function we choose to define the error measure between two distributions with the Mallows distance.

\begin{definition}\label{def3.2}
Consider, for each unit $j$,  $\Psi_{Y(j)}^{-1}(t)$ the quantile function of the observed distribution $Y(j)$ and $\Psi_{\widehat{Y}(j)}^{-1}(t)$ the quantile function that represents the predicted distribution $\widehat{Y}(j)$. The error between $Y(j)$ and $\widehat{Y}(j)$ is defined by:
\vspace{-0.5cm}
\begin{center}
\begin{equation}\label{eq3.4}
SE(j)=D^{2}_M(\Psi_{Y(j)}^{-1}(t),\Psi_{\widehat{Y}(j)}^{-1}(t))
\end{equation}
\end{center}
The total error is the sum of the errors, that according to Property \ref{pr3.1}, may be written as follows:
\vspace{-0.1cm}
\begin{center}
\begin{equation}\label{eq3.4A}
SE=\sum_{j=1}^{m}D^{2}_M(\Psi_{Y(j)}^{-1}(t),\Psi_{\widehat{Y}(j)}^{-1}(t))=\sum_{j=1}^{m}\sum_{i=1}^{n}p_{ji}\left[(c_{Y(j)_i}-c_{\widehat{Y}(j)_{i}})^2+\frac{1}{3}(r_{Y(j)_i}-r_{\widehat{Y}(j)_{i}})^2\right]
\end{equation}
\end{center}
\end{definition}

\subsection{The \textit{DSD Regression Model}} \label{ss3.2}

The first option to define the functional linear relation between histogram data was to adapt the classical model to these data. Consider that we want to predict the distributions that the histogram-valued variable $Y$ takes from $p$ histogram-valued variables $X_{k}$ with $k \in\{1,\ldots,p\}.$  At unit $j$, $j \in \{1,\ldots,m\}$, the predicted distribution $\widehat{Y}(j)$ would than be obtained as follows:
$$
\widehat{Y}(j)=\gamma+\alpha_{1}X_{1}(j)+\alpha_{2}X_{2}(j)+\ldots+\alpha_{p}X_{p}(j).
$$
As already mentioned, in this work we choose to represent the distributions by quantile functions. However, when we multiply a quantile function by a negative number we do not obtain a non-decreasing function. Therefore, it is necessary to impose positivity restrictions on the parameters of the model. Denoting by $\Psi_{\widehat{Y}(j)}^{-1}(t)$ the quantile function of the predicted distribution $\widehat{Y}(j),$ we  obtain the linear regression model as follows:
$$
\Psi_{\widehat{Y}(j)}^{-1}(t)=\beta_{0}+\beta_{1}\Psi_{X_{1}(j)}^{-1}(t)+\beta_{2}\Psi_{X_{2}(j)}^{-1}(t)+\ldots+\beta_{p}\Psi_{X_{p}(j)}^{-1}(t)
$$
\vspace{-0.1cm}
with $\beta_{k} \geq 0$ and $k\in\{1,2,\ldots,p\}.$

The non-negativity constraints imposed on the coefficients force a direct linear relationship, and limitations similar to those present in linear regression models defined for interval-valued variables occur (see, e.g., \cite{netocar10}). Although we did not generalize the model for interval-valued variables to histogram-valued variables, in defining a model that allows to predict a quantile function from other quantile functions, we obtain a model with the same limitations as observed before.

It is not possible to have negative parameters in the previous model. Neverthless, it is fundamental to allow for the possibility of a direct and an inverse linear relation between the variable $Y$ and the variables $X_{k}.$ For this reason, our proposal is to include in the linear regression model both the quantile functions $\Psi_{X_{k}(j)}^{-1}(t),$ that represent the distributions that the histogram-valued variables $X_{k}$ take for each unit $j,$ and the quantile functions that represent the respectively symmetric histograms  $-\Psi_{X_{k}(j)}^{-1}(1-t)$ (see also \textit{Section \ref{ss2.2}}). Therefore we proposed the following model:

\begin{definition}\label{def3.4}

Consider the histogram-valued variables $X_{1}; X_{2}; \ldots; X_{p}.$ The quantile functions that represent the distribution that these histogram-valued variables take for each unit $j$ are $\Psi_{X_1(j)}^{-1}(t),$ $\Psi_{X_2(j)}^{-1}(t),\ldots,$ $\Psi_{X_p(j)}^{-1}(t)$ and the quantile functions that represent the respective symmetric histograms associated to each unit of the refered variables are $-\Psi_{X_1(j)}^{-1}(1-t),-\Psi_{X_2(j)}^{-1}(1-t),\ldots,-\Psi_{X_p(j)}^{-1}(1-t),$ with $t \in [0,1].$ Each quantile function $\Psi_{Y(j)}^{-1},$ can be expressed as follows:

$$
\Psi_{Y(j)}^{-1}(t)=\Psi_{\widehat{Y}(j)}^{-1}(t)+\varepsilon_{j}(t).
$$

where $\Psi_{\widehat{Y}(j)}^{-1}(t)$ is the predicted quantile function for unit $j,$ obtained from

\begin{center}
\begin{eqnarray*}
\Psi_{\widehat{Y}(j)}^{-1}(t)&=&\gamma+\alpha_{1}\Psi_{X_{1}(j)}^{-1}(t)-\beta_{1}\Psi_{X_{1}(j)}^{-1}(1-t)+\alpha_{2}\Psi_{X_{2}(j)}^{-1}(t)-\beta_{2}\Psi_{X_{2}(j)}^{-1}(1-t)+ \nonumber
\\& & + \ldots + \alpha_{p}\Psi_{X_{p}(j)}^{-1}(t)+\beta_{p}\Psi_{X_{p}(j)}^{-1}(1-t).
\end{eqnarray*}
\end{center}
with $t \in \left[0,1\right];$ $\alpha_{k},\beta_{k} \geq 0,$  $k \in \left\{1,2,\ldots,p \right\}$ and $\gamma \in \mathbb{R}.$

The error, for each unit $j$, is the piecewise function given by $\varepsilon_{j}(t)=\Psi_{Y(j)}^{-1}(t)-\Psi_{\widehat{Y}(j)}^{-1}(t).$

For each unit $j,$ the predicted distribution $\widehat{Y}(j)$ can be represented by the quantile function $\Psi_{\widehat{Y}(j)}^{-1}$ or by the respective histogram $H_{\widehat{Y}(j)}.$
This linear regression model will be named \textbf{Distribution and Symmetric Distribution (DSD) Regression Model}.
\end{definition}

Consider the particular case of the linear regression model where there is only one explicative histogram-valued variable $X.$
In this case we can obtain the quantile function $\Psi_{Y(j)}^{-1}(t),$ for each unit $j,$ by the model:
\vspace{-0.5cm}
\begin{center}
\begin{equation}\label{eq3.5}
\Psi_{Y(j)}^{-1}(t)=\gamma+\alpha \Psi_{X(j)}^{-1}(t)-\beta \Psi_{X(j)}^{-1}(1-t)+\varepsilon_{j}(t)
\end{equation}
\end{center}
\vspace{-0.1cm}
with $\alpha,\beta \geq 0,$  and $\gamma \in \mathbb{R}.$

When including in the model both the distribution of the explicative histogram-valued variables, and the respective symmetric distributions, the restrictions on the parameters are imposed; however, this does not imply a direct linear relationship. In the particular case of (\ref{eq3.5}), we consider that the linear regression is direct if $\alpha > \beta$ and inverse if $\alpha < \beta.$

\subsection{Parameters of the \textit{DSD Regression Model}} \label{ss3.3}

In classical statistics, the parameters of the linear regression model are estimated solving the minimization problem  $\displaystyle\sum\limits_{j=1}^m (y_{j}-\widehat{y}_{j})^2,$ where $y_{j}$ are the observed and $\widehat{y}_{j}$ the predicted values, respectively, with $j \in \{1,\ldots, m \}$. To solve this problem the least squares method is used.

For histogram-valued variables the parameters of the \textit{DSD Model}, in \textit{Definition \ref{def3.4}}, are estimated solving a quadratic optimization problem, subject to non-negativity constraints on the unknowns.

\begin{definition}\label{def3.5}
Consider $\Psi_{\widehat{Y}(j)}^{-1}(t)$ obtained by the \textit{DSD Model}. The quadratic optimization problem is written as:
$$
Minimize \qquad SE=\sum_{j=1}^{m} D_{M}^{2} (\Psi_{Y(j)}^{-1}(t),\Psi_{\widehat{Y}(j)}^{-1}(t))
$$

with $\alpha_{k},\beta_{k} \geq 0,$  $k\in\{1,2,\ldots,p\}$ and $\gamma \in \mathbb{R}.$
\end{definition}

To present more specifically the function to minimize, it is important to define all the quantile functions involved in this expression considering the conditions referred in \textit{Section \ref{ss2.1}}. The quantile functions that represent the distributions taken by $X_k$ and the respective symmetric, for a given unit $j$ are, respectively:

\begin{center}
\begin{equation}\label{eq3.6}
\Psi_{X_k(j)}^{-1}(t)=\left\{\begin{array}{lll}
                     c_{X_{k}(j)_{1}}+\left(2\frac{t}{w_{1}}-1\right) r_{X_{k}(j)_{1}} & if & 0 \leq t < w_{1} \\
                    c_{X_{k}(j)_{2}} +\left(2\frac{t-w_{1}}{w_{2}-w_{1}}-1\right) r_{X_{k}(j)_{2}}  & if & w_{1} \leq t < w_{2} \\
                     \vdots & & \\
                   c_{X_{k}(j)_{n}} +\left(2\frac{t-w_{(n-1)}}{1-w_{(n-1)}}-1\right) r_{X_{k}(j)_{n}}  & if & w_{n-1} \leq t \leq
                   1
 \end{array}
  \right.
\end{equation}
\end{center}

\begin{center}
\begin{equation}\label{eq3.7}
-\Psi_{X_{k}(j)}^{-1}(1-t)=\left\{\begin{array}{lll}
                     -c_{X_{k}(j)_{n}}+\left(2\frac{t}{w_{1}}-1\right) r_{X_{k}(j)_{n}}  & if & 0 \leq t < w_{1} \\
                     -c_{X_{k}(j)_{n-1}} +\left(2\frac{t-w_{1}}{w_{2}-w_{1}}-1\right) r_{X_{k}(j)_{n-1}} & if & w_{1} \leq t < w_{2} \\
                     \vdots & & \\
                   -c_{X_{k}(j)_{1}} +\left(2\frac{t-w_{n-1}}{1-w_{n-1}}-1\right) r_{X_{k}(j)_{1}} & if & w_{n-1} \leq t \leq
                   1
 \end{array}
  \right.
\end{equation}
\end{center}

According to the \textit{DSD Model}, the quantile function that represents the distribution taken by the predicted histogram-valued variable $\widehat{Y},$ for a given unit $j$ is:

{\scriptsize
\begin{equation}\label{eq3.8}
 \Psi_{\widehat{Y}(j)}^{-1}(t)=\left\{\begin{array}{lll}
                     \displaystyle\sum\limits_{k=1}^p \left(\alpha_{k}c_{X_{k}(j)_{1}}- \beta_{k}c_{X_{k}(j)_{n}}\right)+\gamma+
                     \left(2\frac{t}{w_{1}}-1\right) \sum_{k=1}^{p} \left(\alpha_{k}r_{X_{k}(j)_{1}}+\beta_{k}r_{X_{k}(j)_{n}}\right) & if & 0 \leq t < w_{1} \\
                     \displaystyle\sum\limits_{k=1}^p \left(\alpha_{k}c_{X_{k}(j)_{2}}- \beta_{k}c_{X_{k}(j)_{n-1}}\right)+\gamma+
                    \left(2\frac{t-w_{1}}{w_{2}-w_{1}}-1\right) \sum_{k=1}^{p} \left(\alpha_{k}r_{X_{k}(j)_{2}}+\beta_{k}r_{X_{k}(j)_{n-1}}\right)   & if & w_{1} \leq t < w_{2} \\
                     \vdots &  &  \\
                     \displaystyle\sum\limits_{k=1}^p \left(\alpha_{k}c_{X_{k}(j)_{n}}- \beta_{k}c_{X_{k}(j)_{1}}\right)+\gamma+
                     \left(2\frac{t-w_{n-1}}{1-w_{n-1}}-1\right) \sum_{k=1}^{p} \left(\alpha_{k}r_{X_{k}(j)_{n}}+\beta_{k}r_{X_{k}(j)_{1}}\right)& if & w_{n-1} \leq t \leq 1
 \end{array}
  \right.
\end{equation}
}

Similarly, for unit $j$, the quantile function that represents the distribution taken by the histogram-valued variable, $Y$ is
\begin{center}
\begin{equation}\label{eq3.9}
\Psi_{Y(j)}^{-1}(t)=\left\{\begin{array}{lll}
                     c_{Y(j)_1}+\left(2\frac{t}{w_{1}}-1\right) r_{Y(j)_1} & if & 0 \leq t < w_{1} \\
                    c_{Y(j)_2} +\left(2\frac{t-w_{1}}{w_{2}-w_{1}}\right) r_{Y(j)_2} & if & w_{1} \leq t < w_{2} \\
                     \vdots & & \\
                   c_{Y(j)_n} +\left(2\frac{t-w_{n-1}}{1-w_{n-1}}\right) r_{Y(j)_n} & if & w_{n-1} \leq t \leq 1
 \end{array} \right.
\end{equation}
\end{center}

Consider these quantile functions and the Mallows distance defined according to \textit{Property \ref{pr3.1}}. The quadratic optimization problem presented in \textit{Definition \ref{def3.5}} can then be rewritten as follows:

\begin{center}
\begin{equation}\label{eq3.10}
\begin{array}{ll}
Minimize \quad SE  & = \displaystyle\sum\limits_{j=1}^m \displaystyle\sum\limits_{i=1}^n p_{i} \left[ \left( c_{Y(j)_i}-\displaystyle\sum\limits_{k=1}^p \left(\alpha_{k} c_{X_{k}(j)_{i}}- \beta_{k} c_{X_{k}(j)_{n-i+1}}\right)-\gamma \right)^{2} \right.\\
 & \left.+\frac{1}{3} \left(r_{Y(j)_i}-\displaystyle\sum\limits_{k=1}^p \left(\alpha_{k} r_{X_{k}(j)_{i}}+ \beta_{k} r_{X_{k}(j)_{n-i+1}}\right)\right)^{2} \right]
\end{array}
\end{equation}
subject to $\alpha_{k},\beta_{k} \geq 0,$  $k\in\{1,2,\ldots,p\}$ and $\gamma \in \mathbb{R}$.
\end{center}

Or, in matricial form:

\begin{center}
\begin{equation}\label{eq3.11}
Minimize \quad SE = \frac{1}{2} B^{T}HB+F^{T}B+C
\end{equation}
subject to $-\alpha_{k},-\beta_{k} \leq 0;$  $k\in\{1,2,\ldots,p\}$ and $\gamma \in \mathbb{R}$.
\end{center}

In this latter case,  $H=\left[h_{lq}\right]$  is the hessian matrix, a symmetric matrix of order $2p+1,$ with $p$ the number of variables $X_k.$ The elements of the symmetric matrix $H$ are defined as follows:

{\scriptsize
$$
h_{lq}=\left\{\begin{array}{lll}
                   \displaystyle\sum\limits_{j=1}^m \displaystyle\sum\limits_{i=1}^n p_{i}\left(2c_{X_{\frac{l+1}{2}}(j)_{i}}
                   c_{X_{\frac{q+1}{2}}(j)_{i}}+\frac{2}{3}r_{X_{\frac{l+1}{2}}(j)_{i}}r_{X_{\frac{q+1}{2}}(j)_{i}} \right)  & if & \textit{ l,q are odd and } l,q \leq 2p \\
                    \displaystyle\sum\limits_{j=1}^m \displaystyle\sum\limits_{i=1}^n p_{i}\left(2c_{X_{\frac{l}{2}}(j)_{n-i+1}}
                   c_{X_{\frac{q}{2}}(j)_{n-i+1}}+\frac{2}{3}r_{X_{\frac{l}{2}}(j)_{n-i+1}}r_{X_{\frac{q}{2}}(j)_{n-i+1}} \right)  & if & \textit{ l,q are even and } l,q \leq 2p \\
                      \displaystyle\sum\limits_{j=1}^m \displaystyle\sum\limits_{i=1}^n p_{i}\left(-2c_{X_{\frac{l}{2}}(j)_{n-i+1}}
                   c_{X_{\frac{q+1}{2}}(j)_{i}}+\frac{2}{3}r_{X_{\frac{l}{2}}(j)_{n-i+1}}r_{X_{\frac{q+1}{2}}(j)_{i}} \right)  & if & \textit{ l is even, q is odd and } l,q \leq 2p \\

                    \displaystyle\sum\limits_{j=1}^m \displaystyle\sum\limits_{i=1}^n 2p_{i} c_{X_{\frac{q+1}{2}}(j)_{i}} & if & \textit{ q is odd and } l=2p+1 \\
                      \displaystyle\sum\limits_{j=1}^m \displaystyle\sum\limits_{i=1}^n -2p_{i} c_{X_{\frac{q}{2}}(j)_{n-i+1}} & if & \textit{ q is even and } l=2p+1
 \end{array} \right.
$$
}

The vector column of independent terms, $F=\left[f_{l}\right]$ with $2p+1$ rows is given by:

{\footnotesize
$$
f_{l}=\left\{\begin{array}{lll}
\displaystyle\sum\limits_{j=1}^m \displaystyle\sum\limits_{i=1}^n p_{i} \left(-2c_{Y(j)_i}c_{X_{\frac{l+1}{2}}(j)_{i}}-\frac{2}{3}r_{Y(j)_i}r_{X_{\frac{l+1}{2}}(j)_{i}} \right) & if & \textit{ l is odd and } l \leq 2p\\
\displaystyle\sum\limits_{j=1}^m \displaystyle\sum\limits_{i=1}^n p_{i} \left(2c_{Y(j)_i}c_{X_{\frac{l+1}{2}}(j)_{n-i+1}}-\frac{2}{3}r_{Y(j)_i}r_{X_{\frac{l+1}{2}}(j)_{n-i+1}} \right) & if & \textit{ l is even and } l \leq 2p\\
 \displaystyle\sum\limits_{j=1}^m \displaystyle\sum\limits_{i=1}^n -2p_{i} c_{Y(j)_i} & if & l=2p+1
 \end{array} \right.
$$
}

The elements of the matrices $H$ and $F$ are computed from the first order partial derivatives of the function $SE$ in (\ref{eq3.10}). These derivatives are presented in \textit{Appendix A}.
Finally, the vector column of the parameters, $B,$ and the real value $C,$ are defined as follows:
$$
B=\left[
    \alpha_{1} \quad  \beta_{1} \quad \alpha_{2} \quad  \beta_{2} \quad \ldots \quad \alpha_{p} \quad  \beta_{p} \quad  \gamma \right]^T
$$

and
$$
C=\displaystyle\sum\limits_{j=1}^m \displaystyle\sum\limits_{i=1}^np_{i} \left(c_{Y(j)_i}^{2}+\frac{1}{3}r_{Y(j)_i}^{2} \right).
$$

For each particular situation, it is possible to solve this quadratic optimization problem, subject to non-negativity on the constraints, and find the optimal solution. Consider the optimal solution for this optimization problem, $$B^{*}=\left[ \alpha_{1}^{*} \quad \beta_{1}^{*} \quad \alpha_{2}^{*} \quad \beta_{2}^{*} \quad \cdots  \quad \alpha_{n}^{*} \quad \beta_{n}^{*} \quad \gamma^{*} \right]^{T}.$$
Afterwards, it is possible to predict the distributions $\widehat{Y}(j),$ for each $j \in \{1,\ldots,m\},$ considering the obtained matrix $B^{*}.$ Each predicted distribution may be represented by the quantile function as in (\ref{eq3.8}) or by the respective histogram

{\small
\begin{eqnarray*}
H_{\widehat{Y}(j)}&=&\left\{\left[\displaystyle\sum\limits_{k=1}^p \left(\alpha_{k}^{*} \underline{I}_{{X_{k}}(j)_{1}}-\beta_{k}^{*} \overline{I}_{{X_{k}}(j)_{n}}\right)+\gamma^{*}, \displaystyle\sum\limits_{k=1}^p \left(\alpha_{k}^{*} \overline{I}_{{X_{k}}(j)_{1}}-\beta_{k}^{*} \underline{I}_{{X_{k}}(j)_{n}}\right)+\gamma ^{*}\right], p_{1};\right.\ldots \\
& &\ldots; \left. \left[\displaystyle\sum\limits_{k=1}^p \left(\alpha_{k}^{*} \underline{I}_{{X_{k}}(j)_{n}}-\beta_{k}^{*} \overline{I}_{{X_{k}}(j)_{1}}\right)+\gamma^{*}, \displaystyle\sum\limits_{k=1}^p \left(\alpha_{k}^{*} \overline{I}_{{X_{k}}(j)_{n}}-\beta_{k}^{*} \underline{I}_{{X_{k}}(j)_{1}}\right)+\gamma^{*} \right], p_{n} \right\}\\
\end{eqnarray*}
}

Consider the minimization problem defined in (\ref{eq3.10}) or matricially in (\ref{eq3.11}). The optimal solution of the quadratic optimization problem, subject to non-negativity constraints, verifies the Kuhn Tucker conditions \cite{win94}.
Therefore, the optimal solution $B^{*}$ for this optimization problem, for all $k\in\left\{1,\ldots,p \right\}$ verifies the following conditions:

\begin{itemize}
  \item $-\alpha_{k}^{*},-\beta_{k}^{*}\leq 0;$
  \item $\frac{\partial SE (B^{*})}{\partial \alpha_{k}} \geq 0;$ $\frac{\partial SE (B^{*})}{\partial \beta_{k}} \geq 0;$ $\frac{\partial SE (B^{*})}{\partial \gamma}=0;$ $\frac{\partial SE (B^{*})}{\partial \alpha_{k}} \alpha_{k}^{*} =0;$ $\frac{\partial SE (B^{*})}{\partial \beta_{k}} \beta_{k}^{*} =0;$
\end{itemize}

From the Kuhn Tucker conditions, it is possible to prove some properties associated with the predicted distribution. Some of these are the counterparts of the corresponding properties in classical statistics, and will allow defining a measure to evaluate the goodness-of-fit of the model. Before describing these properties, it is necessary to present two important definitions of the concept of mean for histogram-valued variables.

\begin{definition}\label{def2.2} \cite{bidi03}
Consider the histogram-valued variable $Y.$ For each unit $j,$ with $j\in \{1,\ldots,m\},$ $Y(j)$ may be represented by the histogram defined in (\ref{eq2.3B}). The mean of variable $Y$ is defined as follows:

$$
\overline{Y}=\frac{1}{m} \sum_{j=1}^{m}
\left(\sum_{i=1}^{n_{j}}
c_{Y(j)_{i}}p_{ji}\right).
$$
where $n_{j}$ is the number of subintervals for the $j^{th}$ unit.
\end{definition}

Irpino and Verde \cite{irve06} defined the \textit{barycentric histogram} as the histogram that is at a minimum distance - in the sense of the Mallows distance - of the $m$ distributions. In this case, a mean distribution is obtained instead of a mean that is a real number.

The quantile function of the \textit{barycentric histogram} is the same as the mean quantile function, that is computed from the average of the $m$ quantile functions that represent the $m$ given distributions. The mean quantile function is defined as follows:

\begin{definition}\label{def2.3} Consider the $m$ quantile functions $\Psi_{Y(j)}^{-1}(t),$ $j \in \{1,\ldots,m\},$ all defined with $n$ pieces. The mean quantile function  $\overline{\Psi_{Y}^{-1}}(t)$ is the function where each piece is the mean of the corresponding $m$ pieces involved. The function is then,

$$
\overline{\Psi_{Y}^{-1}}(t)=\left\{\begin{array}{lll}
                     \displaystyle\sum\limits_{j=1}^m \frac{c_{Y(j)_{1}}}{m}+ \left(2\frac{t}{w_{1}}-1\right)\frac{r_{Y(j)_{1}}}{m} & if & 0 \leq t < w_{1} \\
                    \displaystyle\sum\limits_{j=1}^m \frac{c_{Y(j)_{2}}}{m} +\left(2\frac{t-w_{1}}{w_{2}-w_{1}}-1\right)\frac{r_{Y_{j2}}}{m} & if & w_{1} \leq t < w_{2} \\
                     \vdots & & \\
                   \displaystyle\sum\limits_{j=1}^m \frac{c_{Y(j)_{n}}}{m} + \left(2\frac{t-w_{n-1}}{1-w_{n-1}}-1\right) \frac{r_{Y_{jn}}}{m} & if & w_{n-1} \leq t \leq
                   1
 \end{array}
  \right.
  $$

So, we have  $\overline{\Psi_{Y}^{-1}}(t) =\frac{1}{m}  \displaystyle\sum\limits_{j=1}^m \Psi_{Y(j)}^{-1}(t). $

\end{definition}

These two concepts of mean for histogram-valued variables are related as we can see in the following proposition.

\begin{proposicion}\label{p2.1}
Considering the mean quantile function $\overline{\Psi_{Y}^{-1}}(t)$ of the histogram-valued variable $Y$ and its mean $\overline{Y},$ we have

$$
\overline{Y}=\int_{0}^{1} \overline{\Psi_{Y}^{-1}}(t)  dt.
$$
\end{proposicion}

This result is due to Irpino and Verde \cite{irve08} and may easily be proved considering \textit{Definitions \ref{def2.2}} and \textit{\ref{def2.3}}.

Now, considering the previous results and the Kuhn Tucker conditions, we may prove the following properties.

\begin{property}\label{pr3.2}
For each unit $j,$ let $\widehat{Y}(j)$ be the distribution predicted by the \textit{DSD Model} and consider the parameters obtained for the optimal solution $B^{*}=\left[ \alpha_{1}^{*} \quad \beta_{1}^{*} \quad \alpha_{2}^{*} \quad \beta_{2}^{*} \quad \cdots \quad \alpha_{n}^{*} \quad \beta_{n}^{*} \quad \gamma^{*} \right]^{T}.$  The mean of the predicted histogram-valued variable $\overline{\widehat{Y}}$ is given by:
$$\overline{\widehat{Y}}=\displaystyle\sum\limits_{k=1}^p \left(\alpha_{k}^{*}-\beta_{k}^{*}\right) \overline{X_{k}} + \gamma^{*}.$$
\end{property}

\textbf{Proof:}
Each observation $j,$ of the predicted histogram-valued variable $\widehat{Y}(j),$ can be represented by the quantile function as in (\ref{eq3.8}) considering for parameters the optimal solution $B^{*},$ of the quadratic optimization problem in (\ref{eq3.10}). As such, the mean quantile function $\Psi_{\widehat{Y}}^{-1}$ can be calculated by \textit{Definition \ref{def2.3}}. So, applying \textit{Proposition \ref{p2.1}} we can prove that $\overline{\widehat{Y}}=\displaystyle\sum\limits_{k=1}^p \left(\alpha_{k}^{*}-\beta_{k}^{*}\right) \overline{X_{k}} + \gamma^{*}. \qquad \Box$

\begin{property}\label{pr3.3}
The mean of the predicted  histogram-valued variable $\overline{\widehat{Y}}$ is equal to the mean of the observed histogram-valued variable $\overline{Y}.$
\end{property}

\textbf{Proof:} Consider the function to minimize in (\ref{eq3.10}),
$$\begin{array}{ll}
 SE  & = \displaystyle\sum\limits_{j=1}^m \displaystyle\sum\limits_{i=1}^n p_{i} \left[ \left( c_{Y(j)_i}-\displaystyle\sum\limits_{k=1}^p \left(\alpha_{k} c_{X_{k(j)i}}- \beta_{k} c_{X_{k(j)n-i+1}}\right)-\gamma \right)^{2} \right.\\
 & \left.+\frac{1}{3} \left(r_{Y(j)_i}-\displaystyle\sum\limits_{k=1}^p \left(\alpha_{k} r_{X_{k(j)i}}+ \beta_{k} r_{X_{k(j)n-i+1}}\right)\right)^{2} \right]
\end{array}.$$

For the optimal solution $B^{*}$ we have $\frac{\partial SE (B^{*})}{\partial \gamma}=0.$ Consequently,
{\scriptsize
$$
\begin{array}{l}
 2 \displaystyle\sum\limits_{j=1}^m  \displaystyle\sum\limits_{i=1}^n p_{i} \left( \displaystyle\sum\limits_{k=1}^p \alpha_{k}^{*} c_{X_{k(j)i}} \right)- 2 \displaystyle\sum\limits_{j=1}^m  \displaystyle\sum\limits_{i=1}^n p_{i} \left( \displaystyle\sum\limits_{k=1}^p \beta_{k}^{*} c_{X_{k(j)(n-i+1)}}\right)+2m\gamma^{*}-2\displaystyle\sum\limits_{j=1}^m  \displaystyle\sum\limits_{i=1}^n p_{i} c_{Y(j)_i}=0  \\
 \Longleftrightarrow \displaystyle\sum\limits_{j=1}^m  \displaystyle\sum\limits_{i=1}^n p_{i}  \displaystyle\sum\limits_{k=1}^p \alpha_{k}^{*} \frac{c_{X_{k(j)i}}}{m}-\displaystyle\sum\limits_{(j)=1}^m  \displaystyle\sum\limits_{i=1}^n p_{i}  \displaystyle\sum\limits_{k=1}^p \beta_{k}^{*} \frac{c_{X_{k(j)(n-i+1)}}}{m} +\gamma^{*}=\displaystyle\sum\limits_{(j)=1}^m  \displaystyle\sum\limits_{i=1}^n p_{i} \frac{c_{Y(j)_i}}{m} \\ \Longleftrightarrow \displaystyle\sum\limits_{k=1}^p \left(\alpha_{k}^{*} \overline{X_{k}}-\beta_{k}^{*} \overline{X_{k}}\right)+\gamma^{*}=\overline{Y}
\end{array} $$}
From \textit{Property \ref{pr3.2}}, it follows that $\overline{\widehat{Y}}=\displaystyle\sum\limits_{k=1}^p \left(\alpha_{k}^{*}-\beta_{k}^{*}\right) \overline{X_{k}} + \gamma^{*},$ so $\overline{\widehat{Y}}=\overline{Y}.\qquad \Box$

\begin{property}\label{pr3.4}
For each unit $j,$ the quantile function for the distribution $\widehat{Y}(j)$  predicted by the \textit{DSD Model}, can be rewritten as follows:

$$\Psi_{\widehat{Y}(j)}^{-1}(t)-\overline{Y}=\displaystyle\sum\limits_{k=1}^p \alpha_{k}^{*}\left(\Psi_{X_{k}(j)}^{-1}(t)-\overline{X_{k}}\right)+\beta_{k}^{*}\left(-\Psi_{X_{k}(j)}^{-1}(1-t)+\overline{X_{k}}\right).$$
\end{property}

\textbf{Proof:} In \textit{Property \ref{pr3.3}}, we proved that

$$\overline{Y}=\displaystyle\sum\limits_{k=1}^p \left(\alpha_{k}^{*}-\beta_{k}^{*}\right) \overline{X_{k}} + \gamma^{*} \Longleftrightarrow \gamma^{*} = \overline{Y}-\displaystyle\sum\limits_{k=1}^p \left(\alpha_{k}^{*}-\beta_{k}^{*}\right) \overline{X_{k}}.$$

For the optimal solution $B^{*},$ for each unit $j,$ the quantile function predicted by the linear regression model \textit{DSD}, in \textit{Definition \ref{def3.4}}, is given by
$$\Psi_{\widehat{Y}(j)}(t)=\displaystyle\sum\limits_{k=1}^p \alpha_{k}^{*} \Psi_{X_{k}(j)}^{-1}(t)-\beta_{k}^{*}\Psi_{X_{k}(j)}^{-1}(1-t)+\gamma^{*}$$

which may be rewritten as
$$\Psi_{\widehat{Y}(j)}^{-1}(t)-\overline{Y}=\displaystyle\sum\limits_{k=1}^p \alpha_{k}^{*}\left(\Psi_{X_{k}(j)}^{-1}(t)-\overline{X_{k}}\right)+\beta_{k}^{*}\left(-\Psi_{X_{k}(j)}^{-1}(1-t)+\overline{X_{k}}\right). \qquad \Box $$

\newpage

\begin{property}\label{pr3.5}
For the observed and predicted distributions $Y(j)$ and $\widehat{Y}(j)$, with $j\in \{1,\ldots,m\},$ of the variable $Y,$ we have
$$\displaystyle\sum\limits_{j=1}^m \int_{0}^{1} \left(\Psi_{Y(j)}^{-1}(t)-\Psi_{\widehat{Y}(j)}^{-1}(t)\right)\left(\Psi_{\widehat{Y}(j)}^{-1}(t)-\overline{Y}\right)dt=0 .$$
\end{property}
\textbf{Proof:} The proof is given in \textit{Appendix B}.

\subsection{Goodness-of-fit measure} \label{ss3.5}

To complete the investigation of the linear regression model for histogram-valued variables, a goodness-of-fit measure remains to be deduced.
We define this measure in a similar way as in the classical model for real data.

\begin{proposicion}\label{p3.1} The sum of the square of the Mallows distance between each observed distribution $j,$ $j \in \{1,\ldots,m\},$ of the histogram-valued variable $Y,$ and the mean of the histogram-valued variable $Y,$ $\overline{Y},$ can be decomposed as follows:

\begin{eqnarray*}
\displaystyle\sum\limits_{j=1}^m D_{M}^{2} \left(\Psi_{Y(j)}^{-1}(t),\overline{Y}\right)=\displaystyle\sum\limits_{j=1}^m D_{M}^{2} \left(\Psi_{Y(j)}^{-1}(t), \Psi_{\widehat{Y}(j)}^{-1}(t)\right)+ \displaystyle\sum\limits_{j=1}^m D_{M}^{2} \left(\Psi_{\widehat{Y}(j)}^{-1}(t),\overline{Y}\right)
\end{eqnarray*}

\end{proposicion}

\textbf{Proof:}
Consider each observation $j$ of the  histogram-valued variable $Y,$ represented by its quantile function $\Psi_{Y(j)}^{-1}(t),$ and the mean this histogram-valued variable, $\overline{Y}.$ We have,

$$
\begin{array}{l}
\displaystyle\sum\limits_{j=1}^m D_{M}^{2} \left(\Psi_{Y(j)}^{-1}(t),\overline{Y}\right)=\displaystyle\sum\limits_{j=1}^m \int_{0}^{1} \left(\Psi_{Y(j)}^{-1}(t)-\overline{Y}\right)^2 dt =\\
=\displaystyle\sum\limits_{j=1}^m \int_{0}^{1} \left(\Psi_{Y(j)}^{-1}(t)-\Psi_{\widehat{Y}(j)}^{-1}(t)+\Psi_{\widehat{Y}(j)}^{-1}(t)-\overline{Y}\right)^2 dt=\\
=\displaystyle\sum\limits_{j=1}^m \int_{0}^{1} \left(\Psi_{Y(j)}^{-1}(t)-\Psi_{\widehat{Y}(j)}^{-1}(t)\right)^2 dt+\displaystyle\sum\limits_{j=1}^m \int_{0}^{1} \left(\Psi_{\widehat{Y}(j)}^{-1}(t)-\overline{Y}\right)^2 dt+\\
 \qquad +2\displaystyle\sum\limits_{j=1}^m \int_{0}^{1} \left(\Psi_{Y(j)}^{-1}(t)-\Psi_{\widehat{Y}(j)}^{-1}(t)\right) \left(\Psi_{\widehat{Y}(j)}^{-1}(t)-\overline{Y}\right)  dt\\
\end{array}
$$

From \textit{Property \ref{pr3.5}} we have,
$$\displaystyle\sum\limits_{j=1}^m \int_{0}^{1} \left(\Psi_{Y(j)}^{-1}(t)-\Psi_{\widehat{Y}(j)}^{-1}(t)\right)\left(\Psi_{\widehat{Y}(j)}^{-1}(t)-\overline{Y}\right)dt=0.$$
So, we can write

\begin{center}
\begin{eqnarray}
\lefteqn{\displaystyle\sum\limits_{j=1}^m D_{M}^{2} \left(\Psi_{Y(j)}^{-1}(t),\overline{Y}\right)=}\nonumber \\
&&=\displaystyle\sum\limits_{j=1}^m \int_{0}^{1} \left(\Psi_{Y(j)}^{-1}(t)-\Psi_{\widehat{Y}(j)}^{-1}(t)\right)^2 dt+\displaystyle\sum\limits_{j=1}^m \int_{0}^{1} \left(\Psi_{\widehat{Y}(j)}^{-1}(t)-\overline{Y}\right)^2 dt. \nonumber \qquad \Box
\end{eqnarray}
\end{center}

Therefore, similarly to the classical model, it is possible to define the goodness-of-fit measure of the \textit{DSD Model}.

\begin{definition}\label{def3.6}
Consider the observed and predicted distributions of the histogram-valued variable $Y$ and $\widehat{Y}$ represented, respectively, by their quantile functions $\Psi_{Y(j)}(t)$ and $\Psi_{\widehat{Y}(j)}^{-1}(t)$, and the mean of the histogram-valued variable $Y,$ $\overline{Y}.$ The goodness-of-fit measure is given by
$$
\Omega=\frac{\displaystyle\sum\limits_{j=1}^m D_{M}^{2} \left(\Psi_{\widehat{Y}(j)}^{-1}(t),\overline{Y}\right)}{\displaystyle\sum\limits_{j=1}^m D_{M}^{2} \left(\Psi_{Y(j)}^{-1}(t),\overline{Y}\right)}.
$$
\end{definition}

In classical linear regression, the coefficient of determination $\texttt{R}^2$ ranges from 0 to 1. In this case, the goodness-of-fit measure, $\Omega$, also ranges from 0 to 1.
\begin{proposicion}\label{p3.2}
The goodness-of-fit measure $\Omega$ ranges from 0 to 1.
\end{proposicion}

\textbf{Proof:} Consider the goodness-of-fit measure $\Omega=\frac{\displaystyle\sum\limits_{j=1}^m D_{M}^{2} \left(\Psi_{\widehat{Y}(j)}^{-1}(t),\overline{Y}\right)}{\displaystyle\sum\limits_{j=1}^m D_{M}^{2} \left(\Psi_{Y(j)}^{-1}(t),\overline{Y}\right)}.$ This measure is non-negative. So, $\Omega \geq 0.$

From \textit{ Proposition \ref{p3.1}}, we have
$$\begin{array}{l}
\displaystyle\sum\limits_{j=1}^m D_{M}^{2} \left(\Psi_{Y(j)}^{-1}(t),\overline{Y}\right)= \\
=\displaystyle\sum\limits_{j=1}^m \int_{0}^{1} \left(\Psi_{Y(j)}^{-1}(t)-\Psi_{\widehat{Y}(j)}^{-1}(t)\right)^2 dt+\displaystyle\sum\limits_{j=1}^m \int_{0}^{1}
\left(\Psi_{\widehat{Y}(j)}^{-1}(t)-\overline{Y}\right)^2 dt \Longleftrightarrow\\
\Longleftrightarrow  1 = \frac{\displaystyle\sum\limits_{j=1}^m \int_{0}^{1} \left(\Psi_{Y(j)}^{-1}(t)-\Psi_{\widehat{Y}(j)}^{-1}(t)\right)^2 dt}
 {\displaystyle\sum\limits_{j=1}^m D_{M}^{2} \left(\Psi_{Y(j)}^{-1}(t),\overline{Y}\right)} +
 \frac{\displaystyle\sum\limits_{j=1}^m \int_{0}^{1} \left(\Psi_{\widehat{Y}(j)}^{-1}(t)-\overline{Y}\right)^2 dt}{\displaystyle\sum\limits_{j=1}^m D_{M}^{2} \left(\Psi_{Y(j)}^{-1}(t),\overline{Y}\right)}\\
\end{array}$$

$$\begin{array}{l}
 \Longleftrightarrow  \Omega=1-\frac{\displaystyle\sum\limits_{j=1}^m \int_{0}^{1} \left(\Psi_{Y(j)}^{-1}(t)-\Psi_{\widehat{Y}(j)}^{-1}(t)\right)^2 dt}
 {\displaystyle\sum\limits_{j=1}^m D_{M}^{2} \left(\Psi_{Y(j)}^{-1}(t),\overline{Y}\right)}
\end{array}
$$

Since the term $\frac{\displaystyle\sum\limits_{j=1}^m \int_{0}^{1} \left(\Psi_{Y(j)}^{-1}(t)-\Psi_{\widehat{Y}(j)}^{-1}(t)\right)^2 dt}
 {\displaystyle\sum\limits_{j=1}^m D_{M}^{2} \left(\Psi_{Y(j)}^{-1}(t),\overline{Y}\right)}$ is non-negative, the value of $\Omega$ is always less than or equal to 1.
So, we have that $0 \leq \Omega \leq 1$.

Let us now analyze the extreme situations.

Suppose $\Omega = 0.$ In this case, $$\displaystyle\sum\limits_{j=1}^m D_{M}^{2} \left(\Psi_{\widehat{Y}(j)}^{-1}(t),\overline{Y}\right)=0 \Longleftrightarrow
\displaystyle\sum\limits_{j=1}^m \int_{0}^{1} \left(\Psi_{\widehat{Y}(j)}^{-1}(t)-\overline{Y}\right)^2 dt=0.$$
So, for all $j\in\left\{1,\ldots,m\right\},$ we have $\Psi_{\widehat{Y}(j)}^{-1}(t)-\overline{Y}=0\Longleftrightarrow \Psi_{\widehat{Y}(j)}^{-1}(t)=\overline{Y}$.
In this case the predicted function for all observations $j$ is a constant function.

Suppose now that $\Omega = 1.$ In this case, $$\displaystyle\sum\limits_{j=1}^m D_{M}^{2} \left(\Psi_{\widehat{Y}(j)}^{-1}(t),\overline{Y}\right)=\displaystyle\sum\limits_{j=1}^m D_{M}^{2} \left(\Psi_{Y(j)}^{-1}(t),\overline{Y}\right).$$
From the decomposition obtained in \textit{Proposition \ref{p3.1}} we have,
\vspace{-0.1cm}
$$\begin{array}{l}
\displaystyle\sum\limits_{j=1}^m D_{M}^{2} \left(\Psi_{Y(j)}^{-1}(t),\overline{Y}\right)=\displaystyle\sum\limits_{j=1}^m D_{M}^{2} \left(\Psi_{\widehat{Y}(j)}^{-1}(t),\overline{Y}\right)+\displaystyle\sum\limits_{j=1}^m D_{M}^{2} \left(\Psi_{\widehat{Y}(j)}^{-1}(t),\Psi_{Y(j)}^{-1}(t)\right)\\
\Longleftrightarrow \displaystyle\sum\limits_{j=1}^m D_{M}^{2} \left(\Psi_{\widehat{Y}(j)}^{-1}(t),\Psi_{Y(j)}^{-1}(t)\right)=0.
\end{array}$$

So, for all $j \in \left\{1,\ldots,m\right\},$ $$D_{M}^{2} \left(\Psi_{\widehat{Y}(j)}^{-1}(t),\Psi_{Y(j)}^{-1}(t)\right)=0 \Longleftrightarrow \int_{0}^{1} \left(\Psi_{\widehat{Y}(j)}^{-1}(t)-\Psi_{Y(j)}^{-1}(t)\right)^2 dt=0\Longrightarrow \Psi_{\widehat{Y}(j)}^{-1}(t)=\Psi_{Y(j)}^{-1}(t).$$
In this case, for each observation $j$, the predicted and observed quantile functions are coincident.

In conclusion $0 \leq \Omega \leq 1.$ If $\Omega =0$ there is no linear relationship between the histogram-valued variable $Y$ and the histogram-valued variables $X_{k}.$  If $\Omega =1,$ the linear relation is perfect, so the relationship between the histogram-valued variable $Y$ and histogram-valued variables $X_{k},$ with $k \in \{1,\ldots,p\},$ is exactly the relation defined by the linear regression model. $\qquad \Box$

\newpage

\section {Experiments} \label{s4}

To illustrate and analyze the \textit{DSD Model} we performed a simulation study and applied the method to real datasets.

\subsection{Simulation study} \label{ss4.1}

To analyze the behavior of the parameter estimation and the performance of the \textit{DSD Model} in different situations, we performed a simulation study. The first step was to generate the observations of the histogram-valued variables $X_{k},\, k=\left\{1,\ldots,p\right\}$ and $Y,$ where $Y$ is the variable to be modelized from $X_{k}$ by the linear relationship. Next, the parameters were estimated by the \textit{DSD Model} and goodness-of-fit measures computed, considering symbolic simulated data tables covering different situations. From these results it was possible to analyze the behavior of the model and draw some meaningful conclusions.

\subsubsection{Building symbolic simulated data tables} \label{ss4.1.1}
The observations of the explicative and response histogram-valued variables $X_{k}$ and $Y$ were generated in different ways.
\begin{itemize}
  \item The observations of each histogram-valued variable $X_{k}$ are created.\\
  According to the concept of symbolic variables, to obtain the $m$ observations associated to a histogram-valued variable $X_{k}$, we started by simulating 5000 real values corresponding to each unit. These values are then organized in histograms, that represent the empirical distribution for each unit. It was considered, without loss of generality, that in all observations, the subintervals of each histogram have the same weight (equiprobable) with frequency $0.10.$ This option is not restrictive, and is also supported by the work of Colombo \cite{colombo80}. If we had not considered equiprobable histograms with the same weight in all observations, we would have obtained a large number of different weights and consequently the subintervals would have very low frequencies. It is possible that histograms are not equiprobable, however, the weight in each subinterval has to be the same in all observations (see \textit{Subsection \ref{ss2.1}}). Furthermore diversity of weights would lead to rounding errors that increase the difficulty to work with histograms.

  \item The observations of the histogram-valued variable $Y$ are created.\\
 The histograms that are the observations of the histogram-valued variable $Y$ are obtained in three steps.
 First, we  consider the perfect linear regression, without error, given by $$\Psi_{Y^{*}(j)}^{-1}(t)=\gamma+\displaystyle\sum\limits_{k=1}^p\alpha_{k} \Psi_{X_{k}(j)}^{-1}(t)-\displaystyle\sum\limits_{k=1}^p\beta_{k}\Psi_{X_{k}(j)}^{-1}(1-t),$$ for particular values of the parameters. The histogram-valued variables $X_{k}$ and $Y^{*}$ are in a perfect linear relationship, this is however not what is intended to simulate a symbolic data table. Then, we disturb the perfect linear relationship by introducing an error function in the model $\Psi_{Y(j)}^{-1}(t)=\Psi_{Y^{*}(j)}^{-1}(t)+\varepsilon_{j}(t).$ The error function is a piece-wise linear function (but not necessarily a quantile function) defined by:

{\small
  \begin{center}
\begin{equation}
\varepsilon_{j}(t)=\left\{\begin{array}{lll}
                     a_{(j)_1}+\left(2\frac{t}{w_{1}}-1\right) b_{(j)_1} & if & 0 \leq t < w_{1} \\
                    a_{(j)_1} +b_{(j)_1}+ b_{(j)_2}+\left(2\frac{t-w_{1}}{w_{2}-w_{1}}-1\right) b_{(j)_{2}}  & if & w_{1} \leq t < w_{2} \\
                     \vdots & & \\
                   a_{(j)_1} +b_{(j)_1}+\sum_{i=2}^{n-1}2b_{(j)_{i}}+b_{(j)_{n}}\left(2\frac{t-w_{(n-1)}}{1-w_{(n-1)}}-1\right) b_{(j)_{n}}  & if & w_{n-1} \leq t \leq
                   1
 \end{array}
  \right.
\end{equation}
\end{center}}

Each quantile function $\Psi_{Y^{*}(j)}^{-1}(t)$ is randomly disturbed by the error function for different values of $a_{(j)_1}$ and  $b_{(j)_i},$ $i \in \left\{1,\ldots,n\right\}.$ These values might have a high or low variation depending on whether we want the linear regression between the variables to be better or worse. The selection of these values takes into account the ``magnitude"  of the values considered in each distribution $\Psi_{Y^{*}(j)}^{-1}(t).$ The values of $b_{(j)_i},$ cannot be lower than the minimum value of the half range $-r_{Y(j)_{i}},$ else for this unit $j$ and subinterval $i,$ the half range $r_{Y^{*}(j)_{i}}$ would be negative.
\end{itemize}

To perform the simulation study, symbolic data tables that illustrate different situations were created.
For each situation considered, 1000 data tables were generated. In this study a full factorial design was employed, with the following factors:

\begin{itemize}
    \item Number of explicative histogram-valued variables: $p=1$ and $p=3.$
     \item Parameters of the \textit{DSD Model}.
     \begin{description}
       \item[$\circ$] For $p=1:$
     \begin{description}
        \item[i)] $\alpha=2;$ $\beta=1;$ $\gamma=-1;$ ($\alpha$ and $\beta$ are close)
        \item[ii)]  $\alpha=2;$ $\beta=8;$ $\gamma=3;$ ($\alpha$ is lower than $\beta$)
         \item[iii)]  $\alpha=8;$ $\beta=0;$ $\gamma=4;$ ($\alpha$ is higher than $\beta$)
         \end{description}
        \item[$\circ$] For $p=3:$ $\alpha_1=2;$ $\beta_1=1;$ $\alpha_2=0.5;$ $\beta_2=3;$ $\alpha_3=4;$ $\beta_3=2;$ $\gamma=-1;$
       \end{description}

    \item Distribution of the microdata that allow generating the histograms corresponding to each observation of the variables $X_{k}, \quad k=\left\{1,\ldots,p\right\}:$
     \begin{description}
        \item[i)] Uniform distribution \\
        {\footnotesize ($X_{k}(j)\sim \mathcal{U}(\delta_1(j),\delta_2(j))$ where for each $j \in \left\{1,\ldots,m\right\},$ $\delta_1(j)\sim \mathcal{U}(-2,0)$ and $\delta_2(j)\sim \mathcal{U}(0,2)$)};
        \item[ii)] Normal distribution\\
         {\footnotesize($X_{k}(j)\sim \mathcal{N}(\mu(j),\sigma^2(j))$ where for each $j \in \left\{1,\ldots,m\right\},$ $\mu(j)\sim \mathcal{U}(0,1)$ and $\sigma^2(j) \sim \mathcal{U}(0,2)$)};
         \item[iii)] Log-Normal distribution \\
         {\footnotesize ($X_{k}(j)\sim ln\mathcal{N}(\mu(j),\sigma^2(j))$ where for each $j \in \left\{1,\ldots,m\right\},$ $\mu(j)\sim \mathcal{U}(-0.5,0.5)$ and $\sigma^2(j) \sim \mathcal{U}(0.5,1)$)};
        \item[iv)] Mixture of distributions, randomly selected from $\{$Uniform:$X_{k}(j)\sim \mathcal{U}(1,3)$ ; Normal: $X_{k}(j)\sim \mathcal{N}(1,1)$; Chi-square: $X_{k}(j)\sim \mathcal{X}^2(1)$; Log-normal: $X_{k}(j)\sim ln\mathcal{N}(0,0.5)$; -Log-normal: $X_{k}(j)\sim - ln\mathcal{N}(0,0.5)\}$
       \end{description}
  \item Level of the linearity of the model:
       \begin{description}
        \item[i)] High linearity - In the error $\varepsilon_{j}(t)$, the values of $a_{(j)_1}$ and $b_{(j)_i}$ are randomly generated in \\ $\mathcal{U}_{c1}=\mathcal{U}(-\frac{3}{8}*\ddot{C},\frac{3}{8}*\ddot{C})$\footnote[1]{$\ddot{C}=\frac{1}{n}\displaystyle\sum\limits_{i=1}^n \left|\displaystyle\sum\limits_{j=1}^m \frac{1}{m} c_{Y(j)_i}\right|$} and $\mathcal{U}_{r1}=\mathcal{U}(-\frac{1}{8}*min(r_{Y^{*}(j)_{i}}),\frac{1}{8}*min(r_{Y^{*}(j)_{i}})),$ respectively;
        \item[ii)]  Moderate linearity - In the error $\varepsilon_{j}(t)$, the values of $a_{(j)_1}$ and $b_{(j)_i}$ are randomly generated in $\mathcal{U}_{c2}=\mathcal{U}(-\frac{3}{2}*\ddot{C},\frac{3}{2}*\ddot{C})$ and $\mathcal{U}_{r2}=\mathcal{U}(-\frac{1}{2}*min(r_{Y^{*}(j)_{i}}),\frac{1}{2}*min(r_{Y^{*}(j)_{i}})),$ respectively;
         \item[iii)]  Low linearity -  In  the error $\varepsilon_{j}(t)$, the values of $a_{(j)_1}$ and $b_{(j)_i}$ are randomly generated in $\mathcal{U}_{c3}=\mathcal{U}(-3*\ddot{C},3*\ddot{C})$ and $\mathcal{U}_{r3}=\mathcal{U}(-min(r_{Y^{*}(j)_{i}}),min(r_{Y^{*}(j)_{i}})),$ respectively;
       \end{description}
        \item Sample size: m=10; 30; 100; 250.
\end{itemize}

It is important to underline that in this simulation study, it was only possible to control the type of distributions in observations of the explicative histogram-valued variables. This simulation does not allow selecting the distributions in the observations of the response variable. These distributions depend of the distribution of the variables $Y^{*}$ (that in some situations are known, as we will see later) and the disturbance applied to the histograms $Y^{*}(j).$

\subsubsection{Description of the simulation study} \label{ss4.1.2}

The simulated symbolic data tables include the observations of the histogram-valued variables $X_{k}$ and $Y,$ according to the previous description and factors. For these tables, we computed the estimated parameters for the \textit{DSD Model} and the goodness-of-fit measures. As we considered 1000 replications for each situation, the values presented are the means of the obtained values and the respective standard deviation values (represented by $s$).

The goodness-of-fit measures considered in this study are :
\begin{itemize}
	\item $\overline{\Omega},$ where $\Omega$ is the measure deduced from the \textit{DSD Model} (see \textit{Subsection \ref{ss3.5}});
	\item Root-mean-square error $(RMSE_M)$, a measure defined using the Mallows distance (also used in the \textit{DSD Model}), proposed by Irpino and Verde \cite{irve12}; it is defined by
 $$RMSE_{M}=\sqrt{ \frac{\displaystyle \sum_{j=1}^{m}\displaystyle \int_{0}^{1}\left(\Psi_{\widehat{Y}(j)}^{-1}(t)-\Psi_{Y(j)}^{-1}(t)\right)^2dt}{m}} $$

 \item Adaptations of the lower $(RMSE_{L})$ and the upper bound $(RMSE_{U})$ root-mean-square that Neto and Carvalho \cite{netocar10}, \cite{netocar08}, use to study the performance of the linear regression models defined for interval-valued variables; for histogram-valued variables, the $RMSE_{L}$ and the $RMSE_{U}$ are given by:
 $$RMSE_{L}=\frac{1}{m}\displaystyle \sum_{j=1}^{m}\sqrt{ \displaystyle \sum_{i=1}^{n}(\underline{I}(j)_{i}-\widehat{\underline{I}}(j)_{i})^2p_i} \qquad RMSE_{U}=\frac{1}{m}\displaystyle \sum_{j=1}^{m} \sqrt{ \displaystyle \sum_{i=1}^{n}(\overline{I}(j)_{i}-\widehat{\overline{I}}(j)_{i})^2p_i}$$

with $\left[\underline{I}(j)_{i},\overline{I}(j)_{i}\right[$ and $\left[\widehat{\underline{I}}(j)_{i},\widehat{\overline{I}}(j)_{i}\right[$ the subintervals $i \in \left\{1,\ldots,n\right\}$ of the observed and predicted histograms, for each unit $j.$
\end{itemize}

In \textit{Appendix C} four tables are presented, each of which containing the results obtained with $p=1$ and all distributions used for defining the histogram values of  $X_{j}$, i.e., all observations with Uniform distribution (\textit{Table \ref{table1SA3}}), Normal distribution (\textit{Table \ref{table2SA3}}), Log-Normal distribution (\textit{Table \ref{table3SA3}}) and the observations of $X(j)$ for a mixture of distributions (\textit{Table \ref{table4SA3}}).
In the last two tables, similar results are presented for the cases where $p=3$ (\textit{Table \ref{table5SA3}}, \textit{Table \ref{table6SA3}} ).

\subsubsection{Results and conclusions} \label{ss4.1.2}

The main goals of this study are to analyze the behavior of the parameters' estimation and the performance of the \textit{DSD Model}. The results obtained for the model with one or three explicative variables are similar, and as such in this subsection we will only be analyzing with detail the results obtained when $p=1.$ The results obtained when $p=3$ may be found in \textit{Table \ref{table5SA3}} and \textit{Table \ref{table6SA3}} of \textit{Appendix C.}  For $p=1$ it is also our goal to analyze how the symmetry/asymmetry of the distributions in observations of the explicative histogram-valued variable affect the symmetry/asymmetry of the distributions in observations of the predictive variable.

\begin{figure}[h]
\begin{center}
\includegraphics[width=1\textwidth]{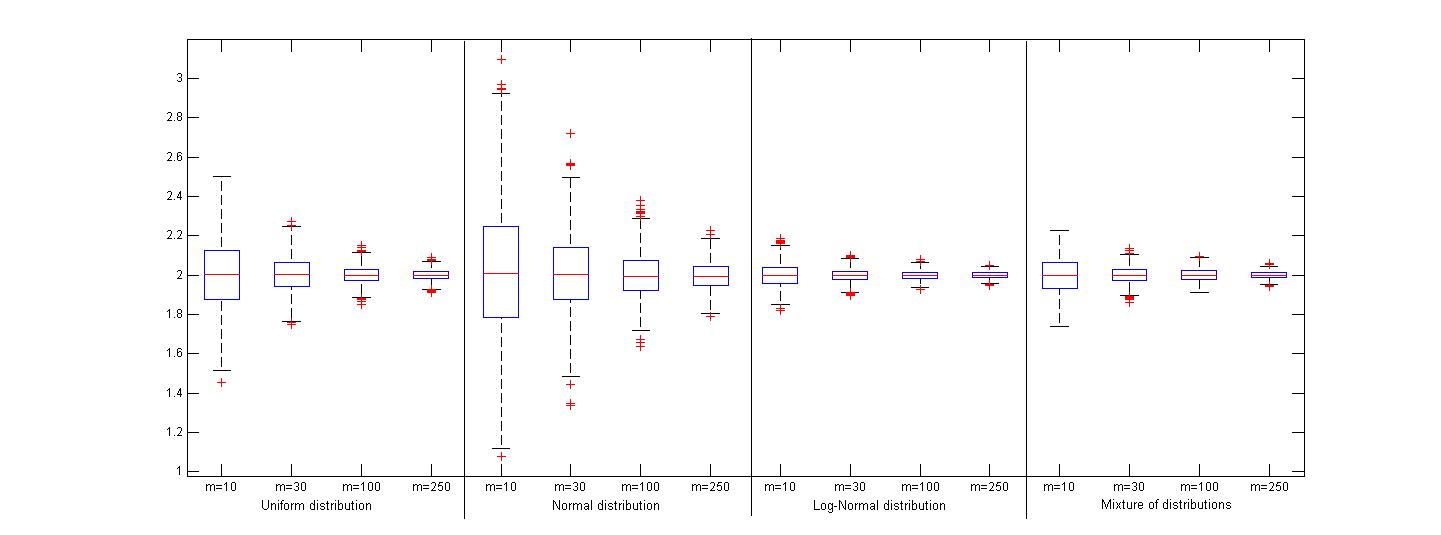}
\caption{Boxplot for the estimated parameter $\widehat{\alpha}$ for a high level of linearity.}
\label{fig1S}
\end{center}
\end{figure}

\begin{figure}[h]
\begin{center}
\includegraphics[width=1\textwidth]{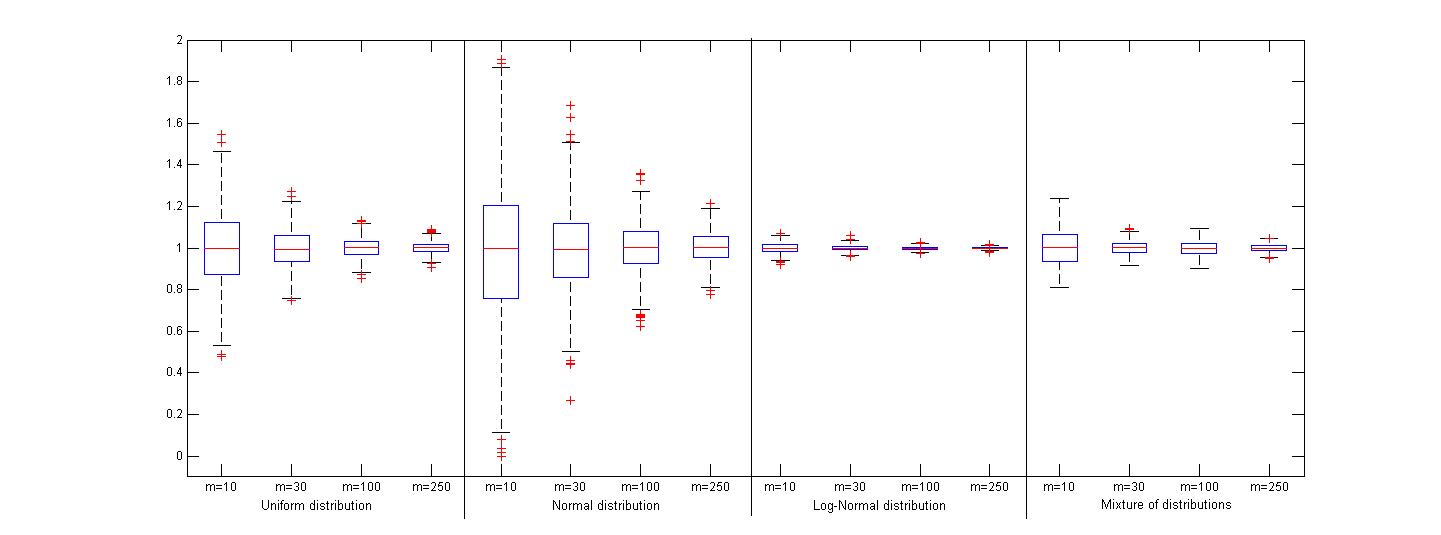}
\caption{Boxplot for the estimated parameter $\widehat{\beta}$ for a high level of linearity.}
\label{fig2S}
\end{center}
\end{figure}

\begin{figure}[h]
\begin{center}
\includegraphics[width=1\textwidth]{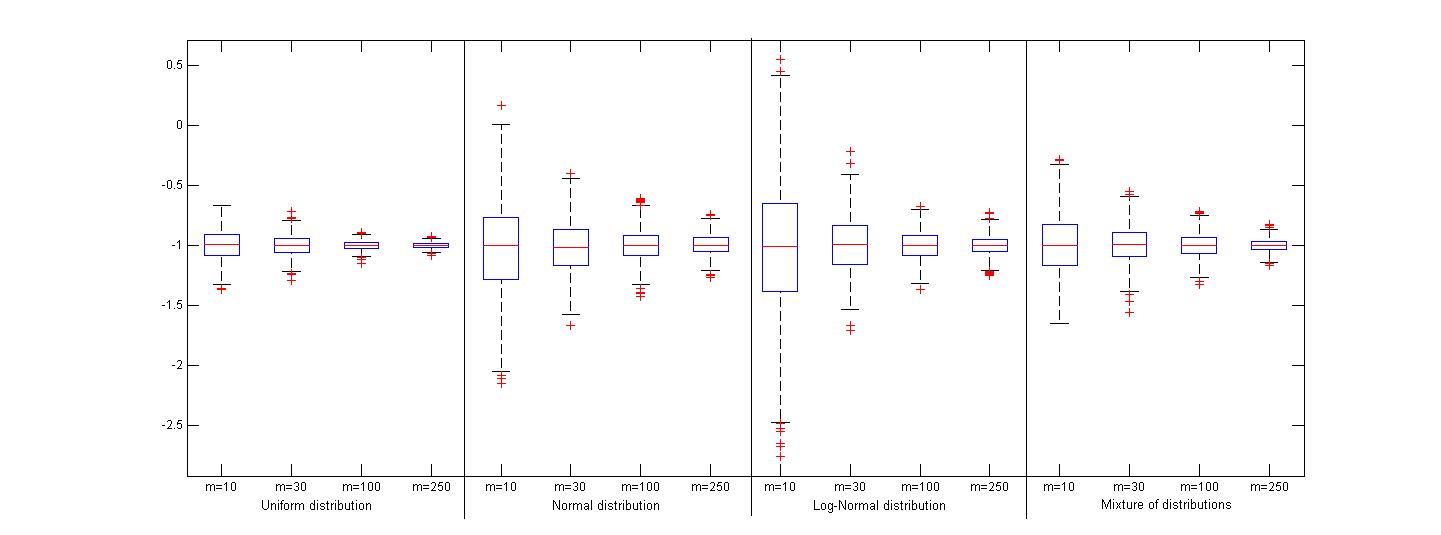}
\caption{Boxplot for the estimated parameter $\widehat{\gamma}$ for a high level of linearity.}
\label{fig3S}
\end{center}
\end{figure}

\vspace{0.5cm}
\textit{Concerning the analysis of the parameters' estimation.}\\
For the simple case with one explicative histogram-valued variable $X,$ we considered the mean of the values obtained for $\widehat{\alpha}, \widehat{\beta},$ $\widehat{\gamma}$ and the mean square error (MSE) \cite{netocar10}. In this case, as we replicated the same situation 1000 times,  $MSE=\frac{1}{1000}\displaystyle \sum_{i=1}^{1000} (\theta-\widehat{\theta}_i)^2$ (with $\theta$ corresponding to each parameter of the model).
Comparing the first four tables in \textit{Appendix C}, we can see that the behavior of  the parameters' estimation is independent of the distribution used to generate the microdata of the explicative variables. Futhermore, the estimated parameters are almost always close to the initial parameter values irrespecive of the level of the linearity. This behavior is expected when the level of linearity is higher but when the level of linearity is moderate or low it would not be surprising if the estimated parameters  were more distant from the original ones, since it seems natural that other models exist that better adjust the symbolic data. This is essentially observed  in \textit{Tables \ref{table1SA3} and \ref{table2SA3}}, when the initial parameters $\alpha$ and $\beta$ are not close and when the number of observations is lower. According to this, the analysis of the behavior of the $MSE$ and the mean of the estimated parameters is essentially applicable in situations where the level of linearity is high. For these cases we observe that the values of the $MSE$ decrease and tend to zero as the number of observations increases and the mean of the estimated parameters becomes very close to the respective parameters of the model. These results confirm the empirical consistency of the estimators. In the boxplots presented in \textit{Figures \ref{fig1S}, \ref{fig2S}} and \textit{\ref{fig3S}} we may observe that, considering the different types of distributions used to generate the histogram values of $X$, the boxes reduced their ranges around the true values of the respective parameters as the number $m$ of observations increases. The figures illustrate only the situation when $\alpha=2; \beta=1; \gamma=-1$, but the behavior for the other values is similar. It can also be observed that the range of the variation of the estimated parameters relatively to its original value is influenced by the distribution of the observations.

\textit{Concerning the study of the goodness-of-fit measures.}\\
The values obtained for $\Omega$ show that this value provides a good evaluation for the level of linearity. The models slightly disturbed presented values of $\Omega$ close to one. On the other hand, when the error function applied to the model presented a high level of variability the values of $\Omega$ are closer to zero. Furthermore, the means of the values $\Omega$ are consistent with the respective values of the measures $RMSE_M;$ $RMSE_L$ and $RMSE_U.$ In general, as expected, the highest values of the $\Omega$ correspond to the lowest values of $RMSE_M.$ In all tables of \textit{Appendix C} we can also verify that in almost all situations, the values of the goodness of fit measure decrease in the same proportion as the levels of linearity. The level of linearity and the mean values associated to the goodness-of-fit measures $RMSE_M;$ $RMSE_L$ and $RMSE_U$ increase approximately four times when we pass from high to moderate linearity and approximately two times when we pass from moderate to low. This increase is an exact reflection of the range of variability tested in this study for the error function (four times from the high to moderate linearity and two times from moderate to low).

\textit{Tables \ref{table7S}} and  \textit{\ref{table8S}} illustrate the results that were obtained in an additional study for the original model with $\alpha=2; \beta=1; \gamma=-1$ and only for samples with 10 and 100 observations. Other situations were tested and the results were similar. The goal of the study was to analyze the level of sensitivity of the measure $\Omega$ to different kinds of error functions, that in some cases affect more the half range of the subintervals of the histograms and in others the centers. To analyze this behavior, the values of $\Omega$ were determined, considering different error functions that use three levels of variability for the values of $a_{(j)_1}:$ $\mathcal{U}_{c1},$ $\mathcal{U}_{c2},$ $\mathcal{U}_{c3}$ as defined in \textit{Subsection \ref{ss4.1.1}} and, for each one, three levels of variability for $b_{(j)_i}:$ $\mathcal{U}_{r1},$ $\mathcal{U}_{r2},$ $\mathcal{U}_{r3}$ as defined in \textit{Subsection \ref{ss4.1.1}}.

\setlength{\extrarowheight}{1.5pt}
\begin{table}[h!]
\begin{center}
{\tiny
\begin{tabular}{|c c|c|c|c|c|c|c|}
  \hline
  & & \multicolumn{3}{|c|}{$\overline{\Omega}_\mathcal{U} \, \, (s)$} & \multicolumn{3}{|c|}{$\overline{\Omega}_\mathcal{N} \, \, (s)$} \\ \hline
    & $m$ & $b_{(j)_i}\hspace{-0.15cm}\sim \mathcal{U}_{r1}$\hspace{-0.15cm} & $b_{(j)_i}\hspace{-0.15cm} \sim \mathcal{U}_{r2}$ & $b_{(j)_i} \hspace{-0.15cm}\sim \mathcal{U}_{r3}$ & $b_{(j)_i} \hspace{-0.15cm}\sim \mathcal{U}_{r1}$ & $b_{(j)_i} \hspace{-0.15cm}\sim \mathcal{U}_{r2}$ & $b_{(j)_i} \hspace{-0.15cm}\sim \mathcal{U}_{r3}$ \\ \hline
\multirow{2} {*}{$a_{(j)_1}\hspace{-0.15cm}\sim \mathcal{U}_{c1}$} & 10  & $0.9741 \, (0.0089)$ & $0.9455  \,  (0.0216)$ & $0.8643 \,   (0.0535)$ & $0.9792 \,   (0.0079)$ &$0.9145 \,   (0.0344)$ & $0.7587  \,  (0.0835)$\\
                                  & 100 & $0.9648 \,   (0.0032)$ & $0.9322 \,  (0.0076)$  & $0.8403  \, (0.0191)$ & $0.9762  \, (0.0025)$ & $0.8982 \,  (0.0122)$& $0.7160 \,  (0.0293)$\\ \hline
\multirow{2} {*}{$a_{(j)_1}\hspace{-0.15cm}\sim \mathcal{U}_{c2}$} & 10 & $0.7323  \, (0.0727)$& $0.7163  \, (0.0786)$ & $0.6690 \,  (0.0906)$ & $0.7980  \, (0.0583)$ & $0.7567  \, (0.0691)$& $0.6555  \, (0.0997)$\\
                            & 100 & $0.6476  \,  (0.0222)$ & $0.6332 \,  (0.0238)$ & $0.5905  \, (0.0288)$ & $0.7701 \,  (0.0165)$ & $0.7217  \, (0.0232)$ & $0.6008 \, (0.0326)$ \\ \hline
\multirow{2} {*}{$a_{(j)_1}\hspace{-0.15cm}\sim \mathcal{U}_{c3}$} & 10 & $0.4422 \,  (0.1098)$ & $0.4320  \, (0.1090)$ & $0.4211  \, (0.1120)$  & $0.5192  \, (0.0931)$ & $0.5017 \,  (0.0944)$ & $0.4587  \,  (0.1013)$\\
                            & 100 & $0.3195  \, (0.0251)$ & $0.3156  \, (0.0265)$ & $0.3054 \,  (0.0268)$ & $0.4627  \, (0.0243)$ & $0.4436  \, (0.0260)$ & $0.3970 \,  (0.0306)$ \\ \hline
\end{tabular}}
\caption{Mean values of $\Omega$ considering different levels of linearity, when the distributions generating observations of $X$ are Uniform $\left(\overline{\Omega}_\mathcal{U}\right)$ and Normal $\left(\overline{\Omega}_\mathcal{N}\right).$}
\label{table7S}
\end{center}
\end{table}

\setlength{\extrarowheight}{1.5pt}
\begin{table}[h!]
\begin{center}
{\tiny
\begin{tabular}{|c c|c|c|c|c|c|c|}
  \hline
  & & \multicolumn{3}{|c|}{$\overline{\Omega}_{Ln\mathcal{N}}  \,  \, (s)$} & \multicolumn{3}{|c|}{$\overline{\Omega}_M  \,  \, (s)$} \\ \hline
    & $m$ & $b_{(j)_i}\hspace{-0.15cm}\sim \mathcal{U}_{r1}$\hspace{-0.15cm}& $b_{(j)_i}\hspace{-0.15cm} \sim \mathcal{U}_{r2}$ & $b_{(j)_i} \hspace{-0.15cm}\sim \mathcal{U}_{r3}$ & $b_{(j)_i} \hspace{-0.15cm}\sim \mathcal{U}_{r1}$ & $b_{(j)_i} \hspace{-0.15cm}\sim \mathcal{U}_{r2}$ & $b_{(j)_i} \hspace{-0.15cm}\sim \mathcal{U}_{r3}$ \\ \hline
\multirow{2} {*}{$a_{(j)_1}\hspace{-0.15cm}\sim \mathcal{U}_{c1}$} & 10 & $0.9843 \,  (0.0054)$ & $0.8848  \, (0.0389)$ & $0.6699  \, (0.0994)$ & $0.9780 \,  (0.0078)$ &$0.9203  \, (0.0275)$ & $0.7789 \,  (0.0721)$\\
                                  & 100 & $0.9822  \, (0.0019)$& $0.8786  \, (0.0146)$  & $0.6571  \, (0.0393)$ & $0.9719  \, (0.0026)$& $0.9042 \,  (0.0095)$ & $0.7403 \,  (0.0220)$\\ \hline
\multirow{2} {*}{$a_{(j)_1}\hspace{-0.15cm}\sim \mathcal{U}_{c2}$} & 10 & $0.8769 \,  (0.0344)$ & $0.8032  \, (0.0587)$ & $0.6130  \, (0.1092)$ &$0.7765 \,  (0.0569)$ & $0.7453  \, (0.0706)$& $0.6568  \, (0.0954)$\\
                            & 100 & $0.8556  \, (0.0112)$ & $0.7765 \,  (0.0229)$ & $0.5982  \, (0.0420)$ &$0.7225  \, (0.0182)$ & $0.6838  \, (0.0223)$& $0.5884  \, (0.0287)$ \\ \hline
\multirow{2} {*}{$a_{(j)_1}\hspace{-0.15cm}\sim \mathcal{U}_{c3}$} & 10 & $0.6542  \, (0.0762)$ & $0.6114 \,  (0.0923)$ & $0.5067  \, (0.1208)$  & $0.4884  \, (0.0948)$& $0.4791  \, (0.0956)$ & $0.4422  \, (0.1024)$\\
                            & 100 & $0.6075  \, (0.0224)$ & $0.5654 \,  (0.0293)$ & $0.4638  \, (0.0418)$ &$0.3979  \, (0.0228)$ & $0.3855  \, (0.0226)$ & $0.3526 \,  (0.0260)$ \\ \hline
\end{tabular}}
\caption{Mean values of $\Omega$ considering different levels of linearity when the distributions generating observations of $X$ are Log-Normal $\left(\overline{\Omega}_{Ln\mathcal{N}}\right)$ and a mixture of distributions $\left(\overline{\Omega}_M\right).$}
\label{table8S}
\end{center}
\end{table}

Based on these results, we can say that, except when the observations of the explicative variables follow a  Log-Normal distribution, the linearity between histogram-valued variables is more affected by disturbances in the center of the subintervals than in the half range. This behavior is not surprising because the distance associated to this model is the the Mallows distance and as we observe in its definition the contribution to the centers of the subintervals is three times more then that of the half-ranges (see \textit{Definition \ref{def3.1}} and \textit{Property \ref{pr3.1}}). On the other hand, when all observations of the explicative histogram-valued variable have asymmetric distributions, the influence of the disturbance in the center and half-range may be similar. This different behavior may be related to the kind of distribution (symmetric/asymmetric) predicted for the observations of the histogram-valued variable $\widehat{Y}(j),$ as we will see next.

\vspace{0.5cm}

\begin{figure}[h]
\begin{center}
\includegraphics[width=1\textwidth]{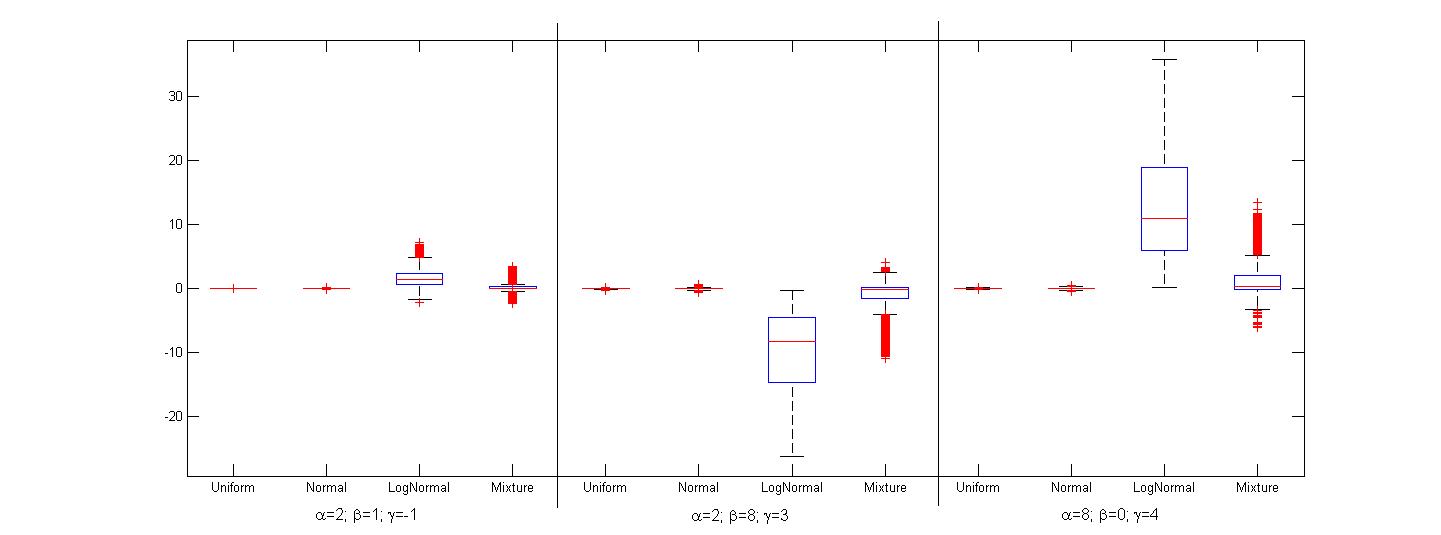}
\caption{Boxplots of the difference between the mean and median of the estimated distributions in all situations with $n=10$ considered in the simulation study, for $p=1.$}
\label{fig4S}
\end{center}
\end{figure}

\textit{ Concerning symmetry/assymetry of $\widehat{Y}(j).$}\\
In this simulation study it was possible to analyze the symmetry/assymetry of the predicted distributions obtained by the \textit{DSD Model}, taking into consideration the symmetry/asymmetry of the distributions in the observations of the histogram-valued variables $X$ and the values of the parameters $\alpha$ and $\beta.$ When the observation of the histogram-valued variable $X$ has a symmetric distribution, represented by $\Psi^{-1}_{X(j)}(t)$, the respective symmetric distribution $-\Psi^{-1}_{X(j)}(1-t)$ is also symmetric, but when the distribution $\Psi^{-1}_{X(j)}(t)$ is asymmetric positive (negative) (Log-Normal, for example), the respective symmetric distribution $-\Psi^{-1}_{X(j)}(1-t)$ is asymmetric negative (positive). In the \textit{DSD Model}, the predicted distributions are obtained from $\Psi_{\widehat{Y}(j)}^{-1}(t)=\gamma+\alpha \Psi_{X(j)}^{-1}(t)-\beta\Psi_{X_(j)}^{-1}(1-t).$ Therefore if the distribution $\Psi^{-1}_{X(j)}(t)$ is symmetric the distribution of $\Psi_{\widehat{Y}(j)}^{-1}(t)$ also tends to be symmetrical, if the distributions $\Psi^{-1}_{X(j)}(t)$ is asymmetric the distribution of $\Psi_{\widehat{Y}(j)}^{-1}(t)$  tends to be symmetrical when the values of $\alpha$ and $\beta$ are close, asymmetrical negative (resp. positive) when the value of $\alpha$ is lower (resp. higher) than the value of $\beta.$ These conclusions are illustrated in \textit{Figure \ref{fig4S}} considering all predicted distributions in the simulation study.

In conclusion, when the distributions of observations $X(j)$ are symmetric; asymmetric positive or assymmetric negative, it is possible to forecast whether the distributions of $\widehat{Y}(j)$ will be symmetric or asymmetric.

\subsection{Applied examples} \label{ss4.2}

\subsubsection{The relation between the hematocrit values and hemoglobin values} \label{ss4.2.1}

This first example was presented in \cite{bidi07} to illustrate their linear regression model for histogram-valued variables. In this case, we have the symbolic data in \textit{Table \ref{table1E1}}, where 10 units are described by two symbolic variables, the hematocrit and the hemoglobin.

\begin{table}[h!]
\begin{center}
{\scriptsize
\begin{tabular}{|c|l|l|}
  \hline
  Obs. & Hematocrit (Y) & Hemoglobin (X) \\
  \hline
  1 & $\left\{\left[33.29;37.52\right[,0.6;\left[37.52;39.61\right],0.4\right\}$ & $\left\{\left[11.54;12.19\right[,0.4;\left[12.19;12.8\right],0.6\right\}$ \\
  2 & $\left\{\left[36.69;39.11\right[,0.3;\left[39.11;45.12\right],0.7\right\}$ & $\left\{\left[12.07;13.32\right[,0.5;\left[13.32;14.17\right],0.5\right\}$ \\
  3 & $\left\{\left[36.69;42.64\right[,0.5;\left[42.64;48.68\right],0.5\right\}$ & $\left\{\left[12.38;14.2\right[,0.3;\left[14.2;16.16\right],0.7\right\}$ \\
  4 & $\left\{\left[36.38;40.87\right[,0.4;\left[40.87;47.41\right],0.6\right\}$ & $\left\{\left[12.38;14.26\right[,0.5;\left[14.26;15.29\right],0.5\right\}$\\
  5 & $\left\{\left[39.19;50.86\right],1\right\}$ & $\left\{\left[13.58;14.28\right[,0.3;\left[14.28;16.24\right],0.7\right\}$ \\
  6 & $\left\{\left[39.7;44.32\right[,0.4;\left[44.32;47.24\right],0.6\right\}$ & $\left\{\left[13.81;14.5\right[,0.4;\left[14.5;15.2\right],0.6\right\}$\\
  7 & $\left\{\left[41.56;46.65\right[,0.6;\left[46.65;48.81\right],0.4\right\}$ & $\left\{\left[14.34;14.81\right[,0.5;\left[14.81;15.55\right],0.5\right\}$\\
  8 & $\left\{\left[38.4;42.93\right[,0.7;\left[42.93;45.22\right],0.3\right\}$ & $\left\{\left[13.27;14.0\right[,0.6;\left[14.0;14.6\right],0.4\right\}$\\
  9 & $\left\{\left[28.83;35.55\right[,0.5;\left[35.55;41.98\right],0.5\right\}$ & $\left\{\left[9.92;11.98\right[,0.4;\left[11.98;13.8\right],0.6\right\}$\\
  10 & $\left\{\left[44.48;52.53\right],1\right\}$ & $\left\{\left[15.37;15.78\right[,0.3;\left[15.78;16.75\right],0.7\right\}$\\
  \hline
\end{tabular}}
\caption{Example of symbolic data table where the two variables hematocrit and hemoglobin are histogram-valued variables.}
\label{table1E1}
\end{center}
\end{table}

We predicted the quantile function representing the distribution taken by the histogram-valued variable $Y$ from the \textit{DSD Model}, and obtained:
$$\Psi_{\widehat{Y}(j)}^{-1}(t)=-1.953+3.5598\Psi_{X(j)}^{-1}(t)-0.4128\Psi_{X(j)}^{-1}(1-t)$$

The value of the goodness-of-fit measure is, for this case, $\Omega=0.96.$

\begin{figure}[h]
\begin{center}
\includegraphics[width=0.9\textwidth]{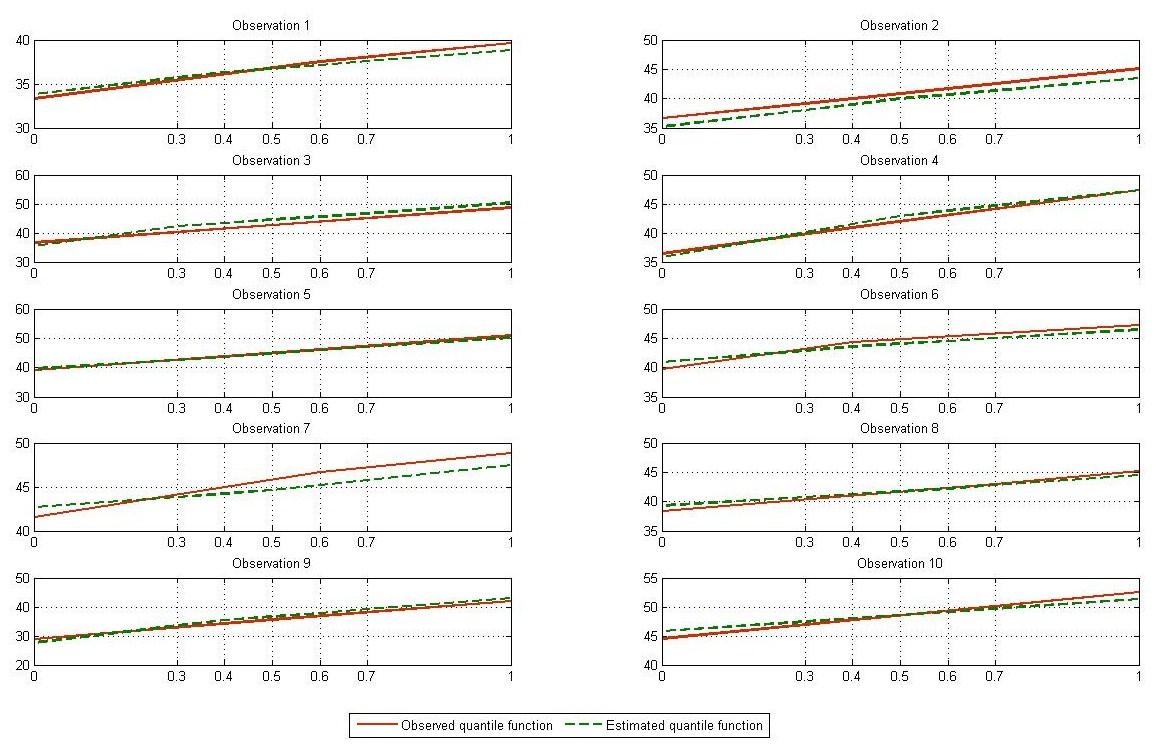}
\caption{Observed and predicted quantile functions of each observation in \textit{Table \ref{table1E1}}.}
\label{fig1E1}
\end{center}
\end{figure}

In \textit{Figure \ref{fig1E1}} we may  compare the quantile functions of the observed and predicted distributions of the histogram-valued variable $Y.$ As it may be observed, the distributions are very similar, in agreement with the value of the coefficient of determination, $\Omega$. The observed and predicted histograms of each observation are presented in \textit{Appendix D}.

\begin{figure}[h]
\begin{center}
 \begin{subfigure}
                \centering
                \includegraphics[width=0.45\textwidth]{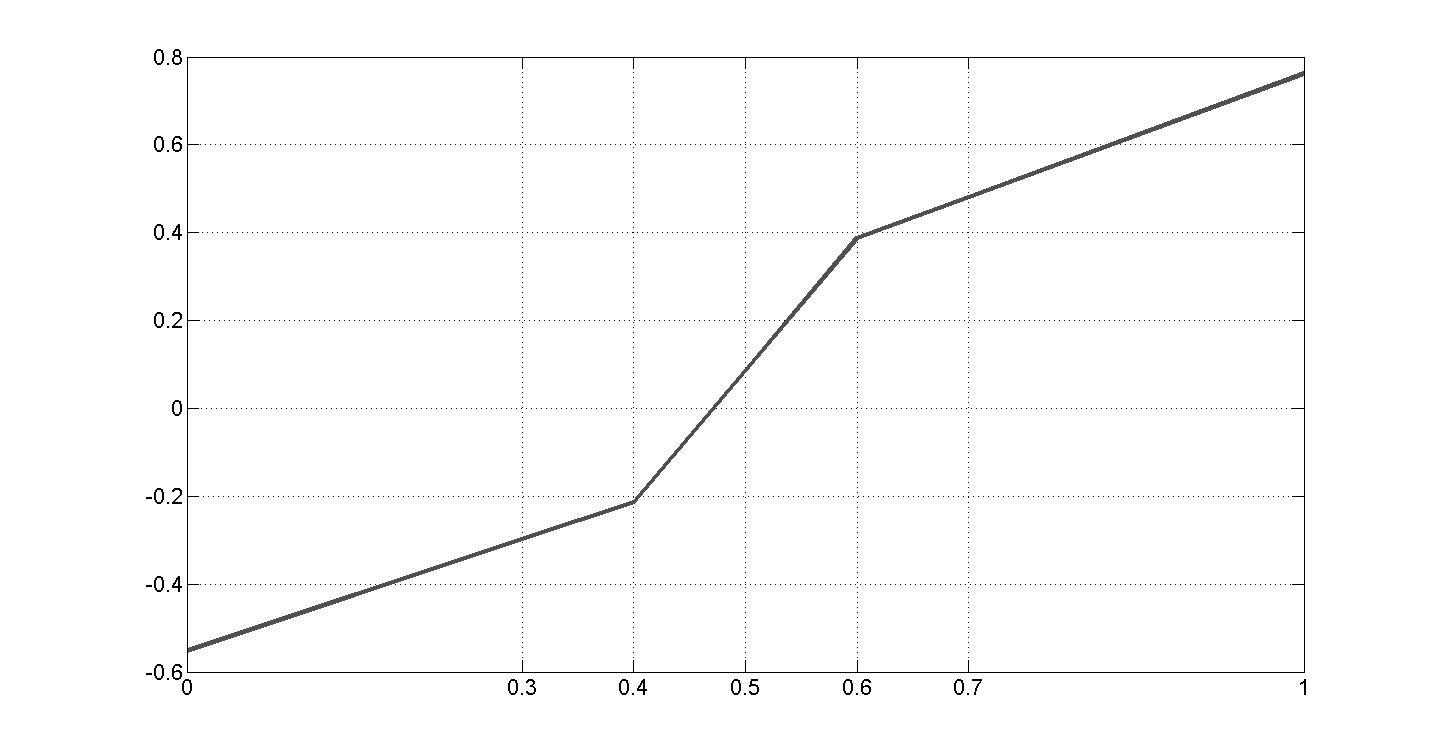}
        \end{subfigure}%
        ~
 \begin{subfigure}
                \centering
                \includegraphics[width=0.45\textwidth]{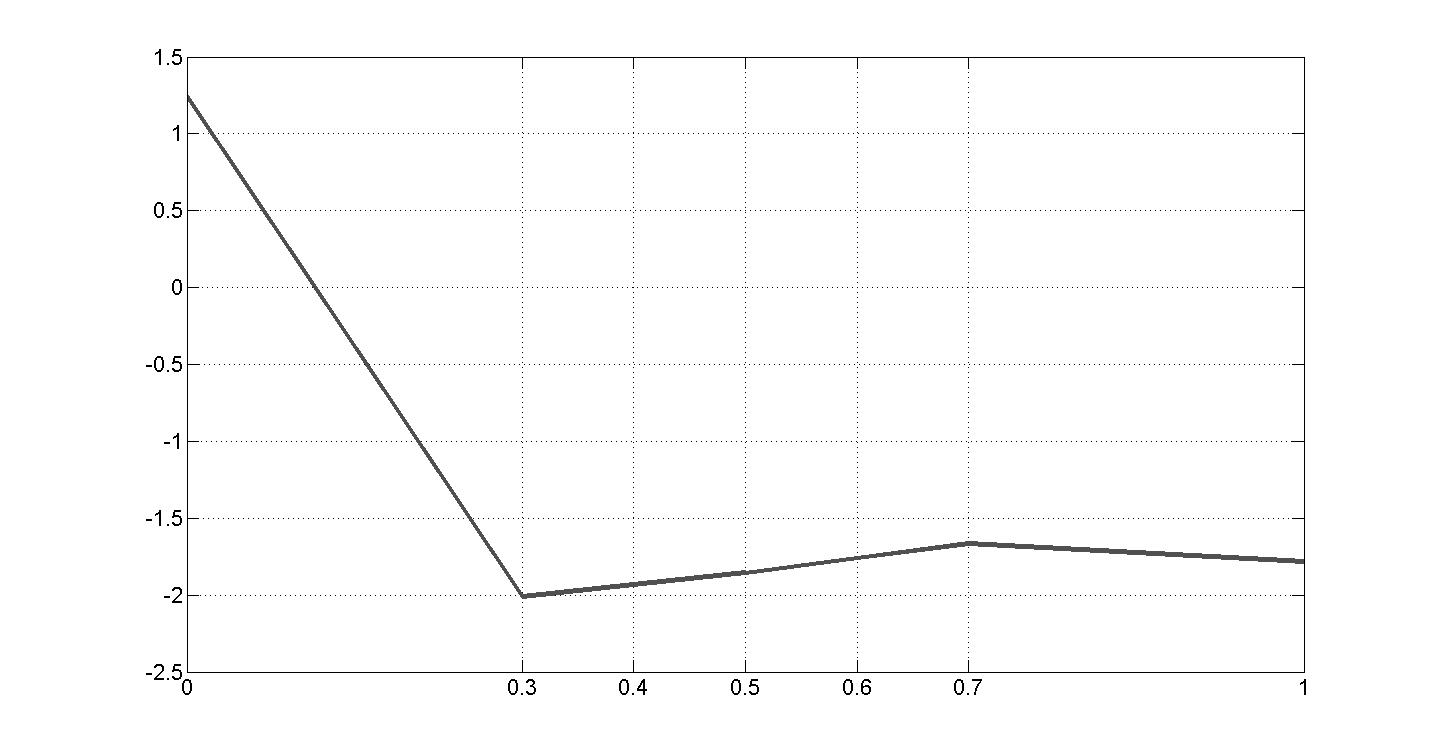}
        \end{subfigure}%
\caption{Error function for the observations $1$ and $3.$}
\label{fig2E1}
\end{center}
\end{figure}

When we predict a histogram value we have always associated an error function defined according to \textit{Definition \ref{def3.4}}. For this example, in \textit{Figure \ref{fig2E1}} we can observe the error function for observations $1$ and $3.$

\begin{figure}[h]
\begin{center}
\includegraphics[width=0.8\textwidth]{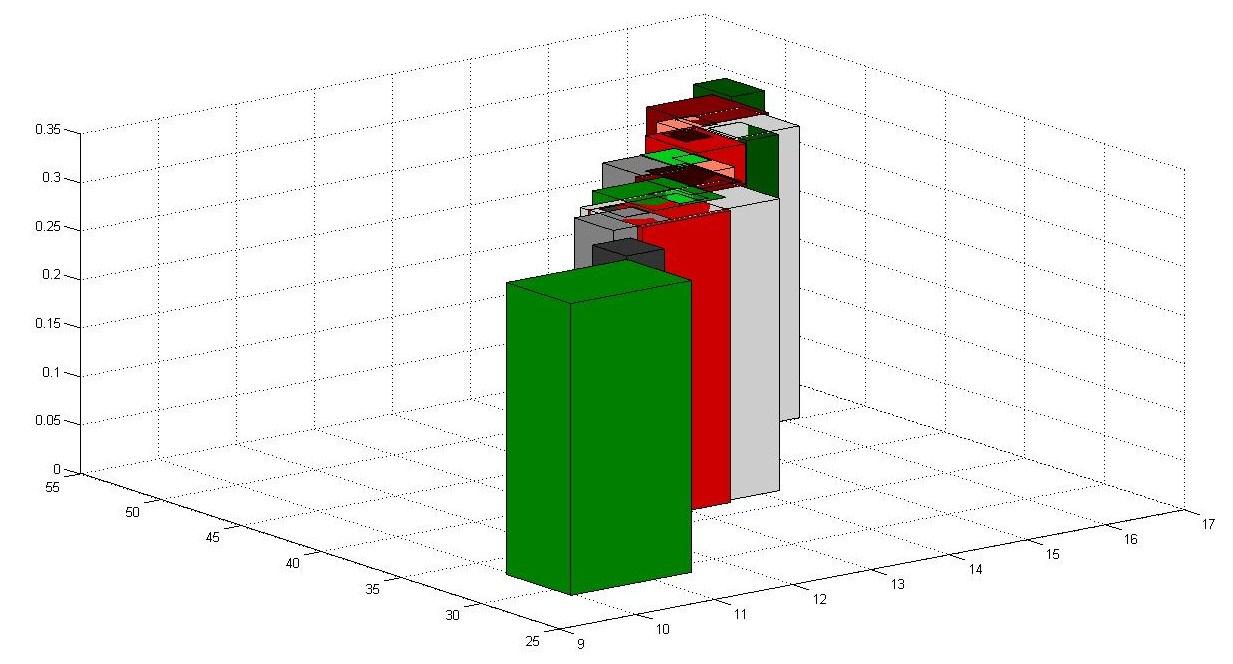}
\caption{Scatter plot of the data in \textit{Table \ref{table1E1}}.}
\label{fig3E1}
\end{center}
\end{figure}

The relationship between the histogram-valued variables in \textit{Table \ref{table1E1}} may be visualized in the scatter plot for histograms in \textit{Figure \ref{fig3E1}}. In this graphic, each of the distributions is represented by a histogram with a different color. These graphics show that a strong linear relation between the histogram-valued variables hematocrit and hemoglobin is observed.

From \textit{Property \ref{pr3.2}} we may conclude that for the set of patients to which the data refers, the symbolic mean of hematocrit increases $\alpha-\beta=3.1470$ for each unit of increase of the symbolic mean of hemoglobin. As this value is positive we may consider that the relationship between the histogram-valued variables is direct.

For this example, we also predicted the hematocrit distributions using the linear regression models proposed by Billard and Diday \cite{bidi02} and Irpino and Verde \cite{irve12},\cite{irve10}.
The hematocrit distributions obtained by these methods are presented in \textit{Appendix D}.
To compare the performance of the methods, the measures  $RMSE_{M}, RMSE_{L}, RMSE_{U}$ (see \textit{Subsection \ref{ss4.1.2}}) were used (see \textit{Table \ref{table2E1}}).

\begin{table}[h!]
\begin{center}
{\scriptsize
\begin{tabular}{|c|c|c|c|}
  \hline
   Measure & {\textit{DSD Model}} & \textit{{Billard-Diday Model}} & \textit{{Verde-Irpino Model}} \\ \hline
  $RMSE_{L}$ & 0.8806 &  1.0288 & 0.9220 \\
  $RMSE_{U}$ & 0.8432 & 1.1064 & 0.8645 \\ \hline
  $RMSE_{M}$ & 0.8946 &  1.0507 & 0.9145 \\ \hline
\end{tabular}}
\caption{Comparison  of the performance between the \textit{DSD Model}, the \textit{Billard-Diday Model} and the \textit{Verde-Irpino Model}.}
\label{table2E1}
\end{center}
\end{table}

\subsubsection{Distributions of Crimes in USA } \label{ss4.2.3}

In this example we consider a real data table (microdata) \cite{basedados11} where we have records related with communities in the USA. The original data combines socio-economic data from the '90 Census and crime data from 1995. For this study we selected the response variable \textit{violent crimes} (total number of violent crimes per 100 000 habitants) and four explicative variables: $X_1$ (percentage of people aged 25 and over with less than 9th grade education);  $X_2$  (percentage of people aged 16 and over who are employed); $X_3$  (percentage of population who are divorced); $X_4$  (percentage of immigrants who immigrated within the last 10 years).
To build the symbolic data table we aggregated the information (contemporary aggregation) for each state. The units (higher units) of this study are the states of USA and their observations for each selected variable are the distributions of the records of the communities of the respective state. To build the initial data table we considered only the states for which the number of records for the variables selected was higher than thirty. Using this criterion, only twenty states were included (AL, CA, CT, FL, GA, IN, MA, MO, NC, NJ, NY, OH, OK, OR, PA, TN, TX, VA, WA, WI). Similarly to the simulation study, we consider, without loss of generality, that in all observations, the subintervals of each histogram have the same weight (equiprobable) with frequency $0.20.$ Furthermore as the response variable \textit{violent crimes} admits only positive values and the distributions of these values are asymmetric, we will consider as response histogram-valued variable, the variable $LVC$ whose observations are the distributions of the logarithm of the number of violent crimes for each USA state. Considering these conditions, the model that allows to predict the distribution of $LVC$ from the distributions of the explicative variables $X_1, X_2, X_3$ and $X_4,$ for each USA state $j$ is as follows:

\begin{center}
\begin{eqnarray}\label{eq1_ex2}
 \Psi_{\widehat{LVC}(j)}^{-1}(t) & = &  3.9321+0.0009\Psi_{X_1(j)}^{-1}(t)-0.0123\Psi_{X_2(j)}^{-1}(1-t)+\nonumber\\
   &&+0.2073\Psi_{X_3(j)}^{-1}(t)-0.0353\Psi_{X_3(j)}^{-1}(1-t) +0.0187\Psi_{X_4(j)}^{-1}(t)
\end{eqnarray}
\end{center}
with $t \in \left[0,1\right]$.
The goodness-of-fit measure associated to this model is $\Omega=0.87.$

The values of the parameters estimated for this situation allow to conclude that the variables $X_1, X_3$ and $X_4$ have a direct influence in the logarithm of the number of violent crimes and the percentage of employed people have an opposite effect. From \textit{Property \ref{pr3.2}} we may conclude that, for the set of states to which the data refer, when the symbolic mean of the percentage of population divorced increases $1\%$ and the other variables remain constant, the symbolic mean of the $LVC$ increases $0.1720.$ The percentage of divorced population is the one that influences the most the predicted histogram-valued variable. This interpretation can be extrapolated for the values of the associated parameter of all other explicative variables.

The advantage of studying a linear relationship between data with variability is the possibility to predict the distribution of the values of the response variable instead of only one real value as in a classical study. In this example, the predicted distribution of the logarithm of the number of violent crimes for a given state is more informative about the criminality in that state than only one descriptive measure (e.g., the mean).

 Consider one state that was not used to build the model, the state of Arkansas (AR). It is possible to predict the distribution of $LVC$ if the distributions of the explicative variables for this state are known. The histogram predicted by the \textit{DSD Model} (\ref{eq1_ex2}) for the state Arkansas is

 \begin{center}
\begin{eqnarray*}
 H_{LVC}(AR) &= &\left\{[4.2250,5.3158],0.2;[5.3158,5.8887],0.2;[5.8887,6.4802],0.2; \right. \\
 &&\left. [6.4802,7.0509],0.2;[7.0509,7.7913],0.2 \right\}
\end{eqnarray*}
\end{center}

\begin{figure}[h]
\begin{center}
\includegraphics[width=0.8\textwidth]{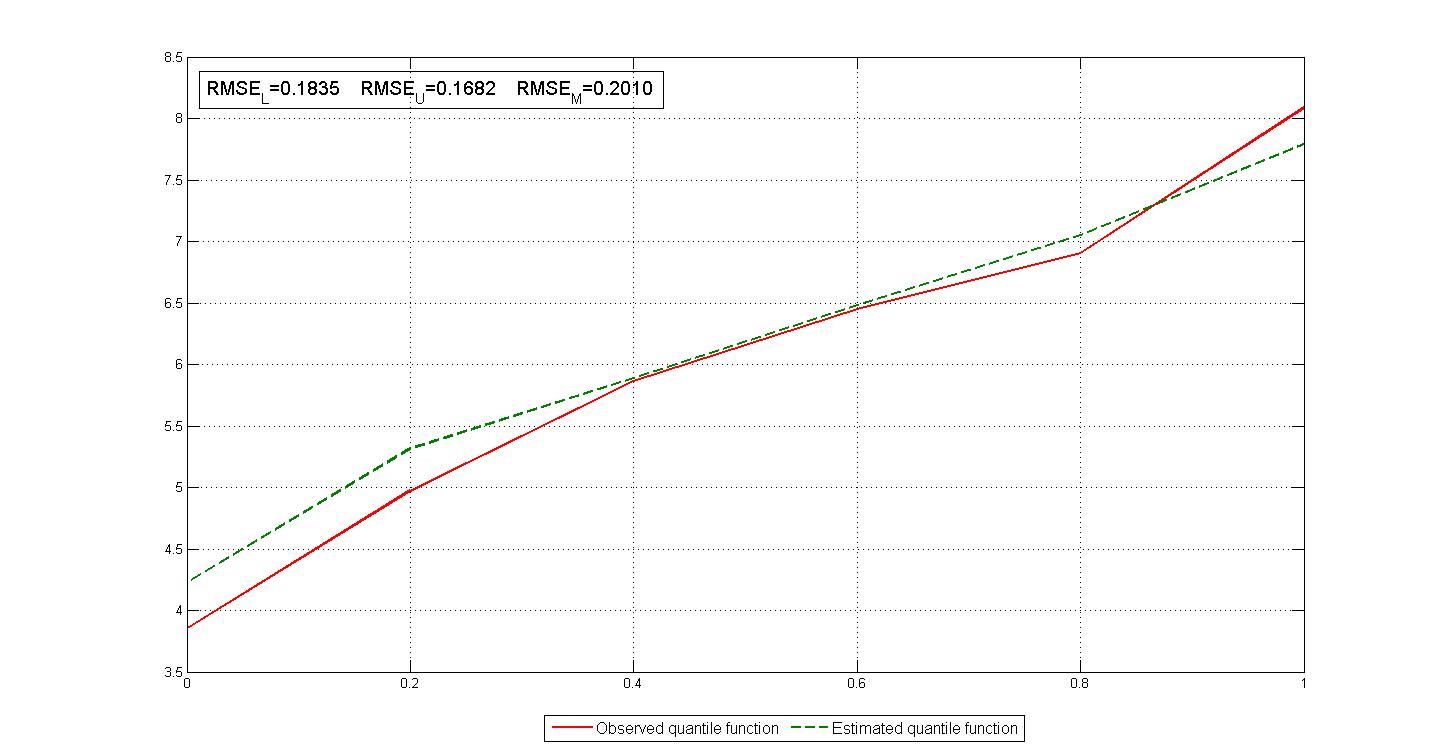}
\caption{Observed and estimated quantile function of the variable $LVC$ in the state of Arkansas}
\label{fig1E2}
\end{center}
\end{figure}

\textit{Figure \ref{fig1E2}} illustrates the estimated and observed quantile function for this state and the values of the measures $RMSE_{M}, RMSE_{L}, RMSE_{U}$ (see \textit{Subsection \ref{ss4.1.2}}). The values of the goodness-of-fit measures prove the closeness between the observed and estimated quantile function that we may see in the figure.

Analyzing the predicted distribution, we may conclude that in the state of Arkansas the estimated distribution tends to an uniform behavior with the values of $LVC$ to range between $4.23$ and $7.79.$

The classical alternative to study the logarithm of the number of violent crimes in each USA state would be to reduce the records of all communities of each state, for example to the mean value and make a classical linear regression study. In this case, the variability of the records would be lost and the predicted results would be less informative. Considering the mean of the records associated to each community, the classical model is the following:

\vspace{-0.2cm}

\begin{center}
\begin{equation}\label{eq2_ex2}
\widehat{\overline{LVC}}(j) = 6.5817+0.0705\overline{X}_1(j)-0.0503\overline{X}_2(j)+0.0933\overline{X}_3(j)+0.0177\overline{X}_4(j)
\end{equation}
\end{center}

For this model the value of $\texttt{R}^2=0.75.$

Considering again the state of Arkansas, with the previous model (\ref{eq2_ex2}) the estimative for  $\widehat{\overline{LVC}}(AR)$ is $6.4511.$  With this approach the information about the behavior of the predicted variable is obviously poorer.

\section{Conclusion and perspectives}

The \textit{DSD Model} allows predicting the distributions taken by one histogram-valued variable from the distributions taken by explicative histogram-valued variables. Moreover, it is possible to deduce a goodness-of-fit measure from the model. This measure appears to have a good behavior: when we compare the representation of the predicted and observed quantile functions for each unit we have good estimates when the value of the goodness-of-fit measure is close to one whereas the predicted and observed quantile functions are more discrepant when the value of the goodness-of-fit measure is lower.
As interval-valued variables are a particular case of histogram-valued variables it is possible to particularize this model for interval-valued variables.
An extension of the \textit{DSD Model}, where instead of a real number we use a quantile function as the independent parameter, is under development. This approach will be applied both to histogram-valued variables and interval-valued variables. With this new approach we expect to obtain a more flexible model.
Finally, and as a future research perspective, other models and methods in Symbolic Data Analysis based on linear relationships between variables may now be developed using this approach.

\vspace{1cm}

\textit{Acknowledgments}\\
This work is funded by the ERDF – European Regional Development Fund through the
COMPETE Programme (operational programme for competitiveness) and by National Funds through the
FCT – Funda\c{c}\~{a}o para a Ci\^{e}ncia e a Tecnologia (Portuguese Foundation for Science and Technology)
within project «FCOMP - 01-0124-FEDER-022701».

\pagebreak{}

\begin{landscape}
\appendix

\section*{\large Appendix A: First order partial derivatives of the function SE} \label{ap1}
\addcontentsline{toc}{chapter}{Appendix A: First order partial derivatives of the function SE}

$$SE = \displaystyle\sum\limits_{j=1}^m \displaystyle\sum\limits_{i=1}^n p_{i} \left[ \left( c_{Y(j)_i}-\displaystyle\sum\limits_{k=1}^p \left(\alpha_{k} c_{X_{k}(j)_{i}}- \beta_{k} c_{X_{k}(j)_{n-i+1}}\right)-\gamma \right)^{2} +\frac{1}{3} \left(r_{Y(j)_i}-\displaystyle\sum\limits_{k=1}^p \left(\alpha_{k} r_{X_{k}(j)_{
i}}+ \beta_{k} r_{X_{k}(j)_{n-i+1}}\right)\right)^{2} \right]$$

In these partial derivatives the subintervals of the histograms are defined from the center and half-range of the intervals.

{\scriptsize
$$\begin{array}{lll}
\frac{\partial SE (B)}{\partial \alpha_{\textbf{k}}} &=& \displaystyle\sum\limits_{j=1}^m \displaystyle\sum\limits_{i=1}^n p_{i} \left[2\left(c_{Y(j)_{i}}-\gamma-\displaystyle\sum\limits_{k=1}^p \left(\alpha_{k}c_{X_{k}(j)_{i}}+\beta_{k} \left(-c_{X_{k}(j)_{n-i+1}}\right)\right)\right)\left(-c_{X_{\textbf{k}}(j)_{i}}\right)+\frac{2}{3}\left(r_{Y(j)_{i}}-\displaystyle\sum\limits_{k=1}^p \left( \alpha_{k} r_{X_{k}(j)_{i}}+\beta_{k} r_{X_{k}(j)_{n-i+1}}\right)\right)\left(-r_{X_{\textbf{k}}(j)_{i}}\right)\right]=\\
&=&\displaystyle\sum\limits_{j=1}^m \displaystyle\sum\limits_{i=1}^n p_{i}\left(2 \displaystyle\sum\limits_{k=1}^p c_{X_{k}(j)_{i}}c_{X_{\textbf{k}}(j)_{i}}+\frac{2}{3} \displaystyle\sum\limits_{k=1}^p r_{X_{k}(j)_{i}}r_{X_{\textbf{k}}(j)_{i}}\right)\alpha_{k}+\displaystyle\sum\limits_{j=1}^m \displaystyle\sum\limits_{i=1}^n p_{i}\left(-2 \displaystyle\sum\limits_{k=1}^p c_{X_{k}(j)_{n-i+1}}c_{X_{\textbf{k}}(j)_{i}}+\frac{2}{3} \displaystyle\sum\limits_{k=1}^p r_{X_{k}(j)_{n-i+1}}r_{X_{\textbf{k}}(j)_{i}}\right)\beta_{k}+\\
 & & \qquad +\displaystyle\sum\limits_{j=1}^m \displaystyle\sum\limits_{i=1}^n 2p_{i}c_{X_{\textbf{k}}(j)_{i}}\gamma+\displaystyle\sum\limits_{j=1}^m \displaystyle\sum\limits_{i=1}^n p_{i}\left(-2c_{Y(j)_{i}}c_{X_{\textbf{k}}(j)_{i}}-\frac{2}{3}r_{Y(j)_{i}}r_{X_{\textbf{k}}(j)_{i}}\right)
\end{array}$$}

{\scriptsize
$$\begin{array}{lll}
\frac{\partial SE (B)}{\partial \beta_{\textbf{k}}} &=& \displaystyle\sum\limits_{j=1}^m \displaystyle\sum\limits_{i=1}^n p_{i} \left[2\left(c_{Y(j)_{i}}-\gamma-\displaystyle\sum\limits_{k=1}^p \left(\alpha_{k}c_{X_{k}(j)_{i}}+\beta_{k} \left(-c_{X_{k}(j)_{n-i+1}}\right)\right)\right)\left(c_{X_{\textbf{k}}(j)_{n-i+1}}\right)+\frac{2}{3}\left(r_{Y(j)_{i}}-\displaystyle\sum\limits_{k=1}^p \left( \alpha_{k} r_{X_{k}(j)_{i}}+\beta_{k} r_{X_{k}(j)_{n-i+1}}\right)\right)\left(-r_{X_{\textbf{k}}(j)_{n-i+1}}\right)\right]=\\
&=&\displaystyle\sum\limits_{j=1}^m \displaystyle\sum\limits_{i=1}^n p_{i}\left(-2 \displaystyle\sum\limits_{k=1}^p c_{X_{k}(j)_{i}}c_{X_{\textbf{k}}(j)_{n-i+1}}+\frac{2}{3} \displaystyle\sum\limits_{k=1}^p r_{X_{k}(j)_{i}}r_{X_{\textbf{k}}(j)_{n-i+1}}\right)\alpha_{k}+\displaystyle\sum\limits_{j=1}^m \displaystyle\sum\limits_{i=1}^n p_{i}\left(2 \displaystyle\sum\limits_{k=1}^p c_{X_{k}(j)_{n-i+1}}c_{X_{\textbf{k}}(j)_{n-i+1}}+\frac{2}{3} \displaystyle\sum\limits_{k=1}^p r_{X_{k}(j)_{n-i+1}}r_{X_{\textbf{k}}(j)_{n-i+1}}\right)\beta_{k}+\\
 & & \qquad +\displaystyle\sum\limits_{j=1}^m \displaystyle\sum\limits_{i=1}^n -2p_{i}c_{X_{\textbf{k}}(j)_{n-i+1}}\gamma+\displaystyle\sum\limits_{j=1}^m \displaystyle\sum\limits_{i=1}^n p_{i}\left(2c_{Y(j)_{i}}c_{X_{\textbf{k}}(j)_{n-i+1}}-\frac{2}{3}r_{Y(j)_{i}}r_{X_{\textbf{k}}(j)_{n-i+1}}\right)
\end{array}$$}

\vspace{1cm}

{\scriptsize
$\begin{array}{lll}
\frac{\partial SE (B)}{\partial \gamma} &=& \displaystyle\sum\limits_{j=1}^m \displaystyle\sum\limits_{i=1}^n p_{i} \left[-2\left(c_{Y(j)_{i}}-
\displaystyle\sum\limits_{k=1}^p \left(\alpha_{k}c_{X_{k}(j)_{i}}+\beta_{k} \left(-c_{X_{k}(j)_{n-i+1}}\right)\right)-\gamma\right)\right]=\\
&=&\displaystyle\sum\limits_{j=1}^m \displaystyle\sum\limits_{i=1}^n p_{i}\left(2 \displaystyle\sum\limits_{k=1}^p \alpha_{k}c_{X_{k}(j)_{i}}\right)+\displaystyle\sum\limits_{j=1}^m \displaystyle\sum\limits_{i=1}^n p_{i}\left(-2 \displaystyle\sum\limits_{k=1}^p \beta_{k}c_{X_{k}(j)_{n-i+1}}\right)+2m\gamma-\displaystyle\sum\limits_{j=1}^m \displaystyle\sum\limits_{i=1}^n p_{i} \left(2c_{Y(j)_{i}}\right)
\end{array}$}

\end{landscape}

\pagebreak
\begin{landscape}

\section*{\large Appendix B: Proof of Property \ref{pr3.5}.}\label{ap2}
\addcontentsline{toc}{chapter}{Appendix B: Proof of Property}

\vspace{1cm}

Defining the quantile functions $\Psi_{Y(j)}^{-1}(t)$ and $\Psi_{\widehat{Y}(j)}^{-1}(t)$ from the centers and half-ranges of the subintervals, according \ref{eq2.1B}, in \textit{Section \ref{ss2.1}} we have,

{\scriptsize
$$
\begin{array}{l}
\displaystyle\sum\limits_{j=1}^m \int_{0}^{1} \left(\Psi_{Y(j)}^{-1}(t)-\Psi_{\widehat{Y}(j)}^{-1}(t)\right)\left(\Psi_{\widehat{Y}(j)}^{-1}(t)-\overline{Y}\right)dt=\\
=\displaystyle\sum\limits_{j=1}^m \displaystyle\sum\limits_{i=1}^n  \int_{w_{i-1}}^{w_{i}} \left[ c_{Y(j)_i}+\left(2\frac{t-w_{i-1}}{w_{i}-w_{i-1}}-1\right)r_{Y(j)_i}-c_{\widehat{Y}(j)_i}-\left(2\frac{t-w_{i-1}}{w_{i}-w_{i-1}}-1\right)r_{\widehat{Y}(j)_i}\right]
\left[c_{\widehat{Y}(j)_i}+\left(2\frac{t-w_{i-1}}{w_{i}-w_{i-1}}-1\right)r_{\widehat{Y}(j)_i}-\overline{Y}\right]dt=\\
=\displaystyle\sum\limits_{j=1}^m \displaystyle\sum\limits_{i=1}^n  \int_{w_{i-1}}^{w_{i}} \left[\left(c_{Y(j)_i}-c_{\widehat{Y}(j)_i}\right)+\left(r_{Y(j)_i}-r_{\widehat{Y}(j)_i}\right)\left(2\frac{t-w_{i-1}}{w_{i}-w_{i-1}}-1\right)\right]\left[\left(c_{\widehat{Y}(j)_i}
-\overline{Y}\right)+r_{\hat{Y}(j)_i}\left(2\frac{t-w_{i-1}}{w_{i}-w_{i-1}}-1\right)\right]dt=\\
=\displaystyle\sum\limits_{j=1}^m \displaystyle\sum\limits_{i=1}^n  \int_{w_{i-1}}^{w_{i}} \left(c_{Y(j)_i}-c_{\widehat{Y}(j)_i}\right)\left(c_{\widehat{Y}(j)_i}-\overline{Y}\right)
+\left(\left(c_{Y(j)_i}-c_{\widehat{Y}(j)_i}\right)r_{\widehat{Y}(j)_i}+\left(r_{Y(j)_i}-r_{\widehat{Y}(j)_i}\right)\left(c_{\widehat{Y}(j)_i}
-\overline{Y}\right)\right)\left(2\frac{t-w_{i-1}}{w_{i}-w_{i-1}}-1\right)+\left(r_{Y(j)_i}-r_{\widehat{Y}(j)_i}\right)r_{\widehat{Y}(j)_i}\left(2\frac{t-w_{i-1}}{w_{i}-w_{i-1}}-1\right)^{2}dt\\
\end{array}
$$}

Solving the definite integral, after some algebra and considering $w_{i}-w_{i-1}=p_{i}$, we obtain,

{\small
$$
\begin{array}{l}
\displaystyle\sum\limits_{j=1}^m \displaystyle\sum\limits_{i=1}^n p_{i}\left[\left(c_{Y(j)_i}-c_{\widehat{Y}(j)_i}\right)c_{\widehat{Y}(j)_i}+\frac{1}{3}\left(r_{Y(j)_i}-r_{\widehat{Y}(j)_i}\right)r_{\widehat{Y}(j)_i}-
\left(c_{Y(j)_i}-c_{\widehat{Y}(j)_i}\right)\overline{Y}\right]
\end{array}
$$}

For the \textit{equation \ref{eq3.8}} in \textit{Section \ref{ss3.2}}, $c_{\widehat{Y}(j)_i}=\displaystyle\sum\limits_{k=1}^p \alpha_{k}^{*}c_{X_{k}(j)_i}-\beta_{k}^{*}c_{X_{k}(j)_{n-i+1}}+\gamma^{*};$  $r_{\widehat{Y}(j)_i}=\displaystyle\sum\limits_{k=1}^p \alpha_{k}^{*}r_{X_{k}(j)_i}+\beta_{k}^{*}r_{X_{k}(j)_{n-i+1}}$ and for the \textit{Property \ref{pr3.3}} also in \textit{Section \ref{ss3.2}} we have $\overline{Y}=\overline{\widehat{Y}},$ so

\newpage

{\scriptsize
$$
\begin{array}{l}
\displaystyle\sum\limits_{j=1}^m \displaystyle\sum\limits_{i=1}^n p_{i}\left[\left(c_{Y(j)_i}-c_{\widehat{Y}(j)_i}\right)\left(\displaystyle\sum\limits_{k=1}^p \alpha_{k}^{*}c_{X_{k}(j)_{i}}-\beta_{k}^{*}c_{X_{k}(j)_{n-i+1}}+\gamma^{*}\right)+\frac{1}{3}\left(r_{Y(j)_i}-r_{\widehat{Y}(j)_i}\right)\left(\displaystyle\sum\limits_{k=1}^p \alpha_{k}^{*}r_{X(j)_i}+\beta_{k}^{*}r_{X(j)_{n-i+1}}\right)-\left(c_{Y(j)_i}-c_{\widehat{Y}(j)_i}\right)\overline{\widehat{Y}}\right]=\\
=\displaystyle\sum\limits_{j=1}^m \displaystyle\sum\limits_{i=1}^n p_{i}\left[\left(c_{Y(j)_i}-c_{\widehat{Y}(j)_i}\right)\displaystyle\sum\limits_{k=1}^p \alpha_{k}^{*}c_{X_{k}(j)_{i}}+\frac{1}{3}\left(r_{Y(j)_i}-r_{\widehat{Y}(j)_i}\right)\displaystyle\sum\limits_{k=1}^p \alpha_{k}^{*}r_{X_{k}(j)_{i}}\right]+\displaystyle\sum\limits_{j=1}^m \displaystyle\sum\limits_{i=1}^n p_{i}\left[-\left(c_{Y(j)_i}-c_{\widehat{Y}(j)_i}\right)\displaystyle\sum\limits_{k=1}^p \beta_{k}^{*}c_{X_{k}(j)_{n-i+1}}+\frac{1}{3}\left(r_{Y(j)_i}-r_{\widehat{Y}(j)_i}\right)\displaystyle\sum\limits_{k=1}^p \beta_{k}^{*}r_{X_{k}(j)_{n-i+1}}\right]+\\
\qquad+\displaystyle\sum\limits_{j=1}^m \displaystyle\sum\limits_{i=1}^n p_{i}\left(c_{Y(j)_i}-c_{\widehat{Y}(j)_i}\right)\left(\gamma^{*}-\overline{\widehat{Y}}
\right)=\\
=\displaystyle\sum\limits_{j=1}^m \displaystyle\sum\limits_{i=1}^n \alpha_{1}^{*}\left[p_{i}\left(c_{Y(j)_i}-c_{\widehat{Y}(j)_i}\right)c_{X_{1}(j)_{i}}+\frac{1}{3}\left(r_{Y(j)_i}-r_{\widehat{Y}(j)_i}\right)r_{X_{1}(j)_{i}}\right]+\ldots+\alpha_{p}^{*}\left[p_{i}\left(c_{Y(j)_i}-c_{\widehat{Y}(j)_i}\right)c_{X_{p}(j)_{i}}+\frac{1}{3}\left(r_{Y(j)_i}-r_{\widehat{Y}(j)_i}\right)r_{X_{p}(j)_{i}}\right]+\\
\qquad+\displaystyle\sum\limits_{j=1}^m \displaystyle\sum\limits_{i=1}^n \beta_{1}^{*}\left[p_{i}\left(c_{Y(j)_i}-c_{\widehat{Y}(j)_{i}}\right)\left(-c_{X_{1}(j)_{n-i+1}}\right)+\frac{1}{3}\left(r_{Y(j)_i}-r_{\widehat{Y}(j)_i}\right)r_{X_{1}(j)_{n-i+1}}\right]+\ldots+
\alpha_{p}^{*}\left[p_{i}\left(c_{Y(j)_i}-c_{\widehat{Y}(j)_i}\right)\left(-c_{X_{p}(j)_{n-i+1}}\right)+\frac{1}{3}\left(r_{Y(j)_i}-r_{\widehat{Y}(j)_i}\right)r_{X_{p}(j)_{n-i+1}}\right]-\\
\qquad-\displaystyle\sum\limits_{j=1}^m \displaystyle\sum\limits_{i=1}^n p_{i}\left(c_{Y(j)_i}-c_{\widehat{Y}(j)_i}\right)\left(\gamma^{*}-\overline{\widehat{Y}}
\right)
\end{array}
$$}

Comparing this expression with the partial derivatives of the function $SE$ (see \textit{Appendix A}) we may write

{\small
$$
\begin{array}{l}
\displaystyle\sum\limits_{j=1}^m \int_{0}^{1} \left(\Psi_{Y(j)}^{-1}(t)-\Psi_{\widehat{Y}(j)}^{-1}(t)\right)\left(\Psi_{\widehat{Y}(j)}^{-1}(t)-\overline{Y}\right)dt=
-\frac{1}{2}\displaystyle\sum\limits_{k=1}^p \alpha_{k}^{*}\frac{\partial SE (B^{*})}{\partial \alpha_{k}} -\frac{1}{2}\displaystyle\sum\limits_{k=1}^p \beta_{k}^{*} \frac{\partial SE (B^{*})}{\partial \beta_{k}}+\frac{1}{2}\displaystyle\sum\limits_{k=1}^p \frac{\partial SE (B^{*})}{\partial \gamma}\left(\gamma^{*}-\overline{\widehat{Y}}
\right)
\end{array}$$}

From the Kuhn Tucker conditions presented in \textit{Section \ref{ss3.3}}, we have $\frac{\partial SE (B^{*})}{\partial \gamma}=0;$ $\frac{\partial SE (B^{*})}{\partial \alpha_{k}}\alpha_{k}^{*}=0$ and $\frac{\partial SE (B^{*})}{\partial \beta_{k}}\beta_{k}^{*}=0$ for all $k \in \left\{1,\ldots,p\right\}$ and $B^{*}=\left[ \alpha_{1}^{*} \quad \beta_{1}^{*} \quad \alpha_{2}^{*} \quad \beta_{2}^{*} \quad \cdots  \quad \alpha_{n}^{*} \quad \beta_{n}^{*} \quad \gamma^{*} \right]^{T}.$ So, $$\displaystyle\sum\limits_{j=1}^m \int_{0}^{1} \left(\Psi_{Y(j)}^{-1}(t)-\Psi_{\widehat{Y}(j)}^{-1}(t)\right)\left(\Psi_{\widehat{Y}(j)}^{-1}(t)-\overline{Y}\right)dt=0. \qquad \Box$$

\pagebreak

\end{landscape}

\pagebreak
\begin{landscape}

\section*{\large Appendix C: Simulation results of the study presented in \textit{Subsection \ref{ss4.1}}.} \label{ap3}
\addcontentsline{toc}{chapter}{Appendix C: Simulation results of the study presented in \textit{Subsection \ref{ss4.1}}.}

\renewcommand{\arraystretch}{1.25}
\begin{table}[h!]
\begin{center}
{\tiny
\begin{tabular}{|p{1.1cm}|p{1cm}|c|c|c|c|c|c|c|c|c|c|c|}
  \hline
   \multirow{2} {1.1cm} {\textit{Parameters}} & \multirow{2} {1cm} {\textit{Degree of linearity }} & \multirow{2} {*} {$n$} & \multicolumn{6}{|c|}{\textit{Parameters estimated}} & \multicolumn{4}{|c|}{\textit{Goodness of fit measures}} \\ \cline{4-13}
      & & & $\overline{\widehat{\alpha}} \, (s)$ & MSE($\alpha$)  & $\overline{\widehat{\beta}} \, (s)$  & MSE($\beta$)&  $\overline{\widehat{\gamma}} \, (s)$ & MSE($\gamma$)&  $\overline{\Omega} \, (s)$ & $\overline{RMSE}_{M} \, (s)$ & $\overline{RMSE}_{L} \, (s)$ & $\overline{RMSE}_{U} \, (s)$ \\ \hline
\multirow{12} {1.1cm}{$\alpha=2$ $\beta=1$ $\gamma=-1$} & \multirow{4} {1cm}{High linearity} & 10 & $1.9986 \, (0.1734)$ & 0.0300 & $1.0000 \, (0.1740)$ & 0.0303 & $-0.9955 \, (0.1208)$ &0.0146 & $0.9741 \, (0.0089)$ & $0.3450 \, (0.0628)$ & $0.2985 \, (0.0585)$ & $0.3001 \, (0.0592)$ \\
                                   &                                 & 30 & $2.0033 \, (0.0881)$ & 0.0078 & $0.9975 \, (0.0871)$ & 0.0076 &$-0.9999 \, (0.0824)$ &0.0068 &  $0.9633 \, (0.0065)$ & $0.3977 \, (0.0362)$ & $0.3448 \, (0.0367)$ & $0.3464 \, (0.0372)$\\
                                   &                                 & 100 & $1.9995 \, (0.0460)$ & 0.0021& $1.0008 \, (0.0464)$ & 0.0021 &$-0.9994 \, (0.0354)$ & 0.0013  &$0.9648 \, (0.0032)$   & $0.3645 \, (0.0172)$ & $0.3176 \, (0.0176)$ & $0.3191 \, (0.0179)$\\
                                   &                                 & 250 & $2.0000 \, (0.0274)$ &7.4928E-4 & $1.0000 \, (0.0273)$ & 7.4639E-4& $-0.9995 \, (0.0232)$ & 5.3692E-4   &$0.9640 \, (0.0020)$  & $0.3604 \, (0.0105)$ & $0.3145 \, (0.0110)$ & $0.3160 \, (0.0111)$ \\
\hline
\end{tabular}}
\caption{Results of the \textit{DSD Model} when the histogram values of $X$ are generated from real values with an Uniform distribution.}
\label{table1SA3}
\end{center}
\end{table}

\end{landscape}

\newpage

\begin{landscape}

\renewcommand{\arraystretch}{1.25}
\begin{table}[h!]
\begin{center}
{\tiny
\begin{tabular}{|p{1.1cm}|p{1cm}|c|c|c|c|c|c|c|c|c|c|c|}
  \hline
   \multirow{2} {*} {\textit{Parameters}} & \multirow{2} {1cm} {\textit{Degree of linearity }} & \multirow{2} {*} {$n$} & \multicolumn{6}{|c|}{\textit{Parameters estimated}} & \multicolumn{4}{|c|}{\textit{Goodness of fit measures}} \\ \cline{4-13}
      & & & $\overline{\widehat{\alpha}} \, (s)$ & MSE($\alpha$)  & $\overline{\widehat{\beta}} \, (s)$  & MSE($\beta$)&  $\overline{\widehat{\gamma}} \, (s)$ & MSE($\gamma$)&  $\overline{\Omega} \, (s)$ & $\overline{RMSE}_{M} \, (s)$ & $\overline{RMSE}_{L} \, (s)$ & $\overline{RMSE}_{U} \, (s)$ \\ \hline
\multirow{12} {1.1cm}{$\alpha=2$ $\beta=1$ $\gamma=-1$} & \multirow{4} {1cm}{High linearity} & 10 &  $2.0178 \, \, (0.3322)$  & 0.1106 &  $0.9834 \, \, (0.3333)$  & 0.1112 &  $-1.0148 \, \, (0.3809)$  & 0.1451 &  $0.9792 \, (0.0079)$  & $0.5101 \, (0.1002)$ & $0.4470 \, (0.0879)$ & $0.4602 \, (0.0918)$\\
                                   &                                 & 30 &  $2.0064 \, (0.1967)$  & 0.0387 &  $0.9931 \, (0.1967)$  & 0.0387 & $-1.0113 \, (0.2222)$  & 0.0486 &  $0.9757 \, (0.0047)$  & $0.6738 \, (0.0669)$ & $0.5868 \, (0.0603)$ & $0.6051 \, (0.0638)$\\
                                   &                                 & 100 &  $1.9977 \, (0.1116)$  & 0.0124 &  $1.0021 \, (0.1118)$  & 0.0125 & $-0.9980 \, (0.1298)$  & 0.0168 &  $0.9762 \, (0.0025)$  & $0.6720 \, (0.0359)$ & $0.5819 \, (0.0322)$ &$0.6002 \, (0.0341)$\\
                                   &                                 & 250 &  $1.9958 \, (0.0728)$  & 0.0053 &  $1.0046 \, (0.0725)$  & 0.0053 & $-0.9948 \, (0.0874)$  & 0.0077 &  $0.9765 \, (0.0016)$  & $0.6635 \, (0.0227)$ & $0.5744 \, (0.0204)$ & $0.5931 \, (0.0219)$ \\
\hline
\end{tabular}}
\caption{Results of the \textit{DSD Model} when the histogram values of $X$ are generated from real values with a Normal distribution.}
\label{table2SA3}
\end{center}
\end{table}

\end{landscape}

\newpage

\begin{landscape}

\renewcommand{\arraystretch}{1.25}
\begin{table}[h!]
\begin{center}
{\tiny
\begin{tabular}{|p{1.1cm}|p{1cm}|c|c|c|c|c|c|c|c|c|c|c|}
  \hline
   \multirow{2} {*} {\textit{Parameters}} & \multirow{2} {1cm} {\textit{Degree of linearity }} & \multirow{2} {*} {$n$} & \multicolumn{6}{|c|}{\textit{Parameters estimated}} & \multicolumn{4}{|c|}{\textit{Goodness of fit measures}} \\ \cline{4-13}
      & & & $\overline{\widehat{\alpha}} \, (s)$ & MSE($\alpha$)  & $\overline{\widehat{\beta}} \, (s)$  & MSE($\beta$)&  $\overline{\widehat{\gamma}} \, (s)$ & MSE($\gamma$)&  $\overline{\Omega} \, (s)$ & $\overline{RMSE}_{M} \, (s)$ & $\overline{RMSE}_{L} \, (s)$ & $\overline{RMSE}_{U} \, (s)$ \\ \hline
\multirow{12} {1.1cm}{$\alpha=2$ $\beta=1$ $\gamma=-1$} & \multirow{4} {1cm}{High linearity} & 10 &  $1.9998 \, (0.0581)$ & 0.0034 & $1.0002 \, (0.0229)$ & 5.2466E-4 & $-1.0104 \, (0.5485)$ &0.3007 & $0.9843 \, (0.0054)$ & $1.7361 \, (0.3009)$ & $1.4034 \, (0.2684)$ & $1.7717 \, (0.2920)$\\
                                   &                                 & 30 & $1.9979 \, (0.0338)$  & 0.0011 & $1.0006 \, (0.0141)$ & 1.9756E-4 & $-0.9938 \, (0.2260)$ & 0.0511 & $0.9821 \, (0.0034)$ & $1.2068 \, (0.1155)$ & $0.9828 \, (0.1020)$ & $1.2178 \, (0.1113)$\\
                                   &                                 & 100 & $2.0001 \, (0.0233)$ & 5.4282E-4 & $1.0002 \, (0.0080)$ & 6.4532E-5 & $-1.0004 \, (0.1154)$ & 0.0133 & $0.9822 \, (0.0019)$ & $1.1460 \, (0.0615)$ & $0.9202 \, (0.0527)$ & $1.1344 \, (0.0561)$\\
                                   &                                 & 250 & $2.0003 \, (0.0168)$ & 2.8347E-4 & $0.9998 \, (0.0050)$ & 2.5440E-5 & $-1.0005 \, (0.0813)$ & 0.0066 & $0.9828 \, (0.0012)$ & $1.3450 \, (0.0472)$ & $1.0631 \, (0.0384)$ & $1.3249 \, (0.0418)$\\ \cline{2-13}
                                   & \multirow{4} {1cm} {Moderate linearity} & 10 & $2.0100 \, (0.2298)$ & 0.0528 & $1.0000 \, (0.0966)$ & 0.0093 & $-1.0167 \, (2.1989)$ & 4.8305 & $0.8032 \, (0.0587)$ & $6.8729 \, (1.1688)$ & $5.5639 \, (1.0222)$ & $7.0662 \, (1.1728)$\\
\hline
\end{tabular}}
\caption{Results of the \textit{DSD Model} when the histogram values of $X$ are generated from real values with a Log-Normal distribution.}
\label{table3SA3}
\end{center}
\end{table}

\end{landscape}

\newpage

\begin{landscape}
\renewcommand{\arraystretch}{1.25}
\begin{table}[h!]
\begin{center}
{\tiny
\begin{tabular}{|p{1.1cm}|p{1cm}|c|c|c|c|c|c|c|c|c|c|c|}
  \hline
   \multirow{2} {*} {\textit{Parameters}} & \multirow{2} {1cm} {\textit{Degree of linearity }} & \multirow{2} {*} {$n$} & \multicolumn{6}{|c|}{\textit{Parameters estimated}} & \multicolumn{4}{|c|}{\textit{Goodness of fit measures}} \\ \cline{4-13}
      & & & $\overline{\widehat{\alpha}} \, (s)$ & MSE($\alpha$)  & $\overline{\widehat{\beta}} \, (s)$  & MSE($\beta$)&  $\overline{\widehat{\gamma}} \, (s)$ & MSE($\gamma$)&  $\overline{\Omega} \, (s)$ & $\overline{RMSE}_{M} \, (s)$ & $\overline{RMSE}_{L} \, (s)$ & $\overline{RMSE}_{U} \, (s)$ \\ \hline
\multirow{12} {1.1cm}{$\alpha=2$ $\beta=1$ $\gamma=-1$} & \multirow{4} {1cm}{High linearity} & 10 &  $1.9956 \, (0.0890)$ & 0.0079 &  $1.0031 \, (0.0835)$ & 0.0070 &  $-0.9941 \, (0.2387)$ & 0.0569 &  $0.9780 \, (0.0078)$ & $0.5395 \, (0.1001)$ & $0.4720 \, (0.0908)$ & $0.4909 \, (0.0945)$\\
                                   &                                 & 30 &  $2.0000 \, (0.0411)$ & 0.0017 &  $1.0009 \, (0.0303)$ & 9.1592E-4 &  $-0.9925 \, (0.1454)$ & 0.0212 &  $0.9768 \, (0.0044)$ & $0.7348 \, (0.0716)$ & $0.6308 \, (0.0663)$ & $0.6854 \, (0.0688)$\\
                                   &                                 & 100 &  $1.995 \, (0.0347)$ & 0.0012 &  $1.0002 \, (0.0337)$ & 0.0011 &  $-0.9991 \, (0.0967)$ & 0.0093 &  $0.9719 \, (0.0026)$ & $0.5956 \, (0.0281)$ & $0.5175 \, (0.0269)$ & $0.5426 \, (0.0280)$\\
                                   &                                 & 250 &  $1.9998 \, (0.0191)$ & 3.6369E-4 &  $1.0003 \, (0.0169)$ & 2.8659E-4 &  $-0.9985 \, (0.0502)$ & 0.0025 &  $0.9736 \, (0.0016)$ & $0.6981 \, (0.0219)$ & $0.6030 \, (0.0206)$ & $0.6334 \, (0.0216)$\\
\hline
\end{tabular}}
\caption{Results of the \textit{DSD Model} when the histogram values of $X$ are generated from real values from a mixture of different distributions.}
\label{table4SA3}
\end{center}
\end{table}

\end{landscape}

\newpage

\begin{landscape}
\renewcommand{\arraystretch}{1.2}
\begin{table}[h!]
\begin{center}
{\tiny
\begin{tabular}{|p{1.8cm}|p{0.8cm}|p{0.25cm}|c|c|c|c|c|c|c|c|c|c|c|c|}
  \hline
   \multirow{2} {1.8cm} {\textit{Distributions in variables $X_1;$ $X_2;$ $X_3$}}  & \multirow{2} {0.8cm} {\textit{Degree of linearity }} & \multirow{2} {0.25cm} {$n$}&  \multicolumn{12}{|c|}{\textit{Parameters estimated}} \\ \cline{4-15}
      & & & $\overline{\widehat{\alpha_1}} \, (s)$ & MSE($\alpha_1$)  & $\overline{\widehat{\beta_1}} \, (s)$  & MSE($\beta_1$)& $\overline{\widehat{\alpha_2}} \, (s)$ & MSE($\alpha_2$)  & $\overline{\widehat{\beta_2}} \, (s)$  & MSE($\beta_2$)& $\overline{\widehat{\alpha_3}} \, (s)$ & MSE($\alpha_3$)  & $\overline{\widehat{\beta_3}} \, (s)$  & MSE($\beta_3$)\\ \hline
\multirow{12} {1.8cm}{Uniform} & \multirow{4} {0.8cm}{High linearity} & 10 &  $1.9790 \, (0.4625)$ & 0.2141 & $0.9664 \, (0.4621)$ & 0.2145 & $0.5375 \, (0.4958)$ & 0.2469 & $3.0505 \, (0.6273)$ & 0.3957 & $3.9859 \, (0.6573)$ & 0.4319 & $1.9784 \, (0.6678)$ & 0.4460 \\
                                   &                                 & 30 & $2.0005 \, (0.3101)$& 0.0961 & $0.9978 \, (0.3096)$ & 0.0958 & $0.4965 \, (0.2887)$ & 0.0833 & $3.0031 \, (0.3088)$ & 0.0953 & $4.0013 \, (0.3722)$ & 0.1384 & $1.9978 \, (0.3676)$ & 0.1350 \\
                                   &                                 & 100 & $1.9930 \, (0.1946)$ & 0.0379 & $1.0054 \, (0.1919)$ & 0.0368 & $0.4935 \, (0.1917)$ & 0.0367 & $3.0049 \, (0.1929)$ & 0.0372 & $3.9966 \, (0.1677)$ & 0.0281 & $2.0063 \, (0.1676)$ & 0.0281 \\
                                   &                                 & 250 & $2.0017 \, (0.1085)$ & 0.0118 & $0.9981 \, (0.1099)$ & 0.0121 & $0.4985 \, (0.1066)$ & 0.0114 & $3.0012 \, (0.1085)$ & 0.0118 & $3.9916 \, (0.1126)$ & 0.0127 & $2.0096 \, (0.1126)$ & 0.0127\\
\hline
\end{tabular}}
\caption{{\footnotesize  Results of the \textit{DSD Model} $\Psi_{\widehat{Y}(j)}^{-1}(t)=-1+2\Psi_{X_{1}(j)}^{-1}(t)-1\Psi_{X_{1}(j)}^{-1}(1-t)+0.5\Psi_{X_{2}(j)}^{-1}(t)-3\Psi_{X_{2}(j)}^{-1}(1-t)+4\Psi_{X_{3}(j)}^{-1}(t)-2\Psi_{X_{3}(j)}^{-1}(1-t)$ in different conditions.}}
\label{table5SA3}
\end{center}
\end{table}

\end{landscape}

\newpage

\begin{landscape}
\renewcommand{\arraystretch}{1.2}
\begin{table}[h!]
\begin{center}
{\tiny
\begin{tabular}{|p{2cm}|p{2cm}|c|c|c|c|c|c|c|}
  \hline
   \multirow{2} {2cm} {\textit{Distributions in variables\\ $X_1;$ $X_2;$ $X_3$}}  & \multirow{2} {2cm} {\textit{Degree of linearity }} & \multirow{2} {*} {$n$}& \multicolumn{2}{|c|}{\textit{Parameter estimated }} &\multicolumn{4}{|c|}{\textit{Goodness-of-fit measures}} \\ \cline{4-9}
      & & & $\overline{\widehat{\gamma}} \, (s)$ & MSE($\gamma$) & $\overline{\Omega} \, (s)$ & $\overline{RMSE}_{M} \, (s)$ & $\overline{RMSE}_{L} \, (s)$ & $\overline{RMSE}_{U} \, (s)$ \\ \hline
\multirow{12} {*}{Uniform} & \multirow{4} {1cm}{High linearity} & 10 & $-0.9954 \, (0.4265)$ & 0.1817 & $0.9802 \, (0.0086)$ & $1.0224 \, (0.2335)$ & $0.8774 \, (0.2080)$ & $0.8810 \, (0.2104$\\
                                   &                                 & 30 & $-1.0036 \, (0.2676)$ & 0.0711 & $0.9713 \, (0.0054)$ & $1.3036 \, (0.1275)$ & $1.1242 \, (0.1248)$ & $1.1303 \, (0.1260$\\
                                   &                                 & 100 & $-1.0029 \, (0.1462)$ & 0.0214  & $0.9682 \, (0.0030)$ & $1.4041 \, (0.0690)$ & $1.2229 \, (0.0701)$ & $1.2293 \, (0.0713$ \\
                                   &                                 & 250 &  $-0.9981 \, (0.0900)$ & 0.0081  & $0.9673 \, (0.0019)$ & $1.3925 \, (0.0411)$ & $1.2150 \, (0.0415)$ & $1.2216 \, (0.0420$ \\ \cline{2-9}
                                   & \multirow{4} {1cm} {Moderate linearity} & 10 &  $-0.8303 \, (1.6879)$ & 2.8749  &$0.7755 \, (0.0760)$ & $4.1952 \, (0.8943)$ & $3.6459 \, (0.7989)$ & $3.6639 \, (0.8095)$\\
                                   &                                   & 30 & $-1.0236 \, (1.0221)$ & 1.0441   &$0.6907 \, (0.0445)$ & $5.2222 \, (0.5082)$& $4.5245 \, (0.4971)$ & $4.5509 \, (0.5021)$ \\
\hline
\end{tabular}}
\caption{\footnotesize Results of the \textit{DSD Model} $\Psi_{\widehat{Y}(j)}^{-1}(t)=-1+2\Psi_{X_{1}(j)}^{-1}(t)-1\Psi_{X_{1}(j)}^{-1}(1-t)+0.5\Psi_{X_{2}(j)}^{-1}(t)-3\Psi_{X_{2}(j)}^{-1}(1-t)+4\Psi_{X_{3}(j)}^{-1}(t)-2\Psi_{X_{3}(j)}^{-1}(1-t)$ in different conditions (continuation of the \textit{Table \ref{table5SA3}}).}
\label{table6SA3}
\end{center}
\end{table}

\end{landscape}

\section*{\large Appendix D: Observed and predicted histograms of the experiments presented in \textit{Subsection \ref{ss4.2.1}}.} \label{ap4}
\addcontentsline{toc}{chapter}{Appendix D: Observed and predicted histograms of the experiments presented in \textit{Subsection \ref{ss4.2.1}}.}

\vspace{1cm}

\textit{Histogram-valued variable Hematocrit in relation with the histogram-valued variable Hemoglobin}
\vspace{0.5cm}

In the example in \textit{Subsection \ref{ss4.2.1}} we performed a comparative study of the \textit{DSD Model} with other existing models. The results of the application of the models proposed by Billard-Diday \cite{bidi02} and Verde-Irpino \cite{irve10} to the data of this example may be found in \textit{Table \ref{table3E1A3}}.
\vspace{0.5cm}

\begin{table}[h!]
\begin{center}
\begin{tabular}{|c|c|}
  \hline
   \rowcolor[gray]{0.8}
 \textit{DSD Model} & $\Psi_{\widehat{Y}(j)}^{-1}(t)=-1.953+3.5598\Psi_{X(j)}^{-1}(t)-0.4128\Psi_{X(j)}^{-1}(1-t)$ \\ \hline
  \rowcolor[gray]{0.6}
 \textit{Billard-Diday Model} & $\widehat{\underline{I}}_{{Y(j)}_{i}}=-2.16+3.16\underline{I}_{{X(j)}_{i}}  \qquad  \widehat{\overline{I}}_{{Y(j)}_{i}}=-2.16+3.16\overline{I}_{{X(j)}_{i}} $  \\ \hline
  \rowcolor[gray]{0.4}
 \textit{Verde-Irpino Model} & $\Psi_{\widehat{Y}(j)}^{-1}(t)=-2.157+3.161\overline{X}(j)+3.918\left(\Psi_{X(j)}^{-1}(t)-\overline{X}(j)\right)$ \\ \hline
\end{tabular}
 \caption{Linear regression models applied to the data in \textit{Table \ref{table1E1}}.}
 \label{table3E1A3}
\end{center}
\end{table}

In \textit{Table \ref{table4E1A3}}, in the white rows we have the observed histograms of each observation of the histogram-valued variable $Y,$ in the light grey rows the histograms $H_{\widehat{Y}_{DSD}(j)}$ predicted using the \textit{DSD Model}, in the grey rows the histograms $H_{\widehat{Y}_{BD}(j)}$  predicted using the model proposed by Billard and Diday \cite{bidi02} and in the dark grey the histograms $H_{\widehat{Y}_{VI}(j)}$  predicting with the Verde and Irpino \cite{irve10}.
\newpage

\begin{table}[h!]
\begin{center}
{\tiny
\begin{tabular}{|c|l|}
  \hline
  Obs. & Distributions of the values of hematocrit  \\
  \hline
 $H_{Y(1)}$ &  $\left\{\left[33.29;35.41\right[,0.3;\left[35.41;36.11\right[,0.1;\left[36.11;36.82\right[,0.1;\left[36.82;37.52\right[,0.1;\left[37.52;38.04\right[,0.1;\left[38.04;39.61\right],0.3 \right\}$ \\
   \rowcolor[gray]{0.8}
   \hline
$H_{\widehat{Y}_{DSD}(1)}$ &  $\left\{\left[33.84;35.70\right[,0.3;\left[35.70;36.32\right[,0.1;\left[36.32;36.73\right[,0.1;\left[36.73;37.13\right[,0.1;\left[37.13;37.56\right[,0.1;\left[37.56;38.85\right],0.3 \right\}$ \\
 \rowcolor[gray]{0.6}
   \hline
$H_{\widehat{Y}_{BD}(1)}$ &  $\left\{\left[34.33;35.87\right[,0.3;\left[35.87;36.38\right[,0.1;\left[36.38;36.70\right[,0.1;\left[36.70;37.02\right[,0.1;\left[37.02;37.35\right[,0.1;\left[37.35;38.31\right],0.3 \right\}$ \\
\rowcolor[gray]{0.4}
      \hline
$H_{\widehat{Y}_{VI}(1)}$ &  $\left\{\left[33.79;35.7\right[,0.3;\left[35.7;36.34\right[,0.1;\left[36.34;36.73\right[,0.1;\left[36.73;37.13\right[,0.1;\left[37.13;37.53\right[,0.1;\left[37.53;38.73\right],0.3 \right\}$ \\
   \hline \hline
   $H_{Y(2)}$ & $\left\{\left[36.69;39.11\right[,0.3;\left[39.11;39.97\right[,0.1;\left[39.97;40.83\right[,0.1;\left[40.83;41.69\right[,0.1;\left[41.69;42.54\right[,0.1;\left[42.54;45.12\right],0.3 \right\}$\\
     \rowcolor[gray]{0.8}
   \hline
   $H_{\widehat{Y}_{DSD}(2)}$ & $\left\{\left[35.16;38.04\right[,0.3;\left[38.04;39.00\right[,0.1;\left[39.00;39.96\right[,0.1;\left[39.96;40.67\right[,0.1;\left[40.67;41.38\right[,0.1;\left[41.38;43.51\right],0.3 \right\}$\\
     \rowcolor[gray]{0.6}
   \hline
    $H_{\widehat{Y}_{BD}(2)}$ & $\left\{\left[36.00;38.37\right[,0.3;\left[38.37;39.16\right[,0.1;\left[39.16;39.95\right[,0.1;\left[39.95;40.49\right[,0.1;\left[40.49;41.03\right[,0.1;\left[41.03;42.64\right],0.3 \right\}$\\
     \rowcolor[gray]{0.4}
   \hline
    $H_{\widehat{Y}_{VI}(2)}$ & $\left\{\left[35.13;38.06\right[,0.3;\left[38.06;39.04\right[,0.1;\left[39.04;40.02\right[,0.1;\left[40.02;40.69\right[,0.1;\left[40.69;41.36\right[,0.1;\left[41.36;43.35\right],0.3 \right\}$\\
   \hline \hline
   $H_{Y(3)}$ & $\left\{\left[36.69;40.26\right[,0.3;\left[40.26;41.45\right[,0.1;\left[41.45;42.64\right[,0.1;\left[42.64;43.85\right[,0.1;\left[43.85;45.06\right[,0.1;\left[45.06;48.68\right],0.3 \right\}$\\
   \rowcolor[gray]{0.8}
   \hline
   $H_{\widehat{Y}_{DSD}(3)}$ &
 $\left\{\left[35.45;42.27\right[,0.3;\left[42.27;43.38\right[,0.1;\left[43.38;44.50\right[,0.1;\left[44.50;45.61\right[,0.1;\left[45.61;46.72\right[,0.1;\left[46.72;50.46\right],0.3 \right\}$\\
   \rowcolor[gray]{0.6}
   \hline
    $H_{\widehat{Y}_{BD}(3)}$ &
 $\left\{\left[36.98;42.74\right[,0.3;\left[42.74;43.62\right[,0.1;\left[43.62;44.51\right[,0.1;\left[44.51;45.39\right[,0.1;\left[45.39;46.28\right[,0.1;\left[46.28;48.93\right],0.3 \right\}$\\
 \rowcolor[gray]{0.4}
   \hline
    $H_{\widehat{Y}_{VI}(3)}$ &
 $\left\{\left[35.29	 42.42\right[,0.3;\left[42.42;43.51\right[,0.1;\left[43.51;44.61\right[,0.1;\left[44.61;45.71\right[,0.1;\left[45.71;46.80\right[,0.1;\left[46.80;50.1\right],0.3 \right\}$\\
   \hline \hline
   $H_{Y(4)}$ &  $\left\{\left[36.38;39.75\right[,0.3;\left[39.75;40.87\right[,0.1;\left[40.87;41.96\right[,0.1;\left[41.96;43.05\right[,0.1;\left[43.05;44.14\right[,0.1;\left[44.14;47.41\right],0.3 \right\}$\\
    \rowcolor[gray]{0.8}
   \hline
   $H_{\widehat{Y}_{DSD}(4)}$&  $\left\{\left[35.80;40.08\right[,0.3;\left[40.08;41.50\right[,0.1;\left[41.50;42.92\right[,0.1;\left[42.92;43.81\right[,0.1;\left[43.81;44.70\right[,0.1;\left[44.70;47.37\right],0.3 \right\}$\\
    \rowcolor[gray]{0.6}
   \hline
    $H_{\widehat{Y}_{BD}(4)}$&  $\left\{\left[36.98;40.55\right[,0.3;\left[40.55;41.74\right[,0.1;\left[41.74;42.93\right[,0.1;\left[42.93;43.58\right[,0.1;\left[43.58;44.23\right[,0.1;\left[44.23;46.18\right],0.3 \right\}$\\
    \rowcolor[gray]{0.4}
     \hline
    $H_{\widehat{Y}_{VI}(4)}$&  $\left\{\left[35.71	 40.13\right[,0.3;\left[40.13;41.61\right[,0.1;\left[41.61;43.08\right[,0.1;\left[43.08;43.89\right[,0.1;\left[43.89;44.69\right[,0.1;\left[44.69;47.12\right],0.3 \right\}$\\
   \hline \hline
   $H_{Y(5)}$ &  $\left\{\left[39.19;42.69\right[,0.3;\left[42.69;43.86\right[,0.1;\left[43.86;45.03\right[,0.1;\left[45.03;46.19\right[,0.1;\left[46.19;47.36\right[,0.1;\left[47.36;50.86\right],0.3 \right\}$ \\
    \rowcolor[gray]{0.8}
   \hline
    $H_{\widehat{Y}_{DSD}(5)}$ &  $\left\{\left[39.68;42.52\right[,0.3;\left[42.52;43.64\right[,0.1;\left[43.64;44.75\right[,0.1;\left[44.75;45.86\right[,0.1;\left[45.86;46.97\right[,0.1;\left[46.97;50.25\right],0.3 \right\}$ \\
   \rowcolor[gray]{0.6}
   \hline
    $H_{\widehat{Y}_{BD}(5)}$ &  $\left\{\left[40.78;42.99\right[,0.3;\left[42.99;43.87\right[,0.1;\left[43.87;44.76\right[,0.1;\left[44.76;45.64\right[,0.1;\left[45.64;46.53\right[,0.1;\left[46.53;49.19\right],0.3 \right\}$ \\
    \rowcolor[gray]{0.4}
   \hline
    $H_{\widehat{Y}_{VI}(5)}$ &  $\left\{\left[39.8	 42.54\right[,0.3;\left[42.54;43.64\right[,0.1;\left[43.64;44.74\right[,0.1;\left[44.74;45.83\right[,0.1;\left[45.83;46.93\right[,0.1;\left[46.93;50.22\right],0.3 \right\}$ \\
   \hline \hline
   $H_{Y(6)}$ & $\left\{\left[39.70;43.17\right[,0.3;\left[43.17;44.32\right[,0.1;\left[44.32;44.81\right[,0.1;\left[44.81;45.29\right[,0.1;\left[45.29;45.78\right[,0.1;\left[45.78;47.24\right],0.3 \right\}$\\
    \rowcolor[gray]{0.8}
   \hline
    $H_{\widehat{Y}_{DSD}(6)}$ & $\left\{\left[40.93;42.92\right[,0.3;\left[42.92;43.58\right[,0.1;\left[43.58;44.04\right[,0.1;\left[44.04;44.51\right[,0.1;\left[44.51;44.99\right[,0.1;\left[44.99;46.45\right],0.3 \right\}$\\
      \rowcolor[gray]{0.6}
   \hline
     $H_{\widehat{Y}_{BD}(6)}$ & $\left\{\left[41.50;43.14\right[,0.3;\left[43.14;43.68\right[,0.1;\left[43.68;44.05\right[,0.1;\left[44.05;44.42\right[,0.1;\left[44.42;44.79\right[,0.1;\left[44.79;45.90\right],0.3 \right\}$\\
     \rowcolor[gray]{0.4}
   \hline
      $H_{\widehat{Y}_{VI}(6)}$ & $\left\{\left[40.92	 42.95\right[,0.3;\left[42.95;43.62\right[,0.1;\left[43.62;44.08\right[,0.1;\left[44.08;44.54\right[,0.1;\left[44.54;44.99\right[,0.1;\left[44.99;46.47\right],0.3 \right\}$\\
   \hline \hline
   $H_{Y(7)}$ & $\left\{\left[41.56;44.11\right[,0.3;\left[44.11;44.95\right[,0.1;\left[44.95;45.80\right[,0.1;\left[45.80;46.65\right[,0.1;\left[46.65;47.19\right[,0.1;\left[47.19;48.81\right],0.3 \right\}$\\
    \rowcolor[gray]{0.8}
   \hline
   $H_{\widehat{Y}_{DSD}(7)}$ & $\left\{\left[42.67;43.86\right[,0.3;\left[43.86;44.26\right[,0.1;\left[44.26;44.65\right[,0.1;\left[44.65;45.22\right[,0.1;\left[45.22;45.78\right[,0.1;\left[45.78;47.48\right],0.3 \right\}$\\
    \rowcolor[gray]{0.6}
   \hline
   $H_{\widehat{Y}_{BD}(7)}$ & $\left\{\left[43.18;44.07\right[,0.3;\left[44.07;44.37\right[,0.1;\left[44.37;44.66\right[,0.1;\left[44.66;45.13\right[,0.1;\left[45.13;45.60\right[,0.1;\left[45.60;47.00\right],0.3 \right\}$\\
   \rowcolor[gray]{0.4}
   \hline
   $H_{\widehat{Y}_{VI}(7)}$ & $\left\{\left[42.76	43.87
\right[,0.3;\left[43.87;44.24\right[,0.1;\left[44.24;44.61\right[,0.1;\left[44.61;45.19\right[,0.1;\left[45.19;45.77\right[,0.1;\left[45.77;47.51\right],0.3 \right\}$\\
   \hline \hline
  $H_{Y(8)}$ &  $\left\{\left[38.4;40.34\right[,0.3;\left[40.34;40.99\right[,0.1;\left[40.99;41.64\right[,0.1;\left[41.64;42.28\right[,0.1;\left[42.28;42.93\right[,0.1;\left[42.93;45.22\right],0.3 \right\}$\\
   \rowcolor[gray]{0.8}
   \hline
   $H_{\widehat{Y}_{DSD}(8)}$ &  $\left\{\left[39.26;40.74\right[,0.3;\left[40.74;41.24\right[,0.1;\left[41.24;41.72\right[,0.1;\left[41.72;42.20\right[,0.1;\left[42.20;42.79\right[,0.1;\left[42.79;44.54\right],0.3 \right\}$\\
     \rowcolor[gray]{0.6}
   \hline
     $H_{\widehat{Y}_{BD}(8)}$ &  $\left\{\left[39.80;40.95\right[,0.3;\left[40.95;41.33\right[,0.1;\left[41.33;41.72\right[,0.1;\left[41.72;42.10\right[,0.1;\left[42.10;42.58\right[,0.1;\left[42.58;44.00\right],0.3 \right\}$\\
 \rowcolor[gray]{0.4}
      $H_{\widehat{Y}_{VI}(8)}$ &  $\left\{\left[39.31	 40.74\right[,0.3;\left[40.74;41.22\right[,0.1;\left[41.22;41.7\right[,0.1;\left[41.7;42.17\right[,0.1;\left[42.17;42.76\right[,0.1;\left[42.76;44.52\right],0.3 \right\}$\\
   \hline \hline
   $H_{Y(9)}$ &  $\left\{\left[28.83;32.86\right[,0.3;\left[32.86;34.21\right[,0.1;\left[34.21;35.55\right[,0.1;\left[35.55;36.84\right[,0.1;\left[36.84;38.12\right[,0.1;\left[38.12;41.98\right],0.3 \right\}$\\
    \rowcolor[gray]{0.8}
   \hline
   $H_{\widehat{Y}_{DSD}(9)}$ &
 $\left\{\left[27.66;33.54\right[,0.3;\left[33.54;35.50\right[,0.1;\left[35.50;36.70\right[,0.1;\left[36.70;37.91\right[,0.1;\left[37.91;39.20\right[,0.1;\left[39.20;43.08\right],0.3 \right\}$\\
  \rowcolor[gray]{0.6}
   \hline
   $H_{\widehat{Y}_{BD}(9)}$ &
 $\left\{\left[29.20;34.09\right[,0.3;\left[34.09;35.72\right[,0.1;\left[35.72;36.68\right[,0.1;\left[36.68;37.63\right[,0.1;\left[37.63;38.59\right[,0.1;\left[38.59;41.47\right],0.3 \right\}$\\
 \rowcolor[gray]{0.4}
  \hline
   $H_{\widehat{Y}_{VI}(9)}$ &
 $\left\{\left[27.54 33.59
\right[,0.3;\left[33.59;35.61\right[,0.1;\left[35.61;36.8\right[,0.1;\left[36.8;37.9\right[,0.1;\left[37.9;39.18\right[,0.1;\left[38.18;42.74\right],0.3 \right\}$\\
   \hline \hline
   $H_{Y(10)}$&  $\left\{\left[44.48;46.90\right[,0.3;\left[46.90;47.70\right[,0.1;\left[47.70;48.51\right[,0.1;\left[48.51;49.31\right[,0.1;\left[49.31;50.12\right[,0.1;\left[50.12;52.53\right],0.3 \right\}$\\
    \rowcolor[gray]{0.8}
   \hline
  $H_{\widehat{Y}_{DSD}(10)}$ &  $\left\{\left[45.85;47.48\right[,0.3;\left[47.48;48.03\right[,0.1;\left[48.03;48.58\right[,0.1;\left[48.58;49.13\right[,0.1;\left[49.13;49.68\right[,0.1;\left[49.68;51.33\right],0.3 \right\}$\\
   \rowcolor[gray]{0.6}
   \hline
  $H_{\widehat{Y}_{BD}(10)}$ &  $\left\{\left[46.43;47.73\right[,0.3;\left[47.73;48.17\right[,0.1;\left[48.17;48.61\right[,0.1;\left[48.61;49.05\right[,0.1;\left[49.05;49.48\right[,0.1;\left[49.48;50.80\right],0.3 \right\}$\\
  \rowcolor[gray]{0.4}
    \hline
  $H_{\widehat{Y}_{VI}(10)}$ &  $\left\{\left[45.91	47.51
\right[,0.3;\left[47.51;48.06\right[,0.1;\left[48.06;48.6\right[,0.1;\left[48.6;49.14\right[,0.1;\left[49.14;49.68\right[,0.1;\left[49.68;51.31\right],0.3 \right\}$\\
  \hline
\end{tabular}}
 \caption{Observed and predicted histograms (using three different methods) of the Hematocrit values for the data in \textit{Table \ref{table1E1}}.}
 \label{table4E1A3}
\end{center}
\end{table}

\end{document}